\documentclass[12pt,a4paper,fleqn,twoside]{report}

\newfont{\form}{cmss10}
\newcommand{\e}{\varepsilon}
\renewcommand{\b}{\beta}
\newcommand{\unity}{1\kern-.65mm \mbox{\form l}}%
\newcommand{\D}{D \raise0.5mm\hbox{\kern-2.0mm /}}
\newcommand{\A}{A \raise0.5mm\hbox{\kern-1.8mm /}}
\def\pmb#1{\leavevmode\setbox0=\hbox{$#1$}\kern-.025em\copy0\kern-\wd0
\kern-.05em\copy0\kern-\wd0\kern-.025em\raise.0433em\box0}

\def\D{\hbox{\hbox{${D}$}}\kern-1.9mm{\hbox{${/}$}}}
\def\kbar{\hbox{$k$}\kern-0.2truecm\hbox{$/$}}
\def\nbar{\hbox{$n$}\kern-0.23truecm\hbox{$/$}}
\def\pbar{\hbox{$p$}\kern-0.18truecm\hbox{$/$}}
\def\nhbar{\hbox{$\hat n$}\kern-0.23truecm\hbox{$/$}}
\newcommand{\dif}{\hspace{-1mm}{\rm d}}
\newcommand{\dil}[1]{{\rm Li}_2\left(#1\right)}
\newcommand{\diff}{{\rm d}}
\usepackage[dvips]{graphicx}
\usepackage{amsmath}
\begin{document}
\pagestyle{empty}
\pagenumbering{roman}                    
\setcounter{page}{1}

\begin{titlepage}

\begin{center}
\bigskip
\bigskip
{\bf \Huge Wilson Loops}\\[2cm]{\bf \Huge in Two-Dimensional}\\[2cm]{\bf \Huge Yang-Mills Theories}
\\[4cm]
{\Large Roberto Begliuomini \\
\bigskip
University of Trento (Italy)}\\[2cm]

\vskip 5 truecm

Thesis submitted for the Ph.D. Degree in Physics \\
at the Faculty of Science, University of Trento.\\
\smallskip
{January 1999}
\end{center}
\end{titlepage}

\newpage 
\
\tableofcontents

\newpage
\chapter*{Acknowledgements}
\addcontentsline{toc}{chapter}{Acknowledgements}

I want to thank all the people who have directly or indirectly contributed
to this work, in particular:
\begin{itemize}
\item my tutor and co--author Giuseppe Nardelli for many discussions and hints, and for proofreading the thesis, 
\item my co--author Prof. Antonio Bassetto for the fruitful collaboration and for proofreading the thesis,
\item Luca Griguolo and Dimitri Colferai, 
\item the Ph.D. Students at the Department of Physics in Trento,
\item Luisa Rossi Doria,
\item my family for their support and encouragement,
\item my graduation thesis advisor at Pisa University Mihail Mintchev,
\item the Italian Ministry of University for its financial support,
\item the Theoretical Group at Department of Physics in Trento University for warm hospitality and for financial support in these last four months,
\item INFN for financial support of my visit in Cambridge at the Summer School
{\it Confinement, Duality, and Nonperturbative Aspects of QCD}.
\item the University of Trento for financial support to cover partly house rent
expenses.

\end{itemize}

\newpage \ 
\newpage

\pagenumbering{arabic}
\pagestyle{myheadings}
\setcounter{page}{1}
\chapter*{Introduction}
\addcontentsline{toc}{chapter}{Introduction}
\markboth{Introduction}{Introduction}

Two--dimensional models have been an extraordinary laboratory to test ideas in quantum field theory (see for a review ref.~\cite{Abdalla}). 
The interest in studying two--dimensional theories is mainly due to 
the possibility of obtaining sometimes exact solutions, which are
believed (or at least hoped!) to share important features with the more realistic (but also much more difficult to face with) situation in four dimensions.

Schwinger's model (massless electrodynamics in two dimensions QED$_2$) 
is the key example,
which can be exactly solved, exhibiting very
interesting and peculiar properties, like fermion confinement,
theta--vacua and the presence of a non--vanishing chiral condensate.

QCD$_2$ is its non--Abelian generalization and has recently received
most attention in many investigations, also thanks to the discovery of
a string picture for it~\cite{gross3}. It is widely believed that
several phenomena that can be fairly easily understood in two
dimensions, could persist when dimensions are increased.

\smallskip
The facts exposed above are the motivations of our perturbative enquiries on pure Yang--Mills theories in $1+1$ dimensions, which are interesting at least for two reasons.

First, the reduction of the dimensions to $D=2$ entails tremendous
simplifications in the theory, so that several important problems can be faced
in this lower dimensional context and, sometimes, solved. We are thinking for instance to the exact evaluation of vacuum to vacuum amplitudes of Wilson loop
operators, that, for a suitable choice of contour and in some specific limit,
provide the potential between two static quarks. Another example is the
spectrum of the Bethe--Salpeter equation, when dynamical fermions are added to
the system.

The second reason, as we will see along this thesis, is that Yang--Mills theories in $D=2$ have several peculiar features that are interesting by their own:

\begin{itemize}

\item[a)] in $D=2$ within the same gauge choice (light-cone gauge) two
inequivalent perturbative formulations of the theory seem to coexist;

\item[b)] $D=2$ is a point of discontinuity for Yang--Mills theories;
this is an intriguing feature whose meaning has not been fully understood so far.

\end{itemize}

All the features we have listed are most conveniently studied
if the light--cone gauge is chosen. In such a gauge the Faddeev--Popov sector
decouples and the unphysical degrees of freedom content of the theory is
minimal. The price to be paid for these nice features is the presence of the so
called "spurious¾ poles in the vector propagator.
To handle the spurious pole at $n \cdot k=0$  is a
delicate matter; basically all difficulties encountered in the past
within the light--cone gauge quantization are related to this problem.
As we shall see, there are two possible, inequivalent, ways for handling the pole: the Cauchy Principal Value (CPV) prescription and the Mandelstam-Leibbrandt (ML) prescription. 

They result from different procedures of quantization. The CPV prescription is
derived in the so--called ``null--frame formalism'', but it is well defined only
in strictly two dimension. The two--dimensional case, for CPV prescription, is a sort of ``happy island'': everything works well, the confinement seems to show itself
already at a perturbative level, the bound--states equation in the large $N$ limit provides a discrete spectrum for mesons. But as soon as we try to escape
towards higher dimensions, we will see that the theory encounters a lot of inconsistencies.

Instead, the ML prescription derives from the equal--time quantization, and is more sincere: it does not provide any ``miracle'',  but it is consistent also in higher dimensions, where it is in
perfect agreement with the milestone represented by Feynman gauge
\footnote{Unfortunately Feynman propagator fails to
be a tempered distribution for $D=2$, then we will not be able to make a comparison with the covariant gauge in strictly two dimensions.}.
Nevertheless, also for the ML prescription, the two--dimensional case is peculiar: in fact, in approaching $D=2$, we will discover the
surprise of its discontinuity.

The coexistence of these two formulations for $D=2$ motivates our perturbative
enquiries, in order to understand their relationship and if there is any
physical reason to have a preference for one of them.

\smallskip
As a technical tool, we will use the Wilson loop, owing to its gauge 
invariance and to its
reasonable infrared properties. It is indeed well known that,
when approaching $D=2$, ultraviolet singularities are no longer
a concern, but wild infrared behaviors usually show up. Besides, this tool
will help us to reduce the computational effort: actually the graphs coming from the perturbative expansion of the Wilson loop, to a given order in the coupling constant, are much easier to compute than the ones of the corresponding $S$-matrix elements. The latter, furthermore, are not well defined
when are computed ``on--shell'', due to their infrared divergences.

\bigskip
This Ph.D. thesis reaches two main results. The first one is represented by a detailed study, in Feynman gauge, of the perturbative ${\cal O}(g^4)$
contribution to a space--time Wilson loop, with respect to its (expected) Abelian--like time exponentiation when the temporal side goes to infinity. As soon as we are
in dimensions greater than two, the expected behavior is found. But if we proceed first to the dimensional limit $D \to 2$, the exponentiation is not
recovered. The limits $T \to \infty$ and $D \to 2$ \emph{do not commute}.
This result has been reached in collaboration with
Antonio Bassetto and Giuseppe Nardelli~\cite{bello}.

\smallskip
The other result is the computation in dimensions $D=2+\epsilon$ and in light-cone gauge with ML prescription of the perturbative ${\cal O}(g^4)$
contribution to the same Wilson loop, coming
from diagrams with a self-energy correction in the vector propagator.
In the limit $\epsilon \to 0$ the result is finite, in spite of
the vanishing of the triple vector vertex in light--cone gauge, and provides
the expected agreement with the analogous calculation in Feynman
gauge.
Also this result has been reached in collaboration with Antonio Bassetto and Giuseppe Nardelli~\cite{bellopr}.

\bigskip
Let us conclude this introduction with a brief outline of the thesis.

\smallskip
In {\bf Chapter \ref{START}} we start reviewing 't Hooft model and its successes, and introduce the main ingredients for the formulation
of two-dimensional Yang--Mills theories, with particular emphasis on
the two different light--cone gauge formulations, CPV and ML. We end with a brief reminder
of Wilson loop, exposing its usefulness as a test of gauge invariance and its physical meaning.

\smallskip
{\bf Chapter \ref{LCR}} is a review of the results contained in
ref.~\cite{Bas1}, that
represents the starting point from which our research begins. A light--like
Wilson loop is computed in perturbation theory up to ${\cal O}(g^4)$ in
$1+1$ dimensions, using Feynman and light--cone gauges to check its gauge
invariance. After dimensional regularization in intermediate steps, a finite
gauge invariant result is obtained, which, however, does not exhibit Abelian
exponentiation. This result is at variance with the common belief that pure Yang--Mills is free in $1+1$ dimensions: only CPV formulation gives an Abelian--like result. Both the discontinuity of ML formulation and the inequivalence of ML and CPV formulations are explicitly shown. 

\smallskip
In {\bf Chapter \ref{XTR}} we report the results obtained from our research and exposed above, turning to study a space--time Wilson loop and reviewing also intermediate results contained in refs.~\cite{Bas7,Stau}.

\smallskip
In the {\bf Conclusions} we comment our results, comparing them to the available
non--perturbative results, and point out some open questions.

\smallskip
At the end, there are some technical appendices, in which we report the main
aspects of the computations leading to the results exposed in
Chapter~\ref{XTR}.

\newpage
\chapter{Preliminaries}
\label{START}

In this first chapter we review some facts
and tools that we will use throughout this thesis. In Section \ref{MODEL} we recall the basic features of `t Hooft model, namely QCD$_2$ in the large $N$ limit. This model is the most famous example of the simplifications that derive from lowering dimensions of the theory from four to two. In Section \ref{2DYMT} we will discuss the possible choices of gauge in treating Yang--Mills theories, with particular attention to their reliability according to the dimensions. We will distinguish between $D>2$ and $D=2$, where $D$ is the dimension in which the theory is defined. Finally, in Section \ref{LOOPS} we will recall the definition of the Wilson loop operator, explaining its main
features and its relevance in the study of Yang--Mills theories, in particular
as a tool to test gauge invariance and to cope with the confinement issue.

\section{`t Hooft Model}
\markboth{Preliminaries}{`t Hooft Model}
\label{MODEL}

\noindent
In 1974, G. `t Hooft proposed a very interesting model~\cite{hoof74} to describe
mesons, starting from a $SU(N)$ Yang--Mills theory in $1+1$ dimensions
in the large $N$ limit.

Quite remarkably in this model quarks look confined, while a discrete
set of quark--antiquark bound states emerges, with squared masses lying
on rising Regge trajectories.
The model is solvable thanks to the ``instantaneous'' character of the potential acting between quark and antiquark.

How does the mechanism of confinement work in 't Hooft model? Following
Coleman~\cite{Cole}, let us start from considering Schwinger model, namely
two--dimensional quantum electrodynamics.  The theory is defined by the following Lagrangian:

\begin{equation}
\mathcal{L}=\frac{1}{2} \left(F_{01}\right)^{2} + \bar{\psi} (i \partial_{\mu}\gamma^{\mu}
- e A_{\mu}\gamma^{\mu}- m)\psi,
\label{lagr}
\end{equation}

where

\begin{equation}
F_{01}=\partial_{0} A_{1} - \partial{_1} A_{0}.
\label{stten}
\end{equation}

If we choose axial gauge

\begin{equation}
\label{axgauge}
A_1=0, 
\end{equation}

the Lagrangian becomes

\begin{equation}
\mathcal{L}=\frac{1}{2} (\partial_1 A_0)^2 + \bar{\psi}
(i \partial_{\mu}\gamma^{\mu}
- e A_{0}\gamma^{0}- m)\psi.
\label{fixedlag}
\end{equation}

This Lagrangian is independent from time derivatives of $A$: then $A$ is not a dynamical variable, but a constrained variable, that obeys to:

\begin{equation}
\partial_{1}^{2}A_0=- e \psi^{\dagger} \psi\equiv -e j_0.
\label{costraint}
\end{equation}

We can eliminate $A_0$ from (\ref{fixedlag}), solving eq. (\ref{costraint}) and obtaining:

\begin{equation}
\mathcal{L}=\mathcal{L}_{0f} + \frac{e^2}{4}\int dx^1 dy^1 j_0(x^0,x^1)
\left|x^1 - y^1\right| j_0(x^0,y^1),
\label{linpot}
\end{equation}

where $\mathcal{L}_{0f}$ is the free fermion Lagrangian.

The interaction between charges is linear: this potential assures confinement, at least for small couplings, {\it i.e.} in the perturbative regime. The interaction between charges can be expressed as the effect of exchange of a photon propagator:

\begin{equation}
\begin{split}
D_{\mu \nu}(k)&=-\frac{i}{2}\delta_{\mu 0}\delta_{\nu 0}\int d^{2}x e^{i k \cdot x} \left|x^1\right|\delta(x^0)\\
&=i \delta_{\mu 0}\delta_{\nu 0} \textrm{P}\frac{1}{(k_1)^2},\\
\end{split}
\label{istprop}
\end{equation}

where P is the Cauchy principal--value symbol,

\begin{equation}
\label{regularized}
\textrm{P}\frac{1}{z^2}=\frac{1}{2}\left[\frac{1}{(z + i\epsilon)^2}
+ \frac{1}{(z - i\epsilon)^2}\right]=-{\partial \over \partial z} \left[
\textrm{P} \left( {1 \over z}\right)\right].
\end{equation}

The presence of $\delta(x^0)$ shows the instantaneous character of this interaction. The principal--value prescription for the pole violates causality, but in strictly two dimensions this is not a problem, because there are no
propagating degrees of freedom at all.

It is straightforward to generalize this simple model to chromodynamics. Considering the simplest case of one flavor, we can easily realize that the non linear terms in $F_{01}$ are proportional to the product of $A_0$ and $A_1$. 
Then in axial gauge the self--coupling terms of the gauge field vanish. The only difference from Schwinger model is the presence of color indices, $SU(N)$ being the color gauge group:

\begin{equation}
\mathcal{L}=\mathcal{L}_{0f} + \frac{g^2}{N}\int dx^1 dy^1 j^{b}_{0a}(x^0,x^1)
\left|x^1 - y^1\right| j^{a}_{0b}(x^0,x^1),
\end{equation}

where

\begin{equation}
j^{b}_{0a}=\psi^{\dagger}_{a} \psi^b -\frac{\delta^b_a}{N} \psi^{\dagger}_{c} \psi^c.
\end{equation}

The original `t Hooft work differs in some technical aspects from this pedagogical presentation. But the confinement is basically due to the reason exposed above.

First, `t Hooft chose light--cone gauge, namely he used light--cone coordinates:

\begin{equation}
x^{\pm}=\frac{\left(x^0 \pm x^1\right)}{\sqrt{2}}
\end{equation}

and put

\begin{equation}
A_{-}=A^{+}=0
\end{equation}

Second, in the original work of `t Hooft an infrared cut--off is used, instead of the principal value prescription.
But a quite remarkable feature of this theory
is that bound state wave functions and related eigenvalues
turn out to be cutoff independent. As a matter of fact in
ref.~\cite{call76}, it has been pointed out that the singularity
at $k_{-}=0$ of the propagator in light--cone gauge can also be regularized by a Cauchy principal
value prescription without finding differences in the resulting meson spectrum.

The $1/N$ expansion of QCD$_{2}$ implies that we have to consider the limit
$N\rightarrow~\infty $ at $g^2 N$ fixed: it corresponds to taking only the planar diagrams with no fermion loops.

In this limit the model was almost exactly soluble; the simplest Green`s functions could be found in closed form and, moreover, the Bethe--Salpeter equation was solved numerically, providing a discrete spectrum for the two--particle states. The quark--antiquark states are only bound states (mesons) and only colourless states can escape the Coulomb--like potential. The physical mass  spectrum consists of a nearly straight ``Regge trajectory''.

In 1977, three years after 't Hooft work, such an approach was criticized by T.T. Wu~\cite{Wu},
who replaced the instantaneous 't Hooft's potential by an expression
with milder analytical properties, choosing
a causal prescription for the infrared singularity in the propagator.

Unfortunately this modified formulation led to a quite involved bound
state equation, which may not be solved. An attempt to treat it
numerically in the zero bare mass case for quarks~\cite{BSW} led only to
partial answers in the form of a completely different physical
scenario. In particular no rising Regge trajectories were found. Another recent investigation~\cite{BNS} confirms these results.

After that pioneering investigation, many interesting papers
followed 't Hooft's approach, pointing out further remarkable
properties of his theory and blooming into the recent achievements
of two--dimensional QCD~\cite{wit1,wit2,BT92,BT93,doug} whereas Wu's approach sank into oblivion.

Still, equal time canonical quantization of Yang--Mills theories
in light--cone gauge~\cite{Bas5} leads precisely in $1+1$ dimensions
to the Wu's expression for the
vector exchange between quarks~\cite{Bas1}.
In the next section we will review in detail the equal time canonical quantization of Yang--Mills theory. We will also compare the vector propagator obtained in this quantization scheme to the one still obtained in light--cone gauge but with a different quantization procedure, the ``null--frame formalism'': in so doing, we will point out the advantages and disadvantages of the two different procedures.

\section{Two-Dimensional Yang-Mills Theories}
\markboth{Preliminaries}{Two-Dimensional Yang-Mills Theories}
\label{2DYMT}

We will use the following conventions for the ``gauge fixed'' Yang--Mills
Lagrangian,

\begin{equation}
{\cal L} = -{1\over 4} F^a_{\mu \nu} F^{a\mu \nu} + {\cal L}_{G.F.} \ ,
\label{lagrangian}
\end{equation}

where, in light--cone gauge:

\begin{equation}
{\cal L}_{G.F.}^{LCG}= -\lambda^a (nA^a) \ , 
\end{equation}

and $\lambda^a$ are Lagrange multipliers enforcing the light-cone gauge
condition

\begin{equation}
n^\mu A_\mu^a= A_-^a=0 \ ,
\end{equation}

$n_\mu=(1/\sqrt{2})(1,1)$ being a constant
(gauge) vector, while in Feynman gauge:

\begin{equation}
{\cal L}_{G.F.}^{F}= +{1\over 2} \partial^{\mu}A_{\mu}^{a} \partial^{\nu}A_{\nu}^{a} \ . 
\end{equation}

Without loss of generality, we consider $SU(N)$ as gauge group, so that
the field strength in eq. (\ref{lagrangian}) is  defined as

\begin{equation}
F_{\mu \nu}^a=
\partial_\mu A_\nu^a - \partial_\nu A_\mu^a + g f^{abc} A_\mu^b A_\nu^c \ ,
\end{equation}

$f^{abc}$ being the structure constants of $SU(N)$.

For later convenience, we recall that the Casimir constants of the
fundamental and adjoint representations for $SU(N)$, $C_F$ and $C_A$, are defined through

\begin{equation}
C_F= \frac{1}{N} {\rm Tr} (T^a T^a)= \frac{N^2-1}{2N}\ , 
\end{equation}

and

\begin{equation}
C_A \delta^{ab} = f^{acd}f^{bcd}= \delta^{ab} N \ .
\end{equation}

In Section \ref{GT2} we focus on the $D\ne 2$ case, and  discuss the so
called ``manifestly unitary'' and ``causal'' formulations of the theory in
light--cone gauge. We shall see that the correct formulation is the causal one: the
manifestly unitary formulation will meet so many inconsistencies to make it
unacceptable. Instead, the causal one is \emph{fully consistent} with covariant gauges. 

In Section \ref{EQ2} we compare the two light--cone formulations at strictly $D=2$. Surprisingly, in
this case both seem to coexist, without obvious inconsistencies.
Thus, a natural question arises: are the two quantization schemes equivalent in
$D=2$? Do they provide us with equal results? This question is the source of our enquiries: at the end of this thesis we will be able to answer in a quite accurate way. Moreover, we will get a deeper insight into the possible formulations of
two--dimensional Yang--Mills theories.

\subsection{Going towards $D=2$: the $D>2$ Case}
\markboth{Preliminaries}{Going towards $D=2$: the $D>2$ Case}
\label{GT2}
\noindent
A formulation of Yang--Mills theories in which no unphysical degrees of freedom are present is called {\it manifestly unitary}. It can be obtained by quantizing the theory in the so called null--frame formalism,
first introduced by Kogut and Soper~\cite{ks}, which implies the use of
light--cone coordinates and interprets $x^+$ as the evolution coordinate
(time) of the system; the remaining components $x^-, x_\perp$ will be
interpreted as  ``space'' coordinates. Within this quantization scheme, one of
the unphysical components of the gauge potential (say $A_-$) is set equal to
zero by the gauge choice whereas the remaining unphysical component ($A_+$)
is no
longer a dynamical variable but  rather a Lagrange multiplier of the secondary
constraint (Gauss' law). Thus, already at the classical level, it is possible
to restrict to the phase space containing only the physical (transverse)
polarization of the gauge fields. Then, canonical quantization on the null
plane provides the answer to the prescription for the spurious pole in the
propagator, the answer being essentially the Cauchy principal value (CPV)
prescription.

Unfortunately, following this scheme, several
inconsistencies arise, all of them being related to the violation of causality
that CPV prescription entails:

\begin{itemize}
\item {\bf non--renormalizability of the theory}: already at the one loop level,
dimensionally regularized Feynman integrals develop loop singularities that
manifest themselves as double poles  at $D=4$~\cite{CAPPER}.
\item {\bf power counting criterion is lost}: the pole structure in the complex $k_0$
plane is such that spurious poles contribute under Wick rotation. As a
consequence euclidean Feynman integrals are not
simply related to Minkowskian ones as an extra contribution shows up
which jeopardizes naive power counting~\cite{BW89}.
\item {\bf gauge invariance is lost}: due to the above mentioned extra
contributions, the $N=4$ supersymmetric version of the theory turns out
not to be finite, at variance with the Feynman gauge result~\cite{CAPPER}.
\end{itemize}

Consequently, manifestly unitary theories do not seem to exist. As explained
above, all the bad features of this formulation have their root in the lack of
causality of the prescription for the spurious pole, and the subsequent
failure of the power counting criterion in the perturbative formulation of the theory.
Thus, a natural way to circumvent
these problems is to choose a causal prescription. It was
precisely following these arguments that Mandelstam~\cite{Man} and
Leibbrandt~\cite{Lei} independently, introduced  the ML prescription
\begin{equation}
{1\over k_-}\equiv ML({1\over k_-})= {k_+\over k_+k_- + i \epsilon}={1\over
k_- +
i \epsilon {\rm sign}(k_+)}\, .
\label{mlp}
\end{equation}
It can be easily realized that with this choice the position of
the spurious pole is always ``coherent'' with that of Feynman ones, no
extra terms appearing after Wick rotation which threaten the power counting
criterion for convergence.
How can one justify such a recipe? One year later Bassetto and
collaborators~\cite{Bas5} filled the gap
by showing that ML prescription naturally arises by quantizing the
theory at equal time,
rather than at equal $x^+$. Eventually they succeeded~\cite{Bas3} in proving
full renormalizability of
the theory and full agreement with Feynman gauge results in
perturbative
calculations~\cite{Bas2}.

At present the level of
accuracy of the light--cone gauge is indeed comparable with that of the
covariant gauges.

An important point to be stressed is that equal time canonical quantization
in light--cone gauge,
leading to the ML prescription for the spurious pole, {\it does not} provide
us with a manifestly
unitary formulation of the theory. In fact in this formalism Gauss' laws do not
hold strongly but, rather, the Gauss' operators obey to a free field equation
and entail the presence in the Fock
space of unphysical degrees of freedom. The causal
nature of the ML prescription for the spurious poles is a consequence of
the causal propagation of those ``ghosts''. A physical
Hilbert space can be selected by imposing the (weakly) vanishing
of Gauss' operators.  This mechanism is similar to the Gupta Bleuler
quantization scheme for electrodynamics in Feynman gauge,
but with the great advantage that it can be naturally extended to the
non--Abelian case without Faddeev Popov ghosts~\cite{Bas4}.

\subsection{The Strictly Two-Dimensional Case}
\markboth{Preliminaries}{The Strictly Two-Dimensional Case}
\label{EQ2}
\noindent
Let us start from considering two--dimensional Yang--Mills theory in light--cone gauge. The causal formulation of the theory can be straightforwardly extended to {\it
any} dimension, including the case $D=2$, since the ML propagator is still a tempered distribution. On the other hand, the
manifestly unitary formulation can {\it only} be defined in $D=2$
without encountering obvious inconsistencies. The reason is simple:
all problems
are related to the lack of causality encoded in the CPV prescription.
But at exactly $D=2$ there are no degrees of
freedom propagating at all, and then causality is no longer a concern.
Moreover, at exactly $D=2$ and within the light--cone gauge, the 3-- and
4--gluon vertices
vanish, so that all the inconsistencies related to the perturbative evaluation
of Feynman integrals are no longer present in this case.
A manifestly unitary formulation provides the following
``instantaneous--Coulomb type'' form for the only non vanishing component of
the propagator:

\begin{equation}
D^{ab}_{++} (x) = - {i\delta^{ab}\over (2\pi)^2}\int d^2 k \, e^{ikx}
{\partial\over \partial k_-} \textrm{P}\left({1\over k_-}\right) = -i\delta^{ab}
{|x^-|\over 2} \delta(x^+)\, , \label{prcpv2}
\end{equation}

where P denotes CPV prescription,
whereas equal time canonical quantization~\cite{Bas1} gives, for the same component of the
propagator,
\begin{equation}
D_{++}^{ab}(x)= {i\delta^{ab}\over \pi^2}\int d^2k\, e^{ikx} {k_+^2\over
(k^2+i\epsilon)^2}={\delta^{ab}(x^-)^2\over \pi ( -x^2 + i \epsilon)}\ .
\label{prml2}
\end{equation}

In fact, starting from the Lagrangian density

\begin{equation}
L = {1 \over 2} F^a_{+-} F^a_{+-} + \lambda^a nA^a,
\end{equation}
$\lambda^a$ being Lagrange multipliers, by imposing the equal time
commutation relations

\begin{equation}
\label{commrel}
[A^a_1 (t, x), F^b_{01} (t,y)] = i \delta (x - y) \delta^{ab},
\end{equation}
we recover for the vector propagator exactly the ML prescription
restricted at $D=2$.

In this context the equation for the Lagrange multipliers
\begin{equation}
\label{mlghosts}
n \cdot \partial \lambda^a = 0
\end{equation}
is to be interpreted as a true equation of motion and the
fields $\lambda^a$ provide propagating degrees of freedom,
although of a ``ghost" type~\cite{Bas5}. The space of states emerging from this
treatment is an indefinite--metric Hilbert space. It is possible to select a
\emph{physical} subspace, with a positive--semidefinite metric, by imposing in
it the vanishing of the Gauss operator.

In fact, the potentials $A_+^a$ have the momentum decomposition

\begin{equation}
\tilde A_+^a (k) = u^a \delta' (k_-) + v^a \delta (k_-),
\end{equation}

$\tilde \lambda^a$ (k) being proportional to $u^a$:
$\tilde \lambda^a = k_+ u^a$.

The canonical algebra (\ref{commrel}) induces on $u^a$ and $v^a$
the algebra
\begin{equation}
[v^a_\pm (k_+), u^b_\mp (q_+)] = \pm \delta (k_+ - q_+)
\delta^{ab},
\end{equation}

$v^a_\pm$ and $u^a_\pm$ being defined as

$$
v^a(k_+) = \theta (k_+) v^a_+ (k_+) + \theta
(- k_+) v^a_- (- k_+),
$$

$$
u^a (k_+) = \theta (k_+) u^a_+ (k_+) - \theta (- k_+)
u^a_- (- k_+),
$$

and with the adjoint operators being:

\begin{equation}
\begin{split}
\left[v^a_+ (k_+)\right]^{\dagger}&=v^a_- (k_+),\\
\left[u^a_+ (k_+)\right]^{\dagger}&=u^a_- (k_+).\\
\end{split}
\end{equation}

All others commutators are vanishing.

This algebra eventually produces the propagator (\ref{prml2}).

Let us consider for instance the state $u^a_+|0\rangle$, $|0\rangle$ being the Fock vacuum. The norm of this state is:

\begin{equation}
\langle 0|u^a_- u^a_+|0\rangle = \langle 0|u^a_+ u^a_-|0\rangle =0
\end{equation}

where the first equality follows from the commutation relations and the second
from the definition of the destruction operator $u^a_-$. Therefore there is a non--null state with vanishing norm, leading to an indefinite-metric Hilbert space.

\smallskip
Thus, it seems we have two different formulation of Yang--Mills theories in
$D=2$, and within the same gauge choice, the light--cone gauge~\cite{Bas1}.
Whether they are equivalent and, in
turn, whether they are equivalent to a different gauge choice, such as
Feynman gauge,  has to be explicitly verified.

We can summarize the situation according to the content of unphysical degrees
of freedom. Since the paper by 't Hooft in 1974~\cite{hoof74},
it is a common belief that
pure Yang--Mills theory in
$D=2$ has no propagating degrees of freedom. This
happens in the
manifestly unitary formulation leading to CPV prescription for the spurious
pole, to the propagator (\ref{prcpv2}) and to 't Hooft bound state equation.
This formulation, however, cannot be
extended outside $D=2$ without inconsistencies. Alternatively, we have the same
gauge choice but with a different quantization scheme, namely at equal time,
leading to the
causal (ML) prescription for the spurious pole, to the propagator
(\ref{prml2}) and to Wu's bound state equation~\cite{Wu}. Here, even in the pure Yang--Mills case, some
degrees of freedom survive, as we have propagating ghosts.
Such a formulation is in a better shape when compared to the
previous one as it can be  smoothly extended to any dimension,
where consistency
with Feynman gauge has been established.

\smallskip
Feynman gauge validity for
any $D\ne 2$ is \emph{unquestionable}, while, at strictly $D=2$, the vector
propagator in this gauge
fails to be a tempered distribution, at variance with the behavior of light--cone gauge propagator both with ML and with CPV prescription. Still, in the spirit of dimensional
regularization, the best one can do is to evaluate the Wilson loop in $D$ dimensions, with $D>2$, and to take
eventually the limit $D\to 2$. In following this attitude, the number of
degrees of freedom
is even bigger as Faddeev--Popov ghosts are also to be taken into account.
In addition,
in the covariant gauge 3- and 4- gluon vertices do not
vanish and the theory does not look free at all.

\smallskip
We have summarized the situation in the following table:

\vskip .5truecm

\begin{center}
\begin{tabular}{|c|c|c|}
\hline
{\bf Formulation} & {\bf $D=2$} & {\bf $D>2$}\\
\hline
Feynman & not viable & OK\\
\hline
Light--Cone (CPV) & OK & not viable\\
\hline
Light--Cone (ML) & OK & OK\\
\hline
\end{tabular}
\end{center}

\vskip .5truecm

Two possible kind of enquiries can be made:

\begin{itemize}

\item
\emph{vertical} enquiry, {\it i.e.} tests of equivalence of two different formulations;

\item
\emph{horizontal} enquiry, {\it i.e.} tests of continuity, comparing a formulation in $D=2$ with the same formulation for $D>2$, taking the limit
$D \rightarrow 2$.

\end{itemize}

We see from the table that we can consider for such enquiries three different couples, namely two vertical couples and one horizontal couple.

But the complete equivalence between Feynman gauge and light--cone gauge with ML prescription for $D>2$ (first vertical couple) has been established in ref.~\cite{Bas2}. In the next chapter we will examine the second vertical couple, namely if there is a complete equivalence between the two different formulations in light--cone gauge for $D=2$, and the only horizontal couple, namely if the ML formulation of light--cone gauge is continuous.

The tool we will use in these enquiries is the Wilson loop, that we are going to describe in the next section.

\section{Wilson Loops: Not Only a Tool}
\markboth{Preliminaries}{Wilson Loops: Not Only a Tool}
\label{LOOPS}

Starting from the second half of the seventies, people began to study the so--called ``phase factors'' (exponentials of integrals over gauge potentials ordered along a closed path)~\cite{poly79}. The main idea was to describe the dynamics of the gauge fields in terms of the phase factors, which are non--local gauge invariant objects~\cite{poly79,Dot,Pol2}.

If we consider the special case of a rectangular path, we have the phase factor
usually known as the Wilson loop~\cite{Wil,Wil2}. Its study may give information on the quark confinement problem, in the sense that the asymptotic behavior of the loop (in a suitable limit which will be clarified in the sequel) gives the structure of the interaction potential between two
quarks~\cite{Kog,Fis,Kog2,Ban}.

To our purposes, the investigation of the Wilson loop will not concern the formulation of gauge theories in terms of phase--factors. Rather, our interest
in it will be twofold:

\begin{itemize}

\item[a)]
{\bf test of gauge invariance}: the Wilson loop, being a gauge invariant quantity, will allow us to perform 
\emph{perturbative} tests of gauge invariance, in order to answer to ``equivalence'' and ``continuity'' issues explained at the end of the previous section. We will also study the consistency of the perturbative formulation of Yang--Mills theory by means of a suitable criterion derived by the asymptotic behavior of the loop plus gauge invariance;

\item[b)]
{\bf confinement issue}: in the (simplified) strictly two dimensional case we will be able to get also some consequences about the interaction potential between two quarks and the bound--state equation. Essentially, we will determine the interaction potential between two quarks \emph{assuming} that we can reconstruct it from a resummation of the perturbative series. In so doing we will follow an approach that, since
the pioneering work of 't Hooft~\cite{hoof74}, regards confinement in QCD$_2$ as a perturbative feature. We will discuss in our Conclusions the validity of this assumption, but we can anticipate that our results lead to a criticism of this approach. 

\end{itemize}

Let us consider first a Yang--Mills theory at the classical level; we define the following functional

\begin{equation}
E[A;dx]=\exp(i g dx^{\mu} A_{\mu});
\label{functional}
\end{equation}

it is easy to check that, under a gauge transformation on the gauge potential,

\begin{equation}
A_{\mu}\rightarrow  A_{\mu}^{\omega}=U A_{\mu} U^{-1} - \frac{i}{g} 
\partial_{\mu}U U^{-1},
\end{equation}

$E[A,dx]$ transforms as

\begin{equation}
E[A;dx]\rightarrow E[A^{\omega};dx]=U(x + dx) E[A;dx] U^{-1}(x),
\label{wltran}
\end{equation}

apart from higher orders in $dx$; the proof of  Eq. (\ref{wltran}) can be obtained by expanding in powers of $dx$ its right--hand side.

Now we want to generalize Eqs. (\ref{functional}) and (\ref{wltran}) to the case of a finite displacement; to this aim we consider an arbitrary oriented path
$\Gamma$ starting from the point $x$ and ending at the point $y$.

The simplest thing to do is to decompose $\Gamma$ as a Riemann sum of infinitesimal displacements and to use, for each of them, Eq. (\ref{wltran}). Let us call $\Delta x^{\Gamma}_{k}$ a small displacement, oriented in the
$\Gamma$ direction, centered at the point $x_{k}\in \Gamma$, $k=1,\ldots,N$.
In this way, the natural generalization of the functional given in Eq.
(\ref{functional}) for a finite displacement $\Gamma$ will be

\begin{equation}
E[A;\Gamma]=\lim_{\stackrel{N \rightarrow \infty}{\Delta x^{\Gamma}_{k}\rightarrow 0}}\quad
\prod_{x_{k}=x}^{y} \exp (i g \Delta x^{\Gamma}_{k} \cdot A(x_{k})).
\label{porder}
\end{equation}

Equation (\ref{porder}) defines the so--called $P$--integral, that is, the
path--ordered integral along the contour $\Gamma$,

\begin{equation}
E[A;\Gamma]=P \exp \left(i g \int_{\Gamma}dx^{\mu} A_{\mu}(x)\right).
\end{equation}

Using Eq. (\ref{wltran}), it is easy to obtain the transformation law for
$E[A;\Gamma]$ under a gauge transformation, namely

\begin{equation}
E[A;\Gamma]\rightarrow U(y) E[A;\Gamma] U^{-1}(x);
\end{equation}

in particular, if we choose a closed path $\Gamma_{0}$, we have $x=y$ so that
$E[A;\Gamma_{0}]$ will transform covariantly according to the corresponding representation,

\begin{equation}
E[A;\Gamma_{0}]\rightarrow U(x) E[A;\Gamma_{0}] U^{-1}(x);
\label{phfac}
\end{equation}

in this case $E[A;\Gamma_{0}]$ is commonly called ``phase factor''. From Eq.
(\ref{phfac}) it follows that the trace of $E[A;\Gamma_{0}]$ will be a gauge invariant object. We define the quantum ``phase factor'' by means of the following vacuum to vacuum expectation value:

\begin{equation}
{\cal W}_{\Gamma{_0}}= {1\over N} \langle 0| {\rm Tr}\left[ {\cal T}{\cal P}
{\rm exp} \left( ig \oint_{\Gamma{_0}} dx^\mu \ A^a_\mu (x) T^a \right)\right]
|0\rangle \ \ ,
\label{wilson}
\end{equation}
where ${\cal T}$ orders gauge fields in time and ${\cal P}$ orders generators
$T^a$ of the gauge group $SU(N)$ along the closed integration path $\Gamma{_0}$.

We stress that in the definition of the quantum phase factor (\ref{wilson}) the ``vacuum'' has to be meant the \emph{true} physical vacuum of the theory, and not the \emph{perturbative} (Fock) vacuum. In general they are different, and we will really exploit this difference to explain our final results about the different formulations of the two--dimensional case.

Nevertheless, since the quantum phase factor is defined as a functional average of a gauge invariant quantity, if we expand the r.h.s. of Eq.(\ref{wilson}), gauge invariance has to hold order--by--order also in the perturbative expansion, that results:

\begin{multline}
{\cal W}_{\Gamma_{0}}= 1 + {1\over N}\sum_{n=2}^\infty (ig)^n \oint_{\Gamma_{0}}
dx_1^{\mu_1} \cdots \oint_{\Gamma_{0}} dx_n^{\mu_n}\theta( x_1 >\cdots >x_n )\\
\times {\rm Tr} [ G_{\mu_1 \cdots \mu_n} (x_1,\cdots ,x_n)]\ ,
\label{wilpert}
\end{multline}
where $ G_{\mu_1 \cdots \mu_n} (x_1,\cdots ,x_n)$ is the Lie algebra valued
$n$-point Green function, and   the Heavyside $\theta$-functions order the
points
$x_1,\cdots ,x_n$ along the integration path $\Gamma_{0}$.

\begin{figure}[h]
\begin{center}
\includegraphics[width=6cm]{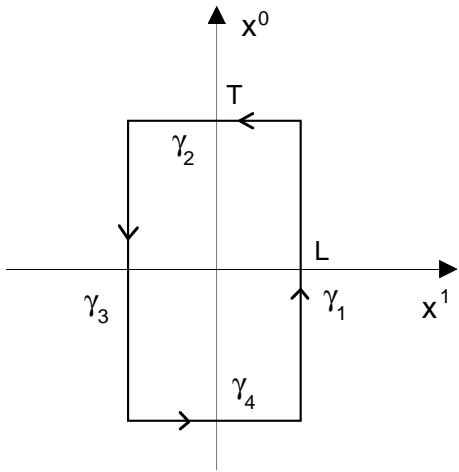}
\caption{Parameterization of the closed rectangular loop $\gamma$ in four
segments $\gamma_i$.}
\label{fig1}
\end{center}
\end{figure}

As a consequence, by comparing the results within different gauge choices, we can obtain a powerful test of gauge invariance. If we choose as closed path
$\Gamma{_0}$ the rectangular one in Fig. \protect\ref{fig1}, the phase factor is usually called {\bf Wilson loop}.

It is easy to show that the perturbative expansion of ${\cal W}_{\Gamma_{0}}$ is an
even power series in the coupling constant, so that we can write
\begin{equation}
{\cal W}_{\Gamma_{0}}=1+g^2 {\cal W}^{(1)} + g^4 {\cal W}^{(2)} + {\cal O}(g^6)
\ .
\label{pert}
\end{equation}

The advantage of using Wilson loop as a test of gauge invariance with respect to another gauge invariant quantity, the scattering amplitude (perturbative $S$-matrix elements), is at least twofold:

\begin{itemize}

\item
in the non--Abelian case, the perturbative $S$-matrix is only formally defined due to the occurrence of severe infrared singularities \emph{on--shell};

\item
to a given order in the coupling constant, the computation of the graphs coming from the perturbative expansion of the Wilson loop is much easier than the ones of the corresponding $S$--matrix elements.

\end{itemize}

Another important symmetry of Yang--Mills theories in the two--dimensional case is their \emph{invariance} under area preserving diffeomorphisms. The general reason for this fact, first emphasized by Witten~\cite{wit1}, is that the field strength is a \emph{two-form} in two dimensions. The action is therefore invariant under all diffeomorphisms that preserve the volume element of the surface. Then, the Wilson loop is not only gauge--invariant, but is also invariant under these diffeomorphisms and depends only, in strictly $1+1$ dimensions, from the area of the loop, no matter the orientation and the shape of the path.

\smallskip
It is possible to derive~\cite{ALTE1,ALTE2}
that the asymptotic behavior of the Wilson loop at fixed $L$ is

\begin{equation}
\lim_{T\to\infty}{\cal W}(L,T)=const.\  e^{-2i T V(2L)} \ ,
\label{potential1}
\end{equation}

where $V(2L)$ is the potential between a
``static" ${\rm q} {\rm\bar q}$ pair in the fundamental representation,
separated by a distance $2L$, and the Wilson loop is the one in Fig.~\protect
\ref{fig1},
one side along the space direction and one side along the time direction,
of length $2L$ and $2T$ respectively.

As a matter of fact, in the strictly two--dimensional case, it is possible to express in closed form the expectation value of the Wilson loop, summing all the contributes in the perturbative series.
When it is not possible, we will investigate its asymptotic behavior up to the $g^4$ order.
The tests we are going to accomplish represent only necessary but, in general, not sufficient conditions to verify the correctness of the quantization scheme within the assigned subsidiary condition and, in particular, of the prescription we use for the spurious poles.
In order to perform the test in the framework of the perturbation theory, we consider the expansion of the exponential:

\begin{equation}
\exp(-2i T V)=1 - 2i T V - 2 T^2 V^2 + \ldots .
\end{equation}

The potential $V(2 L)$ may be developed, in turn, as a perturbative series in
$g^2$, namely

\begin{equation}
V=g^2 V_1 + g^4 V_2 + \ldots .
\end{equation}

A necessary condition to get the exponential behavior of the Wilson loop reads
that the coefficient of $T^2$ at the order $g^4$ is one--half the square of the
linear coefficient in $T$ at the order $g^2$; this is equivalent to the cancellation of the non--Abelian terms increasing like $T^2$ (or worse) in the
large $T$--limit. In order to explain this fact, we first recognize as
non--Abelian terms in the perturbative expansion of the Wilson loop the ones
proportional to the factor $f^{abc} f^{abc}= C_A (N^2 -1)$, where $C_A$ is
the Casimir factor in the adjoint representation of $SU(N)$. The above mentioned
equivalence is clear, once we realize that all the terms linear in $T$ at the
order $g^2$ are Abelian. Hence, for any diagram coming from the perturbative expansion of the Wilson loop, we shall only deal with its non--Abelian
contributions behaving, at the order $g^4$, as $T^2$ (or worse) in the
asymptotic limit. In order to obtain the expected exponential behavior, the sum
of those contribution must vanish up to order $g^4$. As we already pointed out,
the test of the exponential behavior, in the large $T$--limit, of the Wilson
loop ${\cal W}_{\Gamma_{0}}$ is a necessary although not sufficient condition to check
the gauge relativity principle, namely that in the perturbative quantum field
theories different gauge choices must yield equivalent descriptions of the
physical phenomena, {\it i.e.} the same value of gauge--invariant quantities.

The aim of this test is clear: as the exponential behavior of the Wilson loop
(or, better, the cancellation up to $g^4$ of the non--Abelian terms behaving
like $T^2$ or worse) is a necessary condition for the gauge invariance, if it
is not reproduced within a given gauge choice, then the corresponding
formulation of the perturbation theory turns out to be poorly defined. Therefore the results of this test, performed in particular in Chapter \ref{XTR}, will
be crucial to check the \emph{soundness} of the perturbative approach with different choices of gauge and to compare the different results obtained. 

\chapter{Light-like Wilson Loop}
\label{LCR}

We have seen in Chapter \ref{START} that Yang--Mills theory without fermions in two dimensions is, in 't Hooft approach, a \emph{free} theory. This feature is at the root of the possibility of calculating the mesonic spectrum in the
large N approximation~\cite{hoof74,Cavi}, when quarks are introduced.

Still difficulties in performing a
Wick's rotation in those conditions have been pointed out~\cite{Wu}, and
a causal prescription for the infrared singularity has been advocated,
leading to a quite different solution for the vector propagator.
In this context the bound state equation with vanishing bare quark
masses~\cite{BSW} has solutions with quite different properties, when
compared with the ones of refs.~\cite{hoof74,Cavi}.

In view of the above mentioned controversial results and of the
fact that ``pure" Yang--Mills theory does not immediately look free in Feynman gauge,
where degrees of freedom of a ``ghost"--type are present, it was worth performing a test~\cite{Bas1} on a gauge invariant
quantity: the review of this test is the subject of Chapter~\ref{LCR}.

Following~\cite{Bas1}, we choose a rectangular Wilson loop with
light--like sides, directed along the vectors $n_\mu = (T, - T)$ and
$n^*_\mu = (L, L)$, with $L,T>0$, parameterized according to the equations:

\begin{eqnarray}
\label{lcrectangle}
C_1 &:& x^\mu_1 (t) = (tL,-tL)= n^{* \mu} t,\nonumber\\
C_2 &:& x^\mu_2 (t) = (L+tT,-L+tT)= n^{* \mu} + n^\mu t,\nonumber\\
C_3 &:& x^\mu_3 (t) = (T+L-tL,T-L+tL)= n^\mu + n^{* \mu}( 1-t), \nonumber\\
C_4 &:& x^\mu_4 (t) = (T-tT,T-tT)= n^\mu (1 - t), \qquad 0 \leq t \leq 1 \ , 
\end{eqnarray}

and with area 

\begin{equation}
{\cal A}_C=2 L T \ .
\end{equation}

This contour has been considered in refs.~\cite{Korc,Bas2} for
an analogous test of gauge invariance in $1+3$ dimensions. Its
light--like character forces a Minkowski treatment.

We will review a perturbative calculation up to
${\cal O}(g^4)$; in so doing
topological effects will not be considered. The computation was performed in
Feynman gauge (Section \ref{FGT2}) and in light-cone gauge 
(Section \ref{LCG2}), with Cauchy principal value prescription (Section \ref{CPV2}) and Mandelstam--Leibbrandt prescription (Section \ref{ML2}) for the spurious pole in the vector propagator. We can anticipate
the unexpected results that was obtained: the gauge invariant
theory is not free at $d= 1+1$, at variance with the
commonly accepted behavior; the theory in $d= 1+(D - 1)$ is
``discontinuous" in the limit $D \rightarrow 2$. Besides, two different,
inequivalent formulations within the same choice of gauge, the light--cone
gauge, seem to coexist: the Cauchy principal value and the
Mandelstam--Leibbrandt formulations.

\section{Light-like Wilson Loop in Feynman Gauge}
\markboth{Light-like Wilson Loop}{Feynman Gauge}
\label{FGT2}

In Feynman gauge already the free vector propagator
does not exist as a tempered distribution in $1+1$
dimensions. A regularization is thereby mandatory
and we choose to adopt the {\bf dimensional
regularization}, which preserves gauge invariance:

\begin{equation}
\label{feynmanprop}
D^F_{\mu\nu} (x) = - g_{\mu\nu} {\pi^{-D/2} \over 4}
\Gamma (D/2 -1) (- x^2 + i \epsilon)^{1 - D/2},
\end{equation}

The calculation in Feynman gauge of the light--like
Wilson loop (\ref{lcrectangle}), in $1+(D - 1)$ dimensions up to
${\cal O}(g^4)$ has been performed in ref.~\cite{Korc}.
Actually in ref.~\cite{Korc} part of the
contribution from graphs containing three vector
lines, has only been given as a Laurent expansion
around $D=4$, owing to its complexity. In the following
we shall exhibit its general expression in terms of
a generalized hypergeometric series and then we shall
expand it around $D=2$.
We only report the final results
concerning the contributions of the various diagrams.

\smallskip
The single vector exchange $({\cal O}(g^2)$) gives

\begin{equation}
\label{singlexchange}
{\cal W}_F^{(1)} = - \frac{1}{{\pi}^2} C_F {\Gamma(D/2 - 1)
\over (D - 4)^2} {\cal C},
\end{equation}

where $C_F$ is the  Casimir
operator of the fundamental representation of color $SU(N)$ and

\begin{equation}
{\cal C} \equiv \Bigl[( 2\pi {\cal A}_C + i \epsilon)^{2 -D/2}
+ (- 2\pi {\cal A}_C + i \epsilon)^{2-D/2}\Bigr].
\end{equation}

One immediately notices that the propagator pole at $D=2$ is
cancelled after integration over the contour, leading to a finite result.

There are a lot of Feynman diagrams contributing to the Wilson loop expectation value at order ${\cal O}(g^4)$ of perturbation theory. However, the number of diagrams one has to evaluate could be drastically reduced using the non--Abelian
exponentiation theorem~\cite{Gat,Fre}. According to this theorem

\begin{equation}
{\cal W}_C \equiv 1 + \sum_{n=1}^{\infty} g^{2n}{\cal W}^{(n)}=\exp \sum_{n=1}^{\infty} w^{(n)},
\end{equation}

where $w^{(n)}$ is given by the contribution of ${\cal W}^{(n)}$ with the maximal non--Abelian color factor to the order ${\cal O}(g^{2n})$ of perturbation theory, which is equal to $C_F$ for $n=1$ and $C_F C_A$ for $n=2$, where $C_A$ is the Casimir operator of the adjoint representation.

In so doing, at order ${\cal O}(g^4)$ we could restrict ourselves to the
``maximally non--Abelian" contributions proportional to $C_F C_A$.
In fact, the terms proportional to $C_F C_A$ and to $C_F$ are distinct and
cannot mix.
The test reported in this chapter is concerned only with the non--Abelian
terms.
Instead, in the next chapter our attitude will be to \emph{compare} the terms proportional to $C_F C_A$ and to \emph{verify} explicitly that, at order ${\cal O}(g^4)$, the terms proportional to $C_F$ are those that we expect from the
${\cal O}(g^2)$ computation using the non--Abelian exponentiation theorem. The
computation of the (Abelian) terms proportional to $C_F$ will be only a further check of the \emph{internal consistency} of our results, but in all cases the
significant test of gauge invariance will be given by the $C_F C_A$ term.

\smallskip
The diagrams contributing to order ${\cal O}(g^4)$
can be grouped into three distinct families:

\begin{itemize}
\item the ones in which the gluon propagator contains a self--energy
correction;
\item the ones with a double gluon exchange in which the propagators
can either cross or uncross;
\item the ones involving a triple vector vertex.
\end{itemize}

\noindent
In strictly two dimensions and in an axial gauge
only the second family is present. But in Feynman gauge all of them must
be taken into account; moreover we are here considering the problem
in $D$--dimensions. The second family is also the only one contributing
to the Abelian case.

We start by considering the diagrams belonging to the first category.

\smallskip
The self--energy correction to the propagator gives

\begin{equation}
\label{sebubble}
{\cal W}_F^{(2;se)} = {C_F C_A \over
{64 {\pi}^4}} {\Gamma^2 (D/2 - 1) (3D-2) \over (D-4)^3 (D-3)(D-1)}
{\cal E},
\end{equation}

where

\begin{equation}
\label{sellr}
{\cal E} \equiv \Bigl[(2 \pi {\cal A}_C + i \epsilon)^{4-D}
+(- 2\pi {\cal A}_C + i \epsilon)^{4-D}\Bigr]
\end{equation}

and the fermionic loop has not been considered
(pure Yang--Mills theory). Eq. (\ref{sebubble}) exhibits a double pole at $D=2$.

\smallskip
Next we consider the contribution of the so--called ``cross"
graphs, the ones with two non interacting crossed vector
exchanges

\begin{equation}
\label{crcrossed}
{\cal W}_F^{(2;cr)} = -
{C_F C_A \over {16 {\pi}^4}} {\Gamma^2 (D/2 - 1) \over (D - 4)^4}
\left[{\cal E}+ 8 {\cal B} \left(1 - {\Gamma^2 (3 -D/2) \over \Gamma(5-D)}\right)\right],
\end{equation}

where
\begin{equation}
{\cal B} \equiv \Bigl[( 2\pi {\cal A}_C + i \epsilon)
(-2 \pi {\cal A}_C + i \epsilon)\Bigr]^{2 - D/2}.
\end{equation}

Again a double pole occurs at $D=2$.

\smallskip
The contribution coming from graphs with three vector lines
is by far the most complex one. It is convenient to split
it into two parts, one coming from graphs with two vector lines
attached to the same side

\begin{eqnarray}
\label{sspider}
{\cal W}_F^{(2;ss)} &=&{C_F C_A \over {16 {\pi}^4}}
{{\cal E} \over (D - 4)^4} \bigg[2 \Gamma (3 - D/2) \Gamma (D/2 - 1)
\Gamma (D - 3) \nonumber\\
&-&{1 \over D - 3} \Gamma^2 ( D/2 - 1)\bigg]
\end{eqnarray}
and another one in which the three ``gluons" end in three
different rectangle sides

\begin{eqnarray}
\label{dspider}
{\cal W}_F^{(2;ds)}={C_F C_A \over {64 {\pi}^4}}
{\cal E} \left\{{\Gamma^2 (D/2 - 2) \Gamma (4 - D/2) \Gamma
(D - 3) \over \Gamma (D/2)} F (D)\right.\nonumber\\
+\left.{4 \over (D - 4)^4}
\Bigl[\Gamma (3 - D/2) \Gamma (D/2 - 1) \Gamma (D - 3)
- \Gamma^2 (D/2 - 1)\Bigr]\right\}.
\end{eqnarray}

The function $F (D)$ is defined as

\begin{eqnarray}
F(D) = S(D) + {D/2 - 1 \over (3 - D/2) (D - 4)}
\Big[5 \psi (3 - D/2) - \psi (D/2 - 1) \nonumber\\
- 2 \psi (1) - 2 \psi (5 - D)\Big],
\end{eqnarray}

$\psi (D)$ being the digamma function and $S(D)$ the convergent
generalized hypergeometric series

\begin{equation}
S(D) = \sum^\infty_{n=0} {1 \over (n + 1)^2} {1 \over n!}
{\Gamma (n + D - 3) \over \Gamma (D - 3)} {\Gamma (n + 4 - D/2) \over
\Gamma ( 4 - D/2)} {\Gamma (D/2) \over \Gamma (n + D/2)}.
\end{equation}

Both contributions exhibit a double pole at $D=2$.
The Laurent expansion of eq. (\ref{dspider}) around $D=4$, reproduces exactly
the expression given in ref.~\cite{Korc}.

Summing eqs. (\ref{sebubble}), (\ref{crcrossed}), (\ref{sspider}) and (\ref{dspider}) and performing a
careful Laurent expansion around $D=2$, it is tedious but
straightforward to prove that double and single poles cancel,
leaving only the finite contribution

\begin{equation}
\label{resultf}
{\cal W}_F^{(2)} (D = 2) =\frac{C_F C_A {\cal A}_C^2}{16 {\pi}^2}
\left(1 + {\pi^2 \over 3}\right).
\end{equation}

The presence of a non vanishing $C_F C_A$ contribution is
a dramatic result: it means that the theory does not
exponentiate in an Abelian way, as a ``bona fide" free theory
should do. In order to better understand this result, it is worth
turning now our attention to the same Wilson loop calculation,
performed in the light--cone axial gauge $n \cdot A = 0$.

\section{Light-like Wilson Loop in Light-Cone Gauge}
\markboth{Light-like Wilson Loop}{Light-Cone Gauge}
\label{LCG2}

We have seen in Section \ref{GT2} and \ref {EQ2} that, once we have chosen the light-cone gauge, it is possible to adopt two different prescriptions for the pole of the vector propagator in $1+1$ dimensions:
\begin{itemize}
\item the so-called "manifestly unitary¾ formulation, or Cauchy principal value prescription;
\item the causal formulation, or Mandelstam-Leibbrandt prescription.
\end{itemize}
In this section we will compare these two formulations to the result of the previous section, obtained in Feynman gauge. The light--like character of the Wilson loop leads to the vanishing of a lot of Feynman diagrams, thus
simplifying the computation.

\subsection{Cauchy Principal Value Prescription}
\markboth{Light-like Wilson Loop}{Cauchy Principal Value Prescription}
\label{CPV2}

In this section we stick in $1+1$ dimensions, since CPV prescription is
well defined only in this case. If we
interpret $x^+$ as time direction, the field $A_+$ is
not an independent dynamical variable and just provides a non--local
force of Coulomb type between fermions. In momentum space it
can be described by the ``exchange"~$k_-^{-2}$~\cite{hoof74,call76},
where~$k_-^{-2}$ has to be interpreted as in eq.~(\ref{regularized}):

\begin{equation}
\label{propcpv}
D_{++}^{CPV} (x) = -{i \over (2 \pi)^2} \int e^{ikx} d^2k
{\partial \over \partial k_-} \left[\textrm{P} \left( {1 \over k_-}\right)\right] =
- {i \over 2} |x^-| \delta (x^+).
\end{equation}

It is straightforward to check that, by inserting eq. (\ref{propcpv}) in
our Wilson loop, the result (\ref{singlexchange}) at ${\cal O}(g^2)$  is recovered.

At ${\cal O}(g^4)$ in $1+1$ dimensions, the only ``a priori" surviving
non--Abelian contribution, which is due to ``cross" graphs,
vanishes using eq. (\ref{propcpv}) (see Appendix \ref{appcrossedcpv}). Henceforth no $C_F C_A$ term appears, in
agreement with Abelian exponentiation, but at variance with the
result obtained (after regularization!) in Feynman gauge. The sum of the perturbative series easily gives:

\begin{equation}
{\cal W}_{CPV}=\exp \left(-{i \over 2}C_F g^2{\cal A}_C \right) \ .
\end{equation}

On the other hand no fully consistent vector loop calculation
would be feasible in $1+(D-1)$ dimensions, using a CPV prescription
or introducing infrared cutoffs~\cite{CAPPER}.

\subsection{Mandelstam-Leibbrandt Prescription}
\markboth{Light-like Wilson Loop}{Mandelstam-Leibbrandt Prescription}
\label{ML2}

The free vector propagator in light--cone gauge
is very sensitive to the prescription used to
handle the so--called ``spurious" singularity.
The only prescription which allows
to perform a Wick's rotation without extra terms
and to calculate loop diagrams in a consistent
way~\cite{Bas4} is the causal Mandelstam--Leibbrandt (ML)
prescription~\cite{Man,Lei}. In a canonical formalism it is obtained by
imposing equal time commutation relations~\cite{Bas5};
in two dimensions a ``ghost" degree of freedom still
survives, as extensively discussed in Section \ref{EQ2}.
When ML prescription is adopted, the free vector
propagator is indeed a tempered distribution at
$D=2$~\cite{BasPR92}, at variance with its behavior in
Feynman gauge. In particular, when $x_\perp =0$,

\begin{equation}
\label{propLC}
D_{++}^{ML} (x) =
{2 \pi^{-D/2} \Gamma (D/2) \over 4 - D}
{(x^-)^2 \over (- x^2 + i \epsilon)^{D/2}}.
\end{equation}

The calculation of the Wilson loop under consideration at ${\cal O}(g^4)$
in $1 + (D - 1)$ dimensions, using light--cone gauge with Mandelstam--Leibbrandt prescription,
has been performed in ref.~\cite{Bas2}. Here we shall report those results
and then perform their
Laurent expansion around  $D=2$, the value we are interested in.

One might wonder why dimensional regularization should
be introduced at all, as one might presume that single
graph contributions are likely to be finite in this gauge.
On the other hand, while remaining strictly at $D=2$, no
self--interaction should be present. These arguments are true, but the dimensional approach is necessary to make a complete comparison between the Mandelstam--Leibbrandt results and the Feynman ones.

In fact, the calculation ${\cal O}(g^2)$ is easily performed and the result
exactly coincides with eq. (\ref{singlexchange}), for any value of $D$.
Therefore at ${\cal O}(g^4)$ we again confine ourselves to the ``maximally
non--Abelian" contributions, without losing information. The
self--energy graph now gives

\begin{eqnarray}
\label{llse}
{\cal W}_{ML}^{(2; se)}={C_F C_A \over
{16 {\pi}^4}}
{\cal E} \bigg\{{4 \over (4 - D)^4 (D - 3)} \bigg[{\Gamma^2 (3 - D/2) \Gamma
(D - 3) \over \Gamma (5 - D)} \nonumber\\
-\Gamma^2 (D/2 - 1)\bigg] + {\Gamma^2
(D/2 - 1) \over (4 - D)^3 (D-3)} \bigg[3 - {3D - 2 \over 4(D - 1)} -
{D - 2 \over D - 3}\bigg]\bigg\}.
\end{eqnarray}

Its limit at $D=2$

\begin{equation}
\label{sedue}
{\cal W}_{ML}^{(2; se)} (D = 2) =\frac{C_F C_A {\cal A}_C^2}{{16 {\pi}^2}}
\end{equation}

is finite, but it does not vanish, as one might have
naively expected.

We shall discuss this point at the end of the section.
For the time being let us recall that, when $D \not = 2$,
``transverse" vector components are turned on and, although their
contribution is expected to be ${\cal O}(D - 2)$, it can compete
with singularities arising from loop corrections.
This is indeed what happens in the self--energy calculation.

Similarly the contribution from the ``cross" graphs

\begin{equation}
{\cal W}_{ML}^{(2 ; cr)} = -
{C_F C_A \over {16 {\pi}^4}}{\Gamma^2 (D/2 - 1) \over
(D - 4)^4} \left \{2 {\cal E} {D - 2 \over D - 3} +
8{\cal B} \left[1 - 2 {\Gamma^2 (3 - D/2) \over
\Gamma (5 - D)}\right]\right\}
\end{equation}

leads to a finite, non vanishing, result in the
limit $D=2$

\begin{equation}
\label{crdue}
{\cal W}_{ML}^{(2;cr)} (D = 2) = \frac{C_F C_A {\cal A}_C^2}{48} \ .
\end{equation}

Summing eqs. (\ref{sedue}) and (\ref{crdue}) we exactly recover
eq. (\ref{resultf}).

As a matter of fact the contribution due to graphs
with three ``gluon" lines~\cite{Bas2}

\begin{eqnarray}
\label{spidll}
{\cal W}_{ML}^{(2 ; 3g)}&=&\Omega \left\{\Gamma (D/2 - 2)\Gamma (3 - D/2)
+{\Gamma^2 (3 - D/2) \over \Gamma
(5 - D)} {6D - 28 \over (D - 2) (D - 4)}\right.\nonumber\\
&-&\left. {2 \over \Gamma
(2 - D/2)} S_1 (D) - (4 - D) \Gamma (3 - D/2) S_2 (D) \right\}
\end{eqnarray}

where

\begin{equation}
\Omega = {2 C_F C_A \over (2 \pi)^D} (2 {\cal A}_C)^{4-D}
e^{-i \pi D \over 2} \cos \left({\pi D \over 2}\right) {\Gamma (D - 4) \over
(D - 4)^2},
\end{equation}

\begin{multline}
S_1 (D) = \sum^\infty_{n=0} {\Gamma(n +2 - D/2) \over
(n + 3 - D/2)n!} \left[ {2 \over (n + D/2 - 1) (n + D/2)} + {1 \over
n + 3 - D/2}\right.
\\
\left.+ \psi(n + D/2) - \psi (n + 3 - D/2) - {\Gamma (n + D/2 - 1) \Gamma
(5 - D) \over \Gamma (4 + n - D/2)}\right]
\end{multline}
and

\begin{multline}
S_2 (D)= \sum^\infty_{n=0} {\Gamma(n+3 - D/2) \over
\Gamma(n+6 - D)} \left[\Gamma (D/2 - 2) \left({\Gamma (n+5 - D) \over
\Gamma (n+3 - D/2)} - {\Gamma(n+2) \over \Gamma(n+D/2)}\right)\right.
\\
+ \left. 2 {\Gamma(n+1) \Gamma (D/2) \over \Gamma(n+1+D/2)} +
{\Gamma(n+5 - D) \Gamma (D/2 - 1) \over \Gamma (n+4 - D/2)}
- {\Gamma(n+1) \Gamma (3 - D/2) \over \Gamma(n + 4 - D/2)}\right]
\end{multline}

vanishes when $D=2$.

As a consequence the same finite result for the Wilson loop
${\cal O} (g^4)$ at $D=2$ is obtained both in Feynman and in light--cone
gauges. However non--Abelian terms are definitely present; the
theory cannot be considered a free one in quantum loop calculations
at $D=2$, in spite of the quadratic nature of its classical
Lagrangian density in light--cone gauge. From a practical view
point,
in this fully interacting theory, the hope of getting solutions, when
quarks are included, e.g. for the mesonic spectrum, in analogy
with 't Hooft's treatment, seems remote.

If we remain strictly in $1+1$ dimensions, neither self--energy
corrections nor graphs with three vector lines should be considered, and we have only the contribution of crossed graphs {(\ref{crdue}).
The result we obtain neither coincides
with the one in Feynman gauge (the limit $D \rightarrow 2$
being ``discontinuous"), as we have neglected the non--vanishing
self--energy correction, nor obeys Abelian exponentiation as in CPV approach, the reason being rooted in a different
content of the degrees of freedom (the "ghost¾ fields $\lambda^a$ of eq.
(\ref{mlghosts})).

\section{Concluding Remarks}
\markboth{Light-like Wilson Loops}{Concluding Remarks}
\label{CR2}

The computation of the expectation value of the Wilson loop (\ref{lcrectangle}) gave the following unexpected results.

The ${\cal O}(g^4)$ perturbative loop expression in $d= 1+(D - 1)$
dimensions is finite in the limit $D\to 2$. The results in Feynman and
light--cone gauge (with Mandelstam--Leibbrandt prescription)
coincide, as required by gauge invariance. They are function
only of the area ${\cal A}_C$ for any dimension $D$ and exhibit
also a dependence
on $C_A$, the Casimir constant of the adjoint representation.

This dependence, when looked at in the light-cone gauge calculation,
comes from non-planar diagrams with the colour factor $C_F(C_F - C_A/2)$.
Besides, there is a genuine contribution proportional to $C_F C_A$
coming from the one-loop correction to the vector propagator. This is
surprising at first sight, as in strictly $1+1$ dimensions the
triple vector vertex vanishes in axial gauges. What happens is that
transverse degrees of freedom, although coupled with a
vanishing strength at $D=2$, produce finite contributions when
matching with the self-energy loop singularity precisely at $D=2$,
eventually producing a finite result. Such a dimensional ``anomaly-type" phenomenon is responsible of the discontinuity of Yang--Mills theories at $D=2$, namely of the finite, non vanishing result of eq. (\ref{llse}) in the limit $D\to 2$.

Surely this phenomenon could not appear in a strictly
$1+1$ dimensional calculation, which would only lead to the (smooth)
non-planar diagram result. We stress that this
contribution is essential to get agreement with the Feynman gauge
calculation, in other words with gauge invariance.

We notice that no ambiguity affects our light-cone gauge results,
which do not involve infinities; in addition the discrepancy
cannot be accounted for by a simple redefinition of the coupling,
that would also, while unjustified on general grounds, turn out
to be dependent on the area of the loop.

\smallskip

In order to make the argument complete, we recall that a calculation
of the same Wilson loop in strictly $1+1$ dimension in light-cone
gauge with a CPV prescription for the ``spurious'' singularity produces a
vanishing contribution from non--planar graphs.
Only planar diagrams
survive, leading to Abelian--like results depending only
on $C_F$, which can be resummed to all orders in the perturbative
expansion to recover the expected exponentiation of the area.

This result, which is the usual one found in the literature,
although quite transparent, {\it does neither coincide}
with the limit $D\to 2$ of the light--cone gauge result with ML prescription
(which is in agreement with the limit $D\to 2$ in Feynman gauge), nor with the ML result in strictly two dimensions.
The test cannot be generalized to $D\ne 2$ dimensions as
CPV prescription is at odds with causality in this case~\cite{Bas4}.

What we can certainly state is the \emph{perturbative} inequivalence of CPV and Mandelstam--Leibbrandt formulations. At this point, we have two inequivalent formulations of the same theory, the two--dimensional Yang--Mills theory,
within the same gauge choice. Now, a natural question arises: is there
any formulation that is clearly wrong, or are they both legitimate?

To answer to this question, we will study the transition from $D>2$ to $D=2$, both in Feynman gauge and in light--cone gauge with ML prescription,
in order to test the soundness of our perturbative approach when we are going towards the two--dimensional case from higher dimensions.
To do that, we will apply the criterion exposed at the end of Section
\ref{LOOPS}, namely the Wilson loop exponentiation, therefore considering a different Wilson loop, {\it viz} a space--time rectangular loop rather than a light--like one.
We already know that this Wilson loop, from a physical point of view, is even more interesting and provides also information about the interaction potential between a quark and an antiquark.
These computations, that are the subject of the next chapter, will help us to understand the features of Yang--Mills theories in ML formulation.

\chapter{Space-Time Wilson Loop}
\label{XTR}

In order to clarify whether the appearance at ${\cal O}(g^4)$ of the
maximally non--Abelian term (proportional to $C_A$) is indeed a pathology, one should examine the potential $V(2L)$ between a
``static" ${\rm q} {\rm\bar q}$ pair in the fundamental representation,
separated by a distance $2L$. Therefore in this chapter, following the
refs.~\cite{Bas7,bello,bellopr}, we will consider a different
Wilson loop, namely a rectangular loop with
one side along the space direction and one side along the time direction,
of length $2L$ and $2T$ respectively. Eventually the limit $T \to \infty$
at fixed $L$ is to be taken: the potential $V(2L)$  between the
quark and the antiquark is indeed
related to the value of the corresponding Wilson loop
amplitude ${\cal W}(L,T)$ through the equation~\cite{ALTE1,ALTE2}

\begin{equation}
\lim_{T\to\infty}{\cal W}(L,T)=const.\  e^{-2i T V(2L)}\ .
\label{potential}
\end{equation}

The crucial point to notice in eq.(\ref{potential}), as already discussed in Section \ref{LOOPS}, is that the
dependence on the Casimir
constant $C_A$ should cancel at the leading order when $T\to \infty$ in any
coefficient of a perturbative expansion of the potential with
respect to coupling constant. This criterion has often been used
as a check of gauge invariance~\cite{Bas4} and its failure has been considered as the proof of a sick formulation of four--dimensional theories~\cite{ALTE1}.
Actually, at the end of our enquiry we will discover that the criterion is not satisfied just in the two--dimensional case.
But here we will choose to have a more constructive attitude: in fact in the two--dimensional case it is possible to obtain exact \emph{non--perturbative} results. Since they are at our disposal, we can proceed to a comparison with the
results derived from our \emph{perturbative} test.
We could say that, in so doing, we check the behavior of our criterion, varying the dimension in which we are working, in the transition to $D\to 2$. There are some indications~\cite{Basgri,Griguolo} that support the conjecture that the failure of our \emph{perturbative} criterion
in the strictly two--dimensional case depends on the need of taking into
account the non--perturbative contributions,
rather than on inconsistencies in the perturbative ML formulation.
Anyway, we will defer the discussion of this point until the Conclusions.

\bigskip
Then we choose the closed path $\gamma$
parameterized by the following  four segments
$\gamma_i$,
\begin{eqnarray}
\gamma_1 &:& x_1^\mu (s) = (sT, L)\ ,\nonumber\\
\gamma_2 &:& x_2^\mu (s) = (T,-sL)\ ,\nonumber\\
\gamma_3 &:& x_3^\mu (s) = (-sT, -L)\ , \nonumber\\
\gamma_4 &:& x_4^\mu (s) = (-T, sL)\ , \ \ \qquad -1 \leq s \leq 1.
\label{path}
\end{eqnarray}
describing a  (counterclockwise-oriented) rectangle
centered at the origin of the plane ($x^1,x^0$),
with length sides $(2L,2T)$, respectively (see Fig.~\protect\ref{fig1}
at page~\pageref{fig1}) and with area

\begin{equation}
{\cal A}_{\gamma}=4 L T
\end{equation}

To have a sensitive check of gauge invariance, one has to consider at least the
order $g^4$, {\it i.e.} one has to evaluate ${\cal W}^{(2)}$, as this is the
lowest order where genuinely non--Abelian $C_FC_A$ contributions may appear.
In turn, in the calculation of ${\cal W}^{(2)}$, only the so called maximally
non--Abelian contribution needs to be evaluated, that in our case
comes from the terms proportional to $C_F C_A$. The Abelian contribution,
proportional to $C_F^2$, can be easily obtained thanks to the non--Abelian
exponentiation theorem~\cite{Fre} and will be computed explicitly as a further check of consistency.

\smallskip

In Section \ref{FGT3} we will proceed to the computation of the loop (\ref{path}) in Feynman gauge, while in Section \ref{LCG3} we will choose light-cone gauge, respectively with CPV (Section \ref{CPV3}) and with Mandelstam-Leibbrandt (Sections \ref{MLD3} and \ref{MLGT3}) prescriptions for the propagator pole.
We can anticipate the main result of this chapter, namely that,
in the large $T$-limit when $D>2$, the perturbative ${\cal O}(g^4)$
contribution is proportional only to $C_F$, and does not contain terms proportional to $C_A$. If instead we compute the Wilson loop at $D=2$, the result depends also on $C_A$, a pure area law
behavior is recovered, but the term proportional to $C_A$ survives also in the limit $T\to\infty$.

\section{Space-Time Wilson Loop in Feynman Gauge}
\markboth{Space-Time Wilson Loop}{Feynman Gauge}
\label{FGT3}

In this section we will compute the Wilson loop (\ref{path})
in Feynman gauge~\cite{bello}, in the framework of dimensional regularization
($D=2\omega$).

An explicit evaluation of the function  ${\cal W}^{(1)}$ in eq.(\ref{pert})
gives the diagrams contributing to the loop with a single exchange (i.e. one
propagator), namely
\begin{equation}
{\cal W}^{(1)}_F= - {1\over 2} C_F \oint \oint D_{\mu\nu}^F (x-y) dx^\mu dy^\nu \ ,
\label{w2}
\end{equation}

where $D_{\mu\nu}^F (x)$ is the usual free propagator (\ref{feynmanprop}) in Feynman gauge.

A fairly easy calculation leads to the result
\begin{multline}
\label{singleexchange}
{\cal W}^{(1)}_F={{C_F}\over {{\pi}^{\omega}}}(2L)^{2-2\omega} LT \bigg[i
\Gamma(\omega-3/2) \Gamma(1/2)\\
+{{2\beta\Gamma(\omega)}\over {\omega -2}}\left({1\over {3-2\omega}}-
e^{-i\pi\omega}\sum_{n=1}^{\infty}{{\Gamma(n+\omega-2)}\over {\Gamma(\omega
-2)}}{{\beta^{2n+2\omega-4}}\over {(2n-1)(2n+2\omega-3) n!}}\right)\bigg],
\end{multline}
where $\beta=L/T$.

It should be noticed that this result does not coincide with the corresponding
one of eq. (\ref{singlexchange}), which was evaluated with the same gauge choice but with
the loop sides along the  $x^+$ and $x^-$ directions. Contrary to what happens
in eq. (\ref{singlexchange}), eq. (\ref{singleexchange}) exhibits an explicit dependence on the
ratio $\beta=L/T$. Only in the two dimensional limit the two results coincide:
the limit $\omega \to1$ is smooth and restores the pure area dependence

\begin{equation}
{\cal W}^{(1)}_F (D=2)=-\frac{i}{2}C_F {\cal A}_{\gamma}.
\end{equation}

\bigskip

As in Section~\ref{FGT2}, the diagrams contributing to
${\cal W}^{(2)}_F$ can be grouped into three distinct families

\begin{equation}
{\cal W}^{(2)}_F={\cal W}^{(2;2g)}_F+{\cal W}^{(2;se)}_F+{\cal W}^{(2;3g)}_F \ :
\end{equation}

\begin{itemize}
\item[a)] the ones with a double gluon exchange in which the propagators
can either cross or uncross;
\item[b)] the ones in which the gluon propagator contains a self--energy
correction;
\item[c)] the ones involving a triple vector vertex.
\end{itemize}

\noindent
In strictly two dimensions and in an axial gauge only the first family is present, but in Feynman gauge all of them must
be taken into account; moreover we are here considering the problem
in $D$-dimensions, because of the dimensional regularization. The first family is also the only one contributing to the Abelian case.

\newpage

{\bf \centerline{a. Exchange Diagrams in Feynman Gauge}}

\bigskip
\bigskip

We start by considering the diagrams belonging to the first category, namely the ones with a double gluon exchange.
A straightforward calculation gives
\begin{multline}
{\cal W}^{(2;2g)}_F={1\over 8N}\oint\!\!\!\oint\!\!\!\oint\!\!\!\oint{\rm Tr}[{\cal P}
 (T^a_xT^a_yT^b_z T^b_w)]
  D_{\mu\nu}(x-y)D_{\rho\sigma}(z-w)\;
 dx^{\mu}dy^{\nu}dz^{\rho}dw^{\sigma}\;\; ,\\
{ }
\label{w4}
\end{multline}
where subscripts in the matrices have been introduced
to specify their ordering.
From eq. (\ref{w4}), the diagrams with two-gluons exchanges contributing to the
order $g^4$ in the perturbative expansion of the Wilson loop fall into two
distinct classes, depending on the topology of the diagrams:
\begin{enumerate}
\item {\it Non-crossed diagrams}: if the pairs $(x,y)$ and $(z,w)$
are contiguous around the loop the two propagators do not cross (see Fig.
\protect\ref{fig2}a) and the trace in (\ref{w4}) gives 
\begin{equation}
{\rm Tr} [T^aT^aT^bT^b] = N
C_F^2 ;
\end{equation}
\item {\it Crossed diagrams}: if the pairs $(x,y)$ and $(z,w)$
are  not contiguous around the loop the two propagators do cross (see
Fig. \protect\ref{fig2}b) and the trace in (\ref{w4}) gives 
\begin{equation}
{\rm Tr} [T^aT^bT^aT^b] =
{\rm Tr}[ T^a (T^a T^b + [T^b,T^a])T^b]=
N (C_F^2 - (1/2) C_A C_F) ,
\label{weight}
\end{equation}
$C_A$ being
the Casimir constant of the adjoint representation defined by
$f^{abc}f^{d bc}= C_A \delta^{ad}$.
\end{enumerate}

\begin{figure}[h]
\begin{center}
\includegraphics[width=6cm]{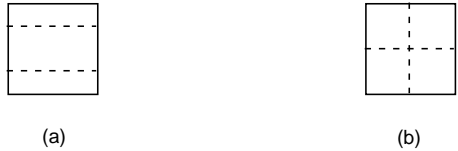}
\caption{Example of non-crossed and crossed diagrams.}
\label{fig2}
\end{center}
\end{figure}

We see that  the $C_F^2$ term is present in both types of diagrams and with
the same coefficient. This term is usually denoted  ``Abelian term'': were the
theory Abelian, only such $C_F^2$ terms would contribute to the loop. On the
other hand, the $C_FC_A$ term is present only in crossed diagrams
and is typical of non--Abelian theories.

Thus, we can decompose  ${\cal W}^{(2;2g)}_F$ as the sum of an Abelian and a
non--Abelian part,
\begin{equation}
{\cal W}^{(2;2g)}_F={\cal W}^{(2;ab)}_F +{\cal W}^{(2;na)}_F \ .
\end{equation}

Moreover, the Abelian part is simply half of the square of the
order-$g^2$ term, i.e.
\begin{eqnarray}
{\cal W}^{(2;ab)}_F &= & {1\over 8} C_F^2
\oint\oint\oint\oint
       D_{\mu\nu}^F(x-y)D_{\rho\sigma}^F(z-w)\;
        dx^{\mu}dy^{\nu}dz^{\rho}dw^{\sigma}\nonumber\\
&=& {1\over 2} \left( - {1\over 2} C_F \oint \oint D_{\mu\nu}^F (x-y)
dx^\mu dy^\nu \right)^2 \ \ .\label{w4ab}
\end{eqnarray}

This result at ${\cal O}(g^4)$ has been explicitly checked calculating
the relevant non-crossed exchange diagrams (see Appendix \ref{appexchange} for details).

Equation (\ref{w4ab}) agrees with the non--Abelian exponentiation theorem, which tells us that the Abelian  terms
(depending only on $C_F$) in
the perturbative expansion of the Wilson loop sum up to reproduce the Abelian
exponential

\begin{equation}
{\cal W}_F^{ab} (L,T) = {\rm exp}\left( - {1\over 2} C_F g^2
\oint\oint D^F_{\mu \nu}(x-y) dx^\mu dy^\nu \right) \  , \label{abexp}
\end{equation}
where the result in eq.(\ref{singleexchange}) can be introduced.

In the limit $D\to 2$ the simple exponentiation of the area is easily
recovered
\begin{equation}
{\cal W}_F^{ab} (L,T)=\exp \left(-{i \over 2}C_F g^2{\cal A}_{\gamma}\right) \ .
\label{area}
\end{equation}

\bigskip

We have now to calculate  loop
integrals of the type given in eq.(\ref{w4}). In view of
the parameterization  (\ref{path}), it is convenient to decompose loop integrals
as sums of integrals  over the segments $\gamma_i$, and to this purpose we
define
\begin{equation}
E_{ij}^F(s,t) = D_{\mu \nu}^F\bigl[ \gamma_i(s) -\gamma_j(t) \bigr]
\dot\gamma_i^\mu(s) \dot\gamma_j^\nu(t) \ , \qquad i,j=1,\dots ,4\ \, ,
\label{e}
\end{equation}
where the dot denotes the
derivative with respect to the variable parameterizing the
segment. In this way, each diagram can be written as integrals of products of
functions of the type (\ref{e}).  Each graph will be labelled by
a set of pairs
$(i,j)$, each pair denoting a gluon propagator joining the segments $\gamma_i$
and $\gamma_j$.

Due to the symmetric choice of the contour $\gamma$ and to the  fact that
propagators are even functions, i.e. $D_{\mu
\nu}^F(x)=D_{\mu \nu}^F(-x)$, we have the following identities that halve the
number of diagrams to be evaluated:
\begin{eqnarray}
E_{ij}^F(s,t) &=& E_{ji}^F(t,s)\ ,\nonumber\\
E_{11}^F(s,t) &=& E_{33}^F(s,t)\ ,\nonumber\\
E_{22}^F(s,t) &=& E_{44}^F(s,t)\ .
\label{sym}
\end{eqnarray}

We remind the reader that in Feynman gauge the propagators can attach
either to the same rectangle side or to opposite sides, but not on a
couple of contiguous ones.

We have now to consider the ${\cal O}(g^4)\  C_F C_A$-terms coming from ``crossed diagrams'' (maximally non--Abelian ones), that need to be evaluated

\begin{eqnarray}
{\cal W}^{(2;na)}_F& =& - {1\over 2} C_AC_F
\sum_{i,j,k,l}{}^{'} \int ds\int dt\int du\int dv \
E_{ij}^F(s,t)E_{kl}^F(u,v)\nonumber\\ &\equiv&- {1\over 2} C_AC_F
\sum_{i,j,k,l}{}^{'} C_{(ij)(kl)}^F \ ,
\label{doppi}
\end{eqnarray}
where the primes mean that we have to sum only over
crossed propagators configurations and over topologically inequivalent
contributions, as carefully explained in the following;
we have not specified the integration
extrema as they depend on the particular type of crossed diagram we are
considering (the extrema
must be chosen in such a way that propagators remain crossed).

The last equality in eq. (\ref{doppi}) defines
the general diagram $C_{(ij)(kl)}^F$: it is a  diagram with two {\it crossed}
propagators joining the sides $(ij)$ and $(kl)$ of the contour (\ref{path}).
In Fig. \protect\ref{fig3} a few examples of diagrams are drawn to get the reader
acquainted with the
notation.

\begin{figure}[h]
\begin{center}
\includegraphics[width=6cm]{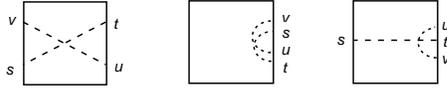}
\caption{Examples of crossed diagrams; they are labelled as $C_{(13)(13)}$,
$C_{(11)(11)}$ and $C_{(13)(11)}$, respectively}
\label{fig3}
\end{center}
\end{figure}

The first of eq.
(\ref{sym})  permits to select just 11 types of topologically distinct
crossed diagrams.
The remaining symmetry
relations (\ref{sym}) further lower the number to 7.
As a matter of fact, although topologically inequivalent, from eq. (\ref{sym})
it is easy to get
\begin{eqnarray}
C_{(11)(11)}^F&= C_{(33)(33)}^F\ \ , \qquad C_{(22)(22)}^F=& C_{(44)(44)}^F\
\ ,\nonumber\\
C_{(11)(13)}^F&= C_{(33)(13)}^F\ \ , \qquad C_{(22)(24)}^F=& C_{(44)(24)}^F.
\label{relations}
\end{eqnarray}
which are the 4
relations needed to lower the number of diagrams to be evaluated
from 11 to 7. Besides the 8 diagrams quoted in eq. (\ref{relations}), there
are three other
crossed diagrams that do not possess any apparent symmetry relation
with other diagrams: $C_{(13)(13)}^F, \ C_{(24)(24)}^F$ and $C_{(13)(24)}^F$
(see Fig. \protect\ref{fig4}), so that the number of topologically inequivalent crossed
diagrams is indeed 11.

\begin{figure}[h]
\begin{center}
\includegraphics[width=6cm]{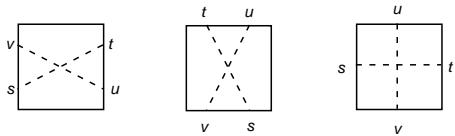}
\caption{The three crossed diagrams that are unrelated to other diagrams through
symmetry relations; they are  $C_{(13)(13)}$,
$C_{(24)(24)}$ and $C_{(13)(24)}$. }
\label{fig4}
\end{center}
\end{figure}

The
calculation of the 7 independent  diagrams needed is lengthy and  not trivial.
The details of such calculation are
sketched in  Appendix \ref{appexchange}. Each diagram depends not only on the area
${\cal A}_{\gamma}=4LT$ of the loop, but also on the dimensionless ratio $\beta = L/T$ through
complicated multiple integrals.

Adding all the contributions as in eq. (\ref{doppi}) we eventually
arrive at the
following result ${\cal O}(g^4)$ for the non--Abelian part of the exchange
diagrams contribution:
\begin{eqnarray}
\label{wmlg4}
 {\cal W}^{(2;na)}_F
&=&C_F C_A{{(2T)^{4-4\omega}}\over {\pi^{2\omega}}}(LT)^2 e^{-2i\pi\omega}\\
&&\times\Big({{\Gamma^2(\omega-1)}\over {(2\omega-4)(2\omega-3)}}\Big[
1+{{1-\omega}\over {(4\omega-5)(2\omega-3)}}+{\cal O}(\beta^{5-4\omega})
\Big]\Big).
\nonumber
\end{eqnarray}

We notice that the expression above exhibits a double and a single
pole at $\omega=1$, whose Laurent expansion gives
\begin{multline}
\label{crocilaurent}
{{\cal W}^{(2;na)}_F\pi^{2\omega} e^{2i\pi  \omega}\over C_FC_A (2T)^{4-4\omega}
 (LT)^2}={1\over 2(\omega-1)^2}+{1-\gamma\over(\omega -1)} -1-2\gamma+\gamma^2
+{\pi^2\over 12} +{\cal O}(\omega -1)\ ,\\
{ }
\end{multline}
$\gamma$ being the Euler-Mascheroni constant.

\bigskip
\bigskip

{\bf \centerline{b. Bubble Diagrams in Feynman Gauge}}

\bigskip

We turn now our attention to the calculation of ${\cal W}^{(2;se)}_F$,
namely of the diagrams with a single gluon exchange in which the
propagator contains a self--energy correction ${\cal O}(g^2)$.
Of course both gluon and ghost contribute to the self--energy.
The color factor is obviously a pure $C_F C_A$.

We call them ``bubble'' diagrams. We denote by $B_{ij}^F$ the
contribution of the diagram in which the propagator connects the
rectangle segments $\gamma_i$, $\gamma_j$ (see Fig. \protect\ref{fig5}).

\begin{figure}[h]
\begin{center}
\includegraphics[width=6cm]{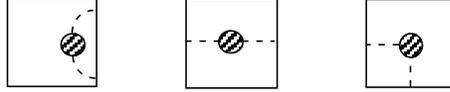}
\caption{Examples of bubble diagrams. They are labelled as $B_{11}$, $B_{13}$
and $B_{34}$,
respectively.}
\label{fig5}
\end{center}
\end{figure}

\smallskip

There are 10 topologically inequivalent diagrams; however, the symmetries
we have already discussed and the symmetric choice of the contour
entails the four conditions
$B_{11}^F=B_{33}^F$, $B_{22}^F=B_{44}^F$, $B_{12}^F=B_{34}^F$,
$B_{14}^F=B_{23}^F$,
whereas the remaining two diagrams $B_{13}^F$ and $B_{24}^F$ are unrelated by any
symmetry relation. In addition, it is easy to see by performing a simple change
of variable that $B_{14}^F$ and $B_{34}^F$ are equal. Thus, there are
5 independent diagrams to be evaluated
\begin{equation}
\label{bubblelist}
{\cal W}^{(2;se)}_F=B_{13}^F +B_{24}^F +2B_{11}^F +2B_{22}^F +4B_{12}^F.
\end{equation}

The calculation is sketched in Appendix \ref{appbubble}.
We here report the final result in the form of an expansion with
respect to the variable $\beta$
\begin{eqnarray}
\label{bubble}
{\cal W}^{(2;se)}_F&=&C_FC_A{{(2T)^{4-4\omega}}\over {\pi^{2\omega}}}(LT)^2
e^{-2i\pi\omega} \times\\ \nonumber
&&\Big[{{(3\omega -1) \Gamma^2(\omega)}\over {2\Gamma(2\omega)
\Gamma(4-\omega)}}\Gamma(1-\omega)\Gamma(2\omega -2)\Big(
{{2\omega -6}\over {5-4\omega}}+{\cal O}(\beta^{5-4\omega})\Big)\Big].
\end{eqnarray}
Again we notice the presence of a double and of a single pole at $\omega =1.$
The relevant Laurent expansion is
\begin{multline}
{{\cal W}^{(2;se)}_F\pi^{2\omega} e^{2i\pi  \omega}\over C_FC_A (2T)^{4-4\omega}
 (LT)^2}=\\
={1\over (\omega-1)^2}+{9-4\gamma\over2(\omega -1)}+{39\over 2}
-9\gamma +2\gamma^2 +{\pi^2\over 6}
  +{\cal O}(\omega -1)\ ,\\
\label{bollelaurent}
\end{multline}

\bigskip
\bigskip

{\bf \centerline{c. Spider Diagrams in Feynman Gauge}}

\bigskip

The third quantity ${\cal W}^{(2;3g)}_F$ is by far the most difficult one
to be evaluated. It comes from ``spider'' diagrams, namely the diagrams
containing the triple gluon vertex.
We denote by $S_{ijk}^F$ the contribution of the diagram in which the
propagators are attached to the segments $\gamma_i$, $\gamma_j$, $\gamma_k$
(see Fig. \protect\ref{fig6}).

\begin{figure}[h]
\begin{center}
\includegraphics[width=6cm]{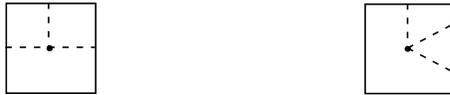}
\caption{Examples of spider diagrams. They are labelled as $S_{123}$, $S_{112}$,
respectively.}
\label{fig6}
\end{center}
\end{figure}

It can be checked that all the spiders with the three legs attached to the same
line vanish, as well as the spiders with two legs on one side and the third leg
attached to the opposite side, {\it i.e.} $S_{111}^F = S_{222}^F =S_{333}^F =S_{444}^F =
S_{113}^F =S_{133}^F =S_{224}^F =S_{244}^F =0$. Thus, there
are 12 non-vanishing topologically inequivalent diagrams; however, their number
is just halved by the symmetric choice of the contour ($S_{124}^F =S_{234}^F $,
$S_{123}^F =S_{134}^F $, $S_{112}^F =S_{334}^F $,
$S_{233}^F =S_{114}^F $, $S_{344}^F =S_{122}^F $, $S_{144}^F =S_{223}^F $), so that ${\cal W}^{(2;3g)}_F$ can be expressed in terms of the remaining 6 independent ones
\begin{equation} \label{spiderlist}
{\cal W}^{(2;3g)}_F=2S_{112}^F+2S_{123}^F +2S_{124}^F +2S_{233}^F +2S_{144}^F +2S_{344}^F.
\end{equation}
Each term is represented by a multiple integral which cannot be
evaluated for a generic dimension $\omega$ in closed form. In
particular it exhibits complicated analyticity properties in the
variable $\beta$, just in the neighborhood of the value $\beta=0$
which is of interest for us. The main aspects of the calculation are again deferred to an Appendix (Appendix \ref{appspider}).

We have succeeded in obtaining the following result for $D>2$
\begin{multline}
\lim_{\beta \to 0}{{\cal W}^{(2;3g)}_F\pi^{2\omega} e^{2i\pi  \omega}
\over C_FC_A (2T)^{4-4\omega}
 (LT)^2}=\\
=-{3\over 2(\omega-1)^2}+{3\gamma -11/2\over(\omega -1)}
-{35\over 2} +11\gamma -3\gamma^2 +{\pi^2\over 12}
 +{\cal O}(\omega -1)\ .\\
\label{ragnilaurent}
\end{multline}

A double and a single pole at $D=2$ again are present in this expression.

\bigskip
{\bf \centerline{* * *}}
\bigskip

Since the Abelian part of our results depends only on the
Casimir constant $C_F$ and smoothly exponentiates in the large--$T$
limit even when $D>2$ (see eqs.(\ref{singleexchange},\ref{abexp})), in
the following
we focus our attention on the non-Abelian part, namely on the quantity
containing the factor $C_F C_A$  $${\cal W}^{(2;na)}_F+
{\cal W}^{(2;se)}_F+{\cal W}^{(2;3g)}_F.$$

It is actually convenient to divide by the square of the loop area,
by introducing the new expression ${\cal N}$
\begin{equation}
\label{rid}
{\cal N} C_F C_A {\cal{A}_{\gamma}}^2=
{\cal W}^{(2;na)}_F+{\cal W}^{(2;se)}_F+{\cal W}^{(2;3g)}_F.
\end{equation}
Then, from eqs. (\ref{wmlg4},\ref{bubble},\ref{ragnilaurent}), it is easy to
conclude that, thanks to the factor $T^{4-4\omega}$, ${\cal N}$
vanishes in the limit $T\to \infty$ when $\omega>1$.

This is precisely the usual necessary condition required
at ${\cal O}(g^4)$ in order to
get agreement with Abelian--like time exponentiation when summing higher
orders~\cite{Bas4}.

We cannot discuss the limit $\omega \to 1$ for generic (small) values
of $\beta$ in our results as we are
only able to master the expressions at $\beta=0$.

Nevertheless if we consider the quantity

\begin{equation}
\lim_{\omega \to 1}\lim_{\beta \to 0} (T^{4\omega -4}{\cal N}) \ ,
\end{equation}

we get a quite interesting result. Indeed double and simple poles
at $\omega =1$ cancel in the sum, leading to

\begin{equation}
\label{dibiagio}
\lim_{\omega \to 1}\lim_{\beta \to 0} (T^{4\omega -4}{\cal N})=
{1\over { 16 \pi^2}}\left(1+{{\pi^2}\over 3}\right) \ ,
\end{equation}

which {\it exactly} coincides with eq. (\ref{resultf}).

This is, first of all, a formidable check
of all our calculations, if we conjecture that the same result
is obtained by performing the limit

\begin{equation}
\lim_{\beta \to 0} \lim_{\omega \to 1} {\cal N} \ .
\end{equation}

But it also entails the quite non trivial
consequences we are going to discuss in Section \ref{CR3}. But first we will review the corresponding calculations in light--cone gauge.

\section{Space-Time Wilson Loop in Light-Cone Gauge}
\markboth{Space-Time Wilson Loop}{Light-Cone Gauge}
\label{LCG3}

In this section we will proceed to the computation of Wilson loop (\ref{path}) in light-cone gauge, in order to compare the results obtained with the ones in
Feynman gauge. We will review the computations done in $D=2$ with Cauchy principal value prescription (Section \ref{CPV3}), and with Mandelstam--Leibbrandt prescription (Section \ref{MLD3}), following
ref.~\cite{Bas7}. At the end we will complete the two--dimensional case, computing the self--energy contribution that arises in $D>2$ and that survives
in the $D\to 2$ limit (Section \ref{MLGT3}).

\subsection{Cauchy Principal Value Prescription}
\markboth{Space-Time Wilson Loop}{Cauchy Principal Value Prescription}
\label{CPV3}
In this section we report the results of the computation of Wilson loop (\ref{path}) in light-cone gauge with Cauchy principal value prescription.
We shall begin with the ${\cal O} (g^2)$ contribution. From  eq. (\ref{w2}), using eqs. (\ref{path}), (\ref{propcpv}), (\ref{e}) and
(\ref{sym})(or, better, the analogous of eq.~(\ref{sym}) for light--cone gauge, see
eq.~(\ref{syml})) one can easily get
\begin{equation}
{\cal W}_{CPV}^{(1)} = - {1\over 2} C_F \sum_{i,j=1}^4 \int_{-1}^1 ds\int_{-1}^1
dt\ E_{ij}^{CPV} (s,t) =-{i\over 2} C_F {\cal A}_{\gamma} \ .
\label{g2cpv}
\end{equation}

Once the ${\cal O}(g^2)$ term is known, simple arguments permit to evaluate 
any order in the perturbative expansion, so that in this case
the loop can be exactly obtained. Let us indeed consider the ${\cal
O}(g^4)$ term. As explained in the previous section, only the ``genuine'' 
non-Abelian part needs to be evaluated, i.e. the crossed diagrams containing the
factor $C_FC_A$, as the Abelian part is already given by eq. (\ref{w4ab}), which
in this case leads to 
\begin{equation}
{\cal W}_{CPV}^{(2;ab)}= -{1\over 8}C_F^2 {\cal A}_{\gamma}^2\ . 
\end{equation}
However, due to the contact nature of the propagator in the CPV case, all the
crossed diagrams trivially vanish so that 
${\cal W}_{CPV}^{(2;na)}=0$: the $\delta(x^+)$ term in the propagator only 
tolerates
diagrams with parallel propagators, which 
therefore cannot cross, both in the Abelian and in the non-Abelian case. 
Obviously, this
argument holds at any order in the perturbative expansion, so that only {\it
Abelian} terms contribute and the sum of all of
them, due to the Abelian exponentiation theorem, reproduces the exponential
\begin{equation}
{\cal W}_{CPV}(L,T) ={\rm exp}\left(-{i\over 2} g^2 C_F {\cal A}_{\gamma}
\right)\ .
\label{wlcpv}
\end{equation}
A detailed discussion of this point can be found in Appendix
\ref{appcrossedcpv}.

Consequently, from eq. (\ref{potential}),  null plane light-cone quantization
provides a linear confining potential for a quark--antiquark pair, with string
tension $\sigma= g^2 C_F/2$. 
This is the very same result one would have obtained in an Abelian theory,
apart from the factor $C_F$.
In this sense null plane light-cone gauge quantization provides a ``free''
theory: the Wilson loop does not feel the  non-Abelian colour structure 
of the theory. This result, obtained in a Minkowskian framework, 
is in agreement with  analogous
Euclidean calculations~\cite{Bra}.

\subsection{Mandelstam-Leibbrandt Prescription ($D=2$)}
\markboth{Space-Time Wilson Loop}{Mandelstam-Leibbrandt Prescription ($D=2$)}
\label{MLD3}

In strictly $D=2$ the contribution to the expectation value of Wilson loop (\ref{path}) is only given by Feynman diagrams with double gluon exchange, in which the propagators are crossed or uncrossed, as already discussed for the light--like contour. The analysis of the diagrams is similar to the one developed in Feynman gauge. The only difference is that we shall use the Mandelstam-Leibbrandt propagator for $D=2$:

\begin{equation}
D_{++}^{(ML)cd}(x) =D_{++}^{ML}(x) \delta^{cd} = 
{i \delta^{cd}\over \pi^2}\int d^2k
\, e^{ikx} {k_+^2\over (k^2 + i \epsilon)^2}= {\delta^{cd}\over \pi}
{(x^-)^2\over (-x^2 + i\epsilon)}\  
\label{propml}
\end{equation}

instead of Feynman propagator (\ref{feynmanprop}) in $D$ dimensions. 

The fact that we are in strictly $1+1$ dimensions greatly simplifies the
perturbative expansion (\ref{wilpert}), as the complete Green functions appearing in it become products of free propagators.

We will again use the definition (\ref{e}), with the same notations of the Feynman case.
Due to the symmetric choice of the contour $\gamma$ and to the  fact that
propagators are even functions, i.e. $D_{\mu
\nu}^{ML}(x)=D_{\mu \nu}^{ML}(-x)$, we have the following identities that halve the
number of diagrams to be evaluated:
\begin{eqnarray}
E_{ij}^{ML}(s,t) &=& E_{ji}^{ML}(t,s)\ ,\nonumber\\
E_{12}^{ML}(s,t) &=& E_{34}^{ML}(s,t)\ ,\nonumber\\
E_{23}^{ML}(s,t) &=& E_{41}^{ML}(s,t)\ ,\nonumber\\
E_{11}^{ML}(s,t) &=& E_{33}^{ML}(s,t)\ ,\nonumber\\
E_{22}^{ML}(s,t) &=& E_{44}^{ML}(s,t)\ .
\label{syml}
\end{eqnarray}

At variance with the Feynman case (\ref{sym}), here also diagrams with propagators attaching on a couple of contiguous sides contribute to the final result, thus leading to a more cumbersome computation.
We begin with the ${\cal O}(g^2)$ terms. Following the notation introduced in
the previous section, ${\cal W}_{ML}^{(1)}$ can be written as the sum of 16 diagrams
\begin{eqnarray}
{\cal W}_{ML}^{(1)}&=& - {1\over 2} C_F \sum_{i,j=1}^4 \int_{-1}^1 ds\int_{-1}^1 dt
E_{ij}^{ML}(s,t)\nonumber\\
 &\equiv &- {1\over 2} C_F \sum_{i,j=1}^4 C_{ij}^{ML}\ .\label{m}
\end{eqnarray}
Thanks to the symmetry properties (\ref{syml}), only 6 of them are independent,
and an explicit evaluation gives

\begin{equation}
\begin{split}
C_{11}^{ML}&=C_{33}^{ML}=\frac{L^2}{\pi}\;\left(-\frac{1}{\b^2}\right)\;\;\\
C_{22}^{ML}&=C_{44}^{ML}=\frac{L^2}{\pi}\;\;\\
C_{12}^{ML}&=C_{21}^{ML}=C_{34}^{ML}=C_{43}^{ML}=\frac{L^2}{\pi}\left[i\pi-\ln(\b)+
             \left(1-\frac{1}{\b^2}\right)\ln(1-\b)\right]\;\;\\
C_{14}^{ML}&=C_{41}^{ML}=C_{23}^{ML}=C_{32}^{ML}=\frac{L^2}{\pi}\left[-\ln(\b)+\left(
           1-\frac{1}{\b^2}\right)\ln(1+\b)\right]\;\;\\
C_{13}^{ML}&=C_{31}^{ML}=\frac{L^2}{\pi}\bigg[\frac{1}{\b^2}+\left(
     \frac{2}{\b}-2\right)i\pi+4\ln(\b)-
        \left(\frac{2}{\b}+2\right)\ln(1+\b)\\
&+\left(\frac{2}{\b}-2\right)\ln(1-\b)\bigg]\;\;\\
C_{24}^{ML}&=C_{42}^{ML}=\frac{L^2}{\pi}\left[-1+\left(\frac{2}{\b^2}
            +\frac{2}{\b}\right)
            \ln(1+\b)+\left(\frac{2}{\b^2}-\frac{2}{\b}\right)
            \ln(1-\b)\right]\;\;\ \label{cij}\\
\end{split}
\end{equation}
Summing up all the coefficients (\ref{cij}) as in (\ref{m}) one gets that the
second--order calculation is

\begin{equation}
{\cal W}_{ML}^{(1)}=-{i\over 2} C_F {\cal A}_{\gamma} \ . 
\label{mlo2}
\end{equation}

The second--order calculation is in agreement with the CPV case (see eq. (\ref{g2cpv})). However, as often happens in Wilson loop calculations, an ${\cal O}(g^2)$ 
computation is too weak a probe to check consistency and gauge invariance. 
Thus, we have to consider the ${\cal O}(g^4)$ terms. Again, only ``crossed
diagrams'' (maximally non--Abelian ones)  need to be evaluated

\begin{eqnarray}
{\cal W}_{ML}^{(2;na)}& =& - {1\over 2} C_AC_F
\sum_{i,j,k,l}{}^{'} \int ds\int dt\int du\int dv \ 
E_{ij}^{ML}(s,t)E_{kl}^{ML}(u,v)\nonumber\\ &\equiv&- {1\over 2} C_AC_F
\sum_{i,j,k,l}{}^{'} C_{(ij)(kl)}^{ML}\ ,
\label{doppiml}
\end{eqnarray}
where the primes mean that we have to sum only over
crossed propagators configurations and over topologically inequivalent
contributions, as carefully explained in the following;
we have not specified the integration
extrema as they depend on the particular type of crossed diagram we are
considering (the extrema
must be chosen in such a way that propagators remain crossed). 

The last equality in eq. (\ref{doppiml}) defines
the general diagram $C_{(ij)(kl)}^{ML}$: it is a  diagram with two {\it crossed}
propagators joining the sides $(ij)$ and $(kl)$ of the contour (\ref{path}).
The notation is the same of the Feynman case (see Fig.~\protect\ref{fig3} at
page~\pageref{fig3}).
The first of eq.
(\ref{syml})  permits to select just 35 types of topologically distinct
crossed diagrams, and to multiply each representative by a factor $8$, which is
the number of permutations of the points $(x,w,y,z)$ leaving the propagators 
$D_{\mu\nu}^{ML}[x(s)-y(t)]D_{\sigma\rho}^{ML}[w(u)-z(v)]$ crossed. The remaining symmetry
relations (\ref{syml}) further lower the number to 19. 
As a matter of fact, although topologically inequivalent, from eq. (\ref{syml})
it is easy to get  
\begin{eqnarray}
C_{(11)(11)}^{ML}&= C_{(33)(33)}^{ML}\ \ ,
\qquad C_{(22)(22)}^{ML}=& C_{(44)(44)}^{ML}\
\ ,\nonumber\\
C_{(11)(13)}^{ML}&= C_{(33)(13)}^{ML}\ \ ,
\qquad C_{(22)(24)}^{ML}=& C_{(44)(24)}^{ML}\
\ ,\nonumber\\
C_{(11)(12)}^{ML}&= C_{(33)(34)}^{ML}\ \ , 
\qquad C_{(22)(23)}^{ML}=& C_{(44)(14)}^{ML}\
\ ,\nonumber\\
C_{(11)(14)}^{ML}&= C_{(33)(23)}^{ML}\ \ , 
\qquad C_{(22)(12)}^{ML}=& C_{(44)(34)}^{ML}\
\ ,\nonumber\\
C_{(13)(12)}^{ML}&= C_{(13)(34)}^{ML}\ \ , 
\qquad C_{(24)(23)}^{ML}=& C_{(24)(14)}^{ML}\
\ ,\nonumber\\
C_{(13)(14)}^{ML}&= C_{(13)(23)}^{ML}\ \ , 
\qquad C_{(24)(12)}^{ML}=& C_{(24)(34)}^{ML}\
\ ,\nonumber\\
C_{(12)(14)}^{ML}&= C_{(23)(34)}^{ML}\ \ , 
\qquad C_{(12)(23)}^{ML}=& C_{(14)(34)}^{ML}\
\ ,\nonumber\\
C_{(12)(12)}^{ML}&= C_{(34)(34)}^{ML}\ \ , 
\qquad C_{(23)(23)}^{ML}=& C_{(14)(14)}^{ML}\ \ ,
\label{relationsml}
\end{eqnarray}
which are the 16 
relations needed to lower the number of diagrams to be evaluated
from 35 to 19. Besides the 32 diagrams quoted in eq. (\ref{relationsml}), there
are three other 
crossed diagrams that do not possess any apparent symmetry relation
with other diagrams: $C_{(13)(13)}^{ML}, \ C_{(24)(24)}^{ML}$ and $C_{(13)(24)}
^{ML}$ (see
Fig.~\protect\ref{fig4} at page~\pageref{fig4}), so that the number of topologically inequivalent crossed diagrams is indeed 35.

The
calculation of the 19 independent  diagrams needed is lengthy and  not trivial.
The details of such calculation are fully
reported in the Appendix \ref{appml}. Each diagram depends not only on the area
${\cal A}_{\gamma}=4LT$ of the loop, but also on the dimensionless ratio $\beta = L/T$ through
complicated functions involving powers, logarithms and dilogarithm functions,
denoted by ${\rm Li}_2(z)$.
Since we shall be interested in the large--$T$
behavior, we always consider the region $\beta<1$
(see Appendix \ref{appml} for details).

Adding all the contributions as in eq. (\ref{doppiml}) we eventually arrive at the
following result for the non--Abelian part of the ${\cal O}(g^4)$ contributions:

\begin{equation}
{\cal W}_{ML}^{(2;na)} = {{C_A C_F {\cal A}_{\gamma}^2}\over {48}}.
\label{wmlg42}
\end{equation}

Several important consequences can be drawn from eq. (\ref{wmlg42}):
\begin{enumerate} 

\item the sum of all non--Abelian terms, proportional to $C_FC_A$, does not
vanish. This fact prevents any possible agreement with the CPV
formulation, where the result is a simple Abelian exponentiation (see eq.
(\ref{wlcpv}));

\item a little thought is enough to realize that the perturbative series
${\cal W}_{ML}$ cannot
sum to a {\it phase factor}, 
even taking into account possible extra non--Abelian terms in the argument of
the exponent.
\end{enumerate}

As the calculation ${\cal O}(g^4)$ is really heavy, a 
consistency check of its accuracy has been performed. The contribution from {\it uncrossed} graphs
${\cal O}(g^4)$ (which only involve $C_F^2$) has been indeed independently
computed  and then it has been summed
to the expression for the corresponding $C_F^2$ terms coming from the {\it crossed}
graphs, which have twice the weight of eq. (\ref{wmlg42})
(see eq.~(\ref{weight})); in so doing
the full ${\cal O}(g^4)$ Abelian result has been correctly recovered.

The result of eq. (\ref{wmlg42}) was confirmed in ref.~\cite{Stau}: in this work the authors managed to do a complete resummation of the perturbative series, obtaining the exact expression for the Wilson loop for any contour $P$ with
the given area ${\cal A}_P$
\footnote{We recall that in the two--dimensional case we have a dependence only
from the area of the loop, and not from the shape of the path.}:

\begin{equation}
\label{contour}
{\cal W}_{ML}({\cal A}_P)=\exp \left(- i \frac{N-1}{2N} g^2 {\cal A }_P \right)
~\frac{1}{N}
\oint\frac{dz}{2 \pi i}~\exp \big( - i g^2 {\cal A }_P z \big)
\bigg( \frac{z+1}{z} \bigg)^N.
\end{equation}
The contour integral, which encloses the multiple pole at $z=0$,
gives a Laguerre polynomial in $g^2 {\cal A }_P$ of order
$N-1$:
$ L^1_{N-1}(i g^2 {\cal A }_P)$.

This last result coincides, at ${\cal O}(g^4)$, with eq. (\ref{wmlg42}), but differs from the CPV result (\ref{wlcpv}). Therefore
we can state that the perturbative series of Wilson loop computed in light--cone gauge with CPV and Mandelstam--Leibbrandt prescriptions are definitely different: this difference, as we will discuss in the Conclusions, can be
related to the different behavior of 't Hooft's and Wu's bound states equations with respect to the confinement issue.

\subsection{Mandelstam--Leibbrandt Prescription ($D \to 2$)}
\markboth{Space-Time Wilson Loop}{Mandelstam-Leibbrandt Prescription ($D \to 2$)}

\label{MLGT3}

The computation of Wilson loop in Mandelstam--Leibbrandt prescription for
$D\to 2$ is performed following the same scheme of the previous sections of this chapter. 
The diagrams contributing to the non--Abelian part of
${\cal W}^{(2)}_{ML}$ can be grouped in the usual three
families, namely
crossed diagrams ${\cal C}_{(ij)(kl)}^{ML}$,
spider diagrams ${\cal S}_{ijk}^{ML}$ and bubble diagrams
${\cal B}_{ij}^{ML}$.

In arbitrary dimensions,  the calculation of the  Wilson loop
is much more awkward in light--cone gauge than in Feynman gauge, due to
a more complicated form of the vector propagator. However, when considering the
$D\to 2$ limit, diagrams in light--cone gauge have much better analyticity properties
in $\omega$ than the ones in Feynman gauge.
In fact the vector propagator in light--cone gauge with
ML prescription is a tempered distribution at $D=2$, at odds with the one
in Feynman gauge. Moreover it is summable along
the (compact) loop contour.

Due to this property, we can conclude that all the maximally non--Abelian
contributions arising from diagrams with crossed propagators
sum to an expression that, in the
limit $D\to 2$, reproduces eq. (\ref{wmlg42}), namely
\begin{equation}
\label{croci}
{\cal W}^{(2;na)}_{ML}(D=2)=\frac{C_A C_F {\cal A}_{\gamma}^2}{48} \ .
\end{equation}

\smallskip

Now we consider the contribution ${\cal W}^{(2;se)}_{ML}$
coming from bubble diagrams.  In light--cone gauge and on the plane $x^0\times x^1$,  the only
non vanishing component of the two point Green function $\Delta_{\mu\nu}^{ML}$ at the
order ${\cal O}(g^2)$ is $ \Delta_{++}^{ML}(x)\equiv \Delta^{ML}(x)$, that reads, at
$x_\perp=0$~\cite{Bas2},
\begin{equation}
\label{self}
\Delta^{ML}(x)=-\frac{g^2}{8\pi^{2\omega}}C_A\frac{(x^-)^2}{(-x^2+i\e)^{2\omega-2}}
f(\omega)\ ,
\end{equation}
where
\begin{equation}
\label{fd}
f(\omega)=\frac{1}{(2-\omega)^3}\left[\frac{\Gamma^2(3-\omega)\Gamma(2\omega
-3)}{\Gamma(5-2\omega)}
 -{\Gamma(\omega-1) \Gamma(\omega) (10\omega^2 -19\omega + 10)\over
4(2\omega-3)(2\omega-1)}\right].
\end{equation}

There are 10 topologically
inequivalent bubble diagrams that contributes to ${\cal W}^{(2;se)}_{ML}$.
However, due to the symmetry of the Green function and to the symmetric choice
of the contour, only six of them  are independent, and the ${\cal O}(g^4)$
contribution  to the Wilson loop arising from bubble diagrams can be
written as
\begin{equation}
\label{bub1}
{\cal W}^{(2;se)}_{ML}= 2({\cal B}_{11}^{ML} + {\cal B}_{22}^{ML} +{\cal B}_{13}^{ML}+ {\cal B}_{24}^{ML}
+ 2{\cal B}_{12}^{ML}+ 2{\cal B}_{14}^{ML}) \ ,
\end{equation}
where each single contribution ${\cal B}_{ij}^{ML}$ can be calculated by replacing
eqs. (\ref{path}), (\ref{self}) in the formula
\begin{equation}
\label{bubbola}
{\cal B}_{ij}^{ML}=-{1\over 2}g^2 C_F \int_{-1}^1 ds \int_{-1}^1 dt 
\Delta_{\mu\nu}^{ML}
(\gamma_i (s)- \gamma_j(t)) \dot \gamma^\mu_i (s)\dot \gamma^\nu_j (t)\ .
\end{equation}

The main details of the calculation are deferred to an appendix
(Appendix~\ref{bubbleml}).

\smallskip
The final result is

\begin{multline}
\label{bub2}
{\cal W}^{(2;se)}_{ML}= {C_FC_A\over \pi^{2\omega} }f(\omega) (LT)^2
(2L)^{4-4\omega}\left\{e^{-2i\pi\omega }\beta^{4\omega-6}\left[ {1\over
(7-4\omega)(8-4\omega)}\right.\right.\\
\times\biggl(1-(8-4\omega)
_2F_1(2\omega-2,2\omega-7/2;2\omega-5/2;\beta^2)+(7-4\omega)
(1-\beta^2)^{3-2\omega}\biggr)\\
\left.
-\frac{1-(1-\beta^2)^{4-2\omega}}{(3-2\omega)(4-2\omega)}
+ {5-2\omega\over(6-4\omega)(4-2\omega)} \left(1-(1-\beta^2)^{3-2\omega}\right)
\right]\\
+e^{-2i\pi\omega}
\beta^{4\omega-4}\left[{(1-\beta^2)^{3-2\omega} \over
(3-2\omega)(4-2\omega)}- { _2F_1(2\omega-2,
2\omega-5/2;2\omega-3/2;\beta^2)\over (5-4\omega)}\right.\\
\left. -  _2F_1(2\omega-2,1/2;3/2;\beta^2){ { } \over  { } }\right]
+i\beta {\sqrt{\pi} (\omega-2)\Gamma(2\omega -7/2)\over
\Gamma(2\omega -2)} \\
\left.  -e^{-2i\pi\omega} {\beta^{4\omega-2}\over 3}
  {_2F_1}(2\omega-2,3/2;5/2;\beta^2) + { \beta^2\over (7-4\omega)}\right\}\ .
\end{multline}

Some comments are here in order. First of all there is a dependence
on the dimensionless ratio $\beta=L/T$, besides the area, at variance with
the analogous result (\ref{llse}) in light--cone gauge (ML) for
the rectangle with light-like sides. However, in the equation above,
one can easily check
that the quantity ${\cal W}^{(2;se)}_{ML}/(LT)^2$ is not singular for $\beta\to 0$.
Actually eq.(\ref{bub2}) exhibits, for $\omega >1$, the expected
damping factor $T^{4-4\omega}$ in the large--$T$ limit.

In the limit $\omega\to 1$  the dependence on $\beta$ disappears and the
pure area law is recovered:

\begin{equation}
\label{sestorto}
{\cal W}^{(2;se)}_{ML}(D=2) =\frac{C_F C_A {\cal A}_{\gamma}^2} {16 {\pi}^2} \ .
\end{equation}

This is exactly the ``missing'' term to be added to the expression of eq. 
(\ref{wmlg42}) to obtain the final result for
the maximally non Abelian contribution to the perturbative ${\cal O}(g^4)$
Wilson loop in the limit $D\to 2$,
\begin{equation}
\label{finale}
{\cal W}^{(2;se)}_{ML}(D=2) + {\cal W}^{(2;na)}_{ML}(D=2)
= \frac{C_F C_A {\cal A}_{\gamma}^2} {16 {\pi}^2} \left( 1 + {\pi^2\over
3}\right)\ .
\end{equation}

Equation (\ref{finale}) is in fact in full agreement with eqs. (\ref{rid}--\ref{dibiagio}),
where the same Wilson loop is calculated in Feynman gauge.

In turn the result above implies that ``spider'' diagrams, namely diagrams with
a triple vector vertex, cannot contribute in the limit $D \to 2$.
This is not surprising, as also the contribution (\ref{spidll}), where the contour of the loop is chosen along the ($x^+$, $x^-$)
directions, does not contribute in the same limit.

In order to support this conclusion,
we can show that the relevant
three point Green
function at ${\cal O}(g)$, vanishes in that limit.

To this aim, let us consider the three point Green function
${\cal V}_{\mu\nu\rho}^{ML} (x,y,z)$. Due to the light--cone gauge choice, its only non vanishing
component when considering the loop in the $x^0\times x^1$ plane is ${\cal
V}^{ML}(x,y,z) = {\cal V}_{+++}^{ML}(x,y,z)$; up to an irrelevant
multiplicative constant, it is given by

\begin{multline}
\label{gamma}
{\cal V}^{ML}(x,y,z)=\int d^{2\omega}\zeta {\partial\over \partial z^\alpha}
\left[{\partial\over \partial x^\alpha}{\partial\over \partial y^+}-
{\partial\over \partial y^\alpha}{\partial\over \partial x^+}\right]
F(x-\zeta)F(y-\zeta) G(z-\zeta)\\
+{\rm cycl.\  perm.} \ \{x,y,z\}\ \equiv ({\cal V}_1-{\cal V}_2)
+{\rm cycl.\  perm.} \ \{x,y,z\}\ .
\end{multline}

Here the index $\alpha$ runs over the transverse components and the functions
$G$ and $F$ are the following Fourier transforms
\begin{equation}
\label{g}
G(x)=\int d^{2\omega} p {e^{ipx}\over p^2 +i\e}=-\pi^\omega
\Gamma(\omega-1) \left(-{x^2\over 4} + i\e\right)^{1-\omega},
 \end{equation}
\begin{equation}
\label{f}
\begin{split}
F(x)&=\int d^{2\omega} p {e^{ipx}\over (p^2 +i\e) (p^++i\e p^-)}\\
&=-i\pi^\omega \Gamma(\omega-1)\int_0^{x_+} d\rho \left({x_\perp^2 - 2 x_-
\rho\over 4} + i\e\right)^{1-\omega}.\\
\end{split}
\end{equation}

Let us consider, for instance, the first term in eq. (\ref{gamma}), that
we call ${\cal V}_1^{ML}$. Using
standard Feynman integrals techniques, integrations over momenta and over the
intermediate point $\zeta$ can be performed, so that ${\cal V}_1^{ML}$ can be
rewritten, after some convenient change of variables, as

\begin{multline}
\label{gamma1}
{\cal V}_1^{ML}={i\pi^\omega (4 \pi)^{3\omega}\over 8}\Gamma(2\omega-1)
(\omega-1)\\
\times
\!\! \int_0^1 \!\! d\xi d\eta d\mu \,  \eta [\mu(1-\mu)]^{\omega -1}
\int_0^\infty \!\! d\tau {[1+\tau (\mu\xi + \eta (1-\mu))]^{2\omega-5}\over
(1+\tau)^{\omega}}\\
\times
{[(x-z)_+ +\tau\eta (1-\mu) (x-y)_+][(y-z)_+ + \tau \mu\xi
(y-x)_+]^2\over[-\mu\xi(x-z)^2 -\eta (1-\mu) (y-z)^2 - \tau\xi\eta\mu(1-\mu)
(x-y)^2 +i\e]^{2\omega-1}}\ .
\end{multline}

Since  ${\cal V}_1^{ML}$ has an explicit zero at $\omega=1$, if we show that the
integral in (\ref{gamma1}) is convergent when evaluated at $\omega=1$, we
have proved that the three point Green function vanishes at $D=2$. The
derivation of this result is exhibited in the Appendix \ref{appgreen}.

\section{Concluding Remarks}
\markboth{Space-Time Wilson Loop}{Concluding Remarks}
\label{CR3}

In this chapter we have proceeded to the computation of the Wilson loop
(\ref{path}) in different gauges and dimensions.

One feature that is confirmed in this series of investigations, as already seen in Chapter~\ref{LCR}, is the
\emph{discontinuity} of Yang--Mills theories in two dimensions. The light--cone results in strictly $1+1$ dimensions are dependent from the different prescription adopted for the vector propagator. But, at a first sight, one could have thought to recover the result 
obtained in strictly $1+1$ dimensions with ML prescription, if we are coming from $D>2$, both with Feynman gauge and with light--cone gauge (ML). This is not the case;
as a matter of fact in strictly $1+1$ dimensions and in light--cone gauge
the contribution from diagrams containing a self--energy insertion
is missing, in spite of the fact that, if calculated first at $D>2$,
it does not vanish in the limit $D\to 2$. This contribution is essential
in order to get agreement with the Feynman gauge computation. This phenomenon, discovered first in ref.~\cite{Bas1}, has been discussed at length in Chapter \ref{LCR}.

What may be surprising is that this extra contribution, required by gauge
invariance, when considered in the limit $D\to 2$, exhibits a pure
area dependence on its own, being the same no matter the orientation of
the loop.
Moreover, in light--cone gauge (ML), different families of diagrams (``crossed'' and ``bubble''
diagrams) give the same contribution ($C_FC_A {(LT)^2 \over 3}$
and $C_FC_A \left({LT\over \pi}\right)^2$ respectively) no matter the
orientation of the loop (compare eqs.~(\ref{sedue}),(\ref{crdue}) with eqs.~(\ref{sestorto}),(\ref{croci})).
Remarkably, the geometrical arguments (invariance under
area-preserving diffeomorphism) which lead to a pure area
dependence in two dimensions, but not in higher ones, are recovered in
the limit $D\to 2$, in spite of the singular nature of this limit,
and of the difference in the two results (with and without self-energy
diagrams).

We consider this point quite intriguing; it seems that, in order to
get beyond two dimensions towards higher ones, the theory
needs further inputs which cannot be {\it a priori} guessed in two
dimensions. On the other hand it is known that operators exist which
are irrelevant at $D>2$, but can be competitive in exactly two 
dimensions~\cite{wit1,doug}.

Our result can therefore be interpreted as a warning when one tries to
extend straightforwardly conclusions obtained in strictly two dimensions
to more realistic situations.

But the new result of this chapter is that we have checked that our findings at
${\cal O}(g^4)$ comply with the Abelian--like time exponentiation
in the large-$T$ limit, as long as $D>2$.
At $D=2$ this is not the case.
 
In fact, while in any dimension $D>2$
\emph{perturbative} Wilson loop calculations are in agreement with
Abelian-like time exponentiation, as all $C_A$ dependent terms
turn out to be depressed in the large-$T$ limit, at $D=2$
neither the result in ref.~\cite{Bas7} nor the one in ref.~\cite{bello} share
this property, as they
both exhibit an explicit $C_A$-dependence in the coefficient of the
leading term when $T \to \infty$. At $D=2$
exponentiation in terms of $C_F$
occurs perturbatively only in light-front formulation (see eq.~(\ref{wlcpv})).

But our investigation is purely perturbative, while the Wilson loop is
defined by means of the ``true'' vacuum, that might have
non--perturbative contributions. There are, in fact, arguments supporting the hypothesis that the expected Abelian--like time exponentiation in the equal--time quantization could be restored by genuine non--perturbative contributions. This point is going to be thoroughly discussed in the Conclusions.

\chapter*{Conclusions}
\addcontentsline{toc}{chapter}{Conclusions}
\markboth{Conclusions}{Conclusions}

In this thesis we have performed a perturbative inquiry on the features of
two--dimensional Yang--Mills theories.

The {\bf first result}, obtained in Chapter~\ref{LCR} and confirmed in
Chapter~\ref{XTR}, is that \emph{two different, inequivalent perturbative
formulations} of two--dimensional Yang--Mills theories within the light--cone
gauge seem to coexist:

\begin{itemize}

\item[a)]
the light--front quantization formulation, leading to (instantaneous) Cauchy
Principal Value (CPV) prescription for the spurious pole of the gluon propagator;

\item[b)]
the equal--time quantization formulation, leading to (causal) Mandelstam--Leibbrandt (ML) prescription for the same pole.

\end{itemize}

The CPV formulation ``lives'' only in $D=2$ and in higher dimensions manifests inconsistencies: in fact it leads to a non--renormalizable theory and does not agree with Feynman gauge results. Instead, the ML formulation is in full agreement with Feynman gauge results for $D>2$, but the comparison cannot be performed in strictly $1+1$
dimensions, since the free propagator (\ref{feynmanprop}) in Feynman gauge is
not a tempered distribution for $D=2$.

Just to compare these two different formulations at $D=2$, in order to decide if one or the other is wrong or if there is a sort of ``compatibility'' (in spite of the different results with respect to our test of gauge invariance), we have performed in Chapter~\ref{XTR} a test on the asymptotic behavior of the
space--time Wilson loop: the dependence on $C_A$ should cancel at the leading
order when $T\to\infty$ in any coefficient of its perturbative expansion. The failure of this criterion has been interpreted in the past
as an evidence of a pathology of four dimensional perturbative formulations of
quantum field theories~\cite{ALTE1}.
The {\bf second result} obtained in this thesis is that the vanishing of the
term proportional to $C_A$ at ${\cal O}(g^4)$, and the subsequent Abelian--like time exponentiation of
the Wilson loop, happens only for $D>2$, but in the limit $D\to 2$ its
asymptotic behavior fails to be the Abelian--like one. In other words, if we take first the limit $D \to 2$, no damping occurs when $T\to \infty$:
the two limits \emph{do not commute}. With respect to this
criterion the CPV formulation seems to be in better shape: it provides the
expected Abelian--like exponentiation in an exact way, once we have summed all the Feynman diagrams contributing to the perturbative series.

\smallskip
Let us summarize in detail the main perturbative features found in our Wilson loop computations, which are common to both Feynman and light-cone gauges.

For $D>2$ the ${\cal O}(g^4)$ result depends on the contour: if it has 
light--like sides, there is a dependence only on the area, no matter the
value of $D$. If instead the contour is a space-time rectangle, for $D>2$
there is also a dependence on the ratio $\beta \equiv L/T$, still
reproducing the expected exponential behavior in the limit $T \to \infty
\ (\beta =0).$

The limit $D \to 2$ of such ${\cal O}(g^4)$ result is finite and depends only
on the area, no matter the orientation and the shape of the contour. It
consists of two addenda (compare eq.~(\ref{resultf}) with eq.~(\ref{dibiagio})). 

In light cone gauge the second addendum comes from ``crossed'' diagrams, 
the first one is due to the
self-energy correction to the gluon propagator. In axial gauges the transverse
degrees of freedom are coupled with a strength of order $D-2$; nevertheless
they produce a finite contribution when matching the self-energy loop
singularity precisely at $D=2$.

Thus the ML formulation exhibits a kind of ``instability'' with respect to
a change of dimensions and looks indeed discontinuous in the limit $D\to 2$.
In fact in exactly $1+1$ dimensions the 
first addendum is missing and one just gets the result of ref.~\cite{Bas7}, 
confirmed in ref.~\cite{Stau} following a different approach.
Its area dependence is hardly surprising in view of the symmetry under
area preserving diffeomorphisms at $D=2$; it is perhaps remarkable that also
the first addendum, which originates at $D>2$, exhibits in the limit
$D \to 2$ a pure area dependence on its own. 

However
both contributions contain the factor $C_F C_A$ and thereby disagree
with the simple area exponentiation, that is obtained only in CPV formulation.
At a perturbative level there is a discontinuity, which 
is not surprising in the light of the argument presented in ref.~\cite{Stau}, namely the possibility of defining (generalized) QCD$_2$~\cite{doug}: it
tells us that one must pay attention in trying to extend
two--dimensional results to higher dimensions.
However it does not explain why agreement with the simple area behavior 
is not recovered.

Moreover, the authors of ref.~\cite{Stau}, working in exactly $1+1$ dimensions,
have succeeded in resumming the perturbative series: the result (\ref{contour}) they get 
does not exhibit the usual purely exponential area law.
Besides the appearance of Laguerre polynomials, the coefficient in the
exponent multiplying the area is different from the usual one.
We are thereby faced with a discrepancy.
It is interesting to point out that they note that the limit for $N \to\infty$ of eq.~(\ref{contour}) gives an expression for the Wilson loop
that is not able to assure confinement, at variance with its behavior at finite
$N$. This feature suggests that Wu's bound state equation~\cite{Wu} is not
expected to lead to a discrete meson spectrum, and is in agreement with the fact
that no rising
Regge trajectories were found in refs.~\cite{BSW,BNS}.

\smallskip

Now we want to discuss the consequences concerning the Abelian--like time exponentiation in the large-$T$ limit. We have checked
in Chapter~\ref{XTR} that the space--time Wilson loop at ${\cal O}
(g^4)$ is in agreement with the expected $C_F$ leading behavior, as long as $D>2$.
At $D=2$, instead, a term proportional to $C_A$ survives; however it is likely that this is not an evidence
of a pathology of two--dimensional ML formulation, as it might appear at a
first glance. Rather, there are some arguments that show that this failure has
his roots in a non--perturbative issue, and that the expected behavior can be
restored by istantonic, non perturbative contributions,
reconciling gauge invariance with basic spectral properties and
solving the paradox.

As a matter of fact
the 't Hooft $D=2$ well established result can most safely be obtained
by means of non perturbative techniques. Actually, confinement in
QCD$_2$ is usually regarded as a perturbative effect: in any axial gauge
the theory is quadratic and it was a wide--accepted belief that the gluon
propagator gave rise to a linear potential, which provides the usual area
behavior for the Wilson loop by resumming the perturbative series. But this
is true only in axial gauges with
a peculiar prescription for the propagator (the Cauchy principal
value, which works wonderfully only in $D=2$, but is certainly incorrect for
$D>2$). The reason why this happens might be deep 
and related to peculiar properties of the $D=2$ light-front vacuum 
in light-front quantization.
In fact, in this latter procedure there are no degrees of freedom at all
(manifestly unitary formulation), at variance with ``ghosts'' (unphysical) degrees of freedom in ML formulation: therefore there is no way to introduce
creation and destruction operators to construct the
Hilbert space of the theory starting from the Fock vacuum. This vacuum is
therefore the only state of the space. Perhaps just this feature could explain why we
obtain the usual, abelian result for Wilson loop already at a perturbative
level. This does not happen with ML prescription: we could say that, also
from this point of view, this latter approach is more \emph{honest}, in the
sense that it does not hide the non--perturbative roots of the confinement.

At a non-perturbative level there are arguments
which show that at large $N$ and on the sphere $S_2$ a phase
transition occurs, induced by instantons
(see refs.~\cite{doug2,boul,gross1,gross2}).
The purely
exponential area behavior occurs only in the strong coupling 
phase, which is 
dominated by instantons. In the weak coupling phase the result is
completely different and agrees (after decompactification) with the
ones of refs.~\cite{Bas7,Stau}. This is the reason why it has been 
suggested that the discrepancy can be solved by taking instantons into
account \cite{Basgri,Griguolo}; they obviously do not contribute to any perturbative
calculation, even when the perturbative series is fully resummed.

\smallskip

One could say that perturbative
results are irrelevant at $D=2$. But just our perturbative results are a clear
evidence of the need of a non--perturbative approach to the confinement issue
also in the two--dimensional case. 't Hooft model works wonderfully in
strictly two dimensions, but it is at least doubtful that it could give
lessons for the real, four dimensional world.
Besides, at $D>2$ perturbative results are the only easily accessible ones;
several perturbative tests of gauge invariance have been performed in the past
at $D>2$, just {\it assuming}
the simple area exponentiation in the large $T$-limit.
In Chapter \ref{XTR} we have \emph{proved} that indeed at $D>2$ such an exponentiation occurs as expected.

We do not think it is immaterial to understand how these nice features behave
in the transition $D \to 2$, also in order to contrast them against
genuine non perturbative results.

\smallskip
Why genuinely non perturbative contributions, which are likely to be
relevant also in higher dimensions, are crucial only in two dimensions
to possibly recover Abelian--like time exponentiation, is at present unknown.
This very interesting issue, together with the difference between the
vacuum structure in ML and CPV formulations, remain open questions to be investigated.

\appendix
\chapter{Exchange Diagrams in Feynman Gauge}
\label{appexchange}
\markboth{Appendix \ref{appexchange}}{Exchange Diagrams in Feynman Gauge}

In order to get the ${\cal O}(g^4)$ contribution to the Wilson Loop arising
from the exchange diagrams, we only have to evaluate the maximally non Abelian
diagrams, that in the present case are the so called crossed diagrams. In fact,
the contribution coming from planar diagrams
can be easily obtained through the Abelian exponentiation theorem, as explained
in the main text. In this appendix we shall give the main sketch of computation
of the seven independent crossed diagrams needed:

\footnotesize

\begin{align}
\label{crociapp1}
C_{(13)(13)}^F=&\!\!\int_{-1}^{1} \!\!du\int_{-1}^{1} \!\!dv {\alpha T^4\over[4
L^2 - (u+v)^2 T^2 + i \epsilon]^{\omega -1}} \int_u^1\!\! dt \int_v^1 \!\!ds
{1\over [4 L^2 - (t + s)^2 T^2 + i \epsilon]^{\omega - 1}},\\
C_{(24)(24)}^F=&\!\!\int_{-1}^{1} \!\!du\int_{-1}^{1} \!\!dv {\alpha L^4\over[L^2
(u+v)^2- 4T^2 + i \epsilon]^{\omega -1}} \int_u^1\!\! dt \int_v^1 \!\!ds {1\over
[ L^2 (t+s)^2 - 4 T^2 + i \epsilon]^{\omega - 1}},\\
C_{(13)(24)}^F=&\!\!\int_{-1}^{1} \!\!du\int_{-1}^{1} \!\!dv {(-
\alpha L^2T^2)\over[ L^2 (u+v)^2 - 4 T^2 + i \epsilon]^{\omega -1}}
\int_{-1}^1\!\! dt \int_{-1}^1 \!\!ds {1\over [4 L^2 - (t + s)^2 T^2 + i
\epsilon]^{\omega - 1}},\\
C_{(11)(13)}^F=&\!\!\int_{-1}^{1} \!\!du\int_{-1}^{1}
\!\!dv {(-2\alpha T^4)\over[4 L^2 - (u+v)^2 T^2 + i \epsilon]^{\omega -1}}
\int_{-1}^u\!\! dt \int_u^1 \!\!ds {1\over [ - (t - s)^2 T^2 + i
\epsilon]^{\omega - 1}},\\
C_{(22)(24)}^F=&\!\!\int_{-1}^{1} \!\!du\int_{-1}^{1}
\!\!dv {(-2\alpha L^4)\over[ L^2 (u+v)^2- 4 T^2 + i \epsilon]^{\omega -1}}
\int_{-1}^u\!\! dt \int_u^1 \!\!ds {1\over [(t - s)^2 L^2 + i \epsilon]^{\omega
- 1}},\\
 C_{(11)(11)}^F=&\!\!\int_{-1}^{1} \!\! dt\int_{t}^{1} \!\! du \int_{u}^1
\!\! ds\int_s^1 \!\! dv {(2\alpha T^4)\over [ - (t-s)^2 T^2 + i\epsilon]^{\omega
- 1} [ - (u - v)^2 T^2 + i \epsilon]^{\omega -1}},\\
C_{(22)(22)}^F=&\!\!\int_{-1}^{1} \!\! dt\int_{t}^{1} \!\! du \int_{u}^1 \!\!
ds\int_s^1 \!\! dv {(2\alpha L^4)\over [  (t-s)^2 L^2 + i\epsilon]^{\omega - 1}
[  (u - v)^2 L^2 + i \epsilon]^{\omega -1}},
\end{align}

\normalsize

where  $\alpha =[\Gamma (\omega-1)]^2/(16\pi^{2\omega})$.
As an example, we
report the main sketch of computation of $C_{(13)(13)}^F$.

In order to perform the integrations a series expansion of the
denominators in (A1) is not enough as the series does not converge in the
entire integration domains. The necessary analytic continuations
are provided by
two Mellin-Barnes transformations:

\footnotesize

\begin{multline}
\label{MB1}
C_{(13)(13)}^F=\alpha T^4 (4 L^{2})^{2 - 2 \omega}
{1 \over{[\Gamma(\omega-1)]^2}} {1 \over {(2 \pi i)^2}}\\
\int_{-i \infty}^{+i
\infty} dy \Gamma(\omega -1 + y) \Gamma(-y) \left(
{1 \over{- \beta^2 - i
\epsilon}}\right)^y \int_{-i \infty}^{+i \infty} dz
\Gamma(\omega
-1 + z) \Gamma(-z) \left( {1 \over{- \beta^2 - i \epsilon}}\right)^z \\ 
\times \int_{-1}^{1} du \int_{-1}^{1} dv \left |{u + v}\over 2 \right|^{2 y}
\int_{u}^{1} dt \int_{v}^{1} ds \left |{s + t}\over 2 \right|^{2 z} \ ,
\end{multline}

\normalsize

where the path of integration over $z$ is chosen in such a
way that the poles
of the function $\Gamma(\omega -1 + z)$ lie to the left of
the path of integration and the poles of the function
$\Gamma(- z)$ lie to the
right of it (the same for the integration over $y$).

After the integration over $s$, $t$, $v$ and $u$ we have
 the following expression:

\footnotesize

\begin{multline}
\label{MB2}
C_{(13)(13)}^F= \alpha T^4 (4 L^{2})^{2 - 2 \omega}
{1 \over{[\Gamma(\omega-1)]^2}} {1 \over {(2 \pi i)^2}} \times\\
\int_{-i \infty}^{+i
\infty} dy \Gamma(\omega -1 + y) \Gamma(-y) \left( {1 \over{- \beta^2 - i
\epsilon}}\right)^y \int_{-i \infty}^{+i \infty} dz
\Gamma(\omega
-1 + z) \Gamma(-z) \left( {1 \over{- \beta^2 - i \epsilon}}\right)^z \\ 
\times {2^5 \over{(2 z + 1) (2 z + 2)}} \left[{1\over{(2 y + 1)
(2 y + 2)}} -
{1\over{(2 y + 1) (2 y + 2 z + 4)}}\right. \\ 
\left. +{1\over{(2 y
+ 2 z + 3) (2 y + 2 z + 4)}} -  {{\Gamma(2 z + 3) \Gamma(2 y + 1)}
\over{\Gamma(2
z + 2 y + 5)}}\right] \ .
\end{multline}

\normalsize

Then one has to integrate over $z$ and $y$; the integration contours have to
be suitably chosen: for instance, in the present example, in order to apply
Jordan's lemma, the integration paths  must be closed with
half-circles lying in the half planes ${\rm Re} z<0$ and   ${\rm Re} y<0$.
These integrations produce several double power series in the variable
$\beta^2$ with finite convergence radii, which are  particularly suited
for a large $T$ (small $\beta$) expansion.
For instance,  the last term in
the square bracket of  eq. (\ref{MB2}) gives the following contribution

\footnotesize

\begin{multline}
\alpha T^4 (4 L^{2})^{2 - 2 \omega}
{1 \over{[\Gamma(\omega-1)]^2}} {1 \over {(2 \pi i)^2}} \int_{-i \infty}^{+i
\infty} dy \Gamma(\omega -1 + y) \Gamma(-y) \left( {1 \over{- \beta^2 - i
\epsilon}}\right)^y \\
\times \int_{-i \infty}^{+i \infty} dz
\Gamma(\omega
-1 + z) \Gamma(-z) \left( {1 \over{- \beta^2 - i \epsilon}}\right)^z
{2^5 \over{(2 z + 1) (2 z + 2)}} \left[
 -  {{\Gamma(2 z + 3) \Gamma(2 y + 1)}\over{\Gamma(2
z + 2 y + 5)}}\right] \\
={{(2T)^{4-4\omega}}\over {\pi^{2\omega}}}(LT)^2
\Bigg\{e^{-2i\pi\omega}
{{\beta}^{-2}}(-\frac{1}{8}) \left({\pi \over{\sin(\pi\omega)}}\right)^2
\frac{\Gamma^2(\frac{3}{2}-\omega)}{\Gamma(\frac{1}{2})
\Gamma(\frac{9}{2} - 2 \omega) \Gamma(5 - 2\omega)} \\
\times F_4(2 \omega - \frac{7}{2},2 \omega - 4, \omega -\frac{1}{2},
\omega -\frac{1}{2};\beta^2,\beta^2)
+ e^{-i\pi\omega} {\beta}^{1 - 2 \omega}
\frac{i}{4} {\pi \over{\sin(\pi\omega)}}
\frac{\Gamma(\frac{1}{2})\Gamma(\frac{3}{2}-\omega)\Gamma(\omega -
\frac{3}{2})}{\Gamma(3-\omega)\Gamma(\frac{7}{2}-\omega)} \\
\times F_4(\omega - \frac{5}{2},\omega - 2, \omega -\frac{1}{2},
\frac{5}{2}-\omega;\beta^2,\beta^2)
+ e^{-i\pi\omega} {\beta}^{2 - 2 \omega}
\frac{1}{4} {\pi \over{\sin(\pi\omega)}} \Gamma(\frac{1}{2})\times \\
\sum_{l=0}^\infty \sum_{m=0}^\infty \frac
{\Gamma(m+1)\Gamma(l+m+\omega-\frac{3}{2})\Gamma(l+m+\omega-2)}
{\Gamma(l+\omega-\frac{1}{2})\Gamma(m+3-\omega)\Gamma(m+\frac{3}{2})}
\frac{(\beta^2)^l}{l!}
\frac{(\beta^2)^m}{m!}+{\beta}^{4 - 4 \omega}
\frac{\Gamma^2(\omega-\frac{3}{2})\Gamma^2(\frac{1}{2})}{4} \\
+ i {\beta}^{5 - 4 \omega} \Gamma(\frac{1}{2})\Gamma(\omega-2)
\Gamma(\omega-\frac{3}{2}) -\frac{{\beta}^{6 - 4 \omega}}{2}
\left[\frac{1}{2} \Gamma^2(\frac{1}{2})\Gamma(\omega-\frac{3}{2})
\Gamma(\omega-\frac{5}{2}) + \Gamma^2(\omega - 2)\right]\Bigg\}
\label{MB3}
\end{multline}

\normalsize

with the notations as in~\cite{erd}.

Repeating this procedure for each integral (A1)-(A7), 
one eventually recover eq.
(\ref{wmlg4}).

As a check of our calculations, we have explicitly verified the Abelian
exponentiation theorem. The sum of all the crossed diagrams, which are
proportional to $C_F^2- (1/2) C_FC_A$, behaves like  $(LT)^2 T^{4 - 4\omega}$,
and therefore is depressed in the large $T$ limit as long as $\omega >1$. This
means that only planar diagrams should contribute to the Abelian exponentiation
in the large $T$ limit. As a matter of fact this is in fact what happens:
there is a single planar
diagram that,
alone, provides the dominant term for the  Abelian exponentiation;
it is the one
in Fig.~(\ref{fig2}a) at page~\pageref{fig2}. It can be  checked  that for such a diagram the leading
term in the large $T$ expansion is
$(-1/2\pi^{2\omega})((2L)^{4-4\omega}(C_FLT)^2
(\Gamma(\omega-3/2)\Gamma(1/2))^2$, which is precisely  the half of the square
of the corresponding leading term of the sum of the single exchange diagrams.

\chapter{Bubble Diagrams in Feynman Gauge}
\label{appbubble}
\markboth{Appendix \ref{appbubble}}{Bubble Diagrams in Feynman Gauge}

The complete one loop propagator (including ghost interaction) is given by
\begin{eqnarray}
D_{\mu\nu}^{(2)F,ab}&=& \delta^{ab} {g^2 C_A\over 16 \pi^{2\omega}}{(1 -
3\omega)(2-\omega)(3 - 2 \omega)\Gamma (1-\omega)\Gamma^2(\omega) \Gamma
(2\omega - 4)\over \Gamma(2 \omega)\Gamma(4 - \omega)(-x^2 + i
\epsilon)^{2\omega - 3}}\nonumber\\
&\times&\left[ {x_\mu x_\nu\over (- x^2 + i\epsilon)} - g_{\mu\nu}{2\omega - 5\over 2
(3 - 2\omega)}\right].
  \end{eqnarray}

Writing explicitly all the possible bubble diagrams $B^F_{ij}$, it is not
difficult to realize that  $B^F_{11}=B^F_{33}$, $B^F_{22}=B^F_{44}$, $B^F_{12}=B^F_{34}$,
 $B^F_{14}=B^F_{23}$. In addition, the two last pairs of bubbles are in turn equal
after a trivial change of variables, so that eq. (\ref{bubblelist}) follows.
The five independent bubbles are then

\footnotesize

\begin{align}
B^F_{11}&=\int_{-1}^1\!\! ds \int_{-1}^1 \!\! dt { (\sigma T^2)\over [-(s-t)^2
T^2 + i\epsilon]^{2\omega -3}}\left[{2\omega -1 \over 2 (3 -
2\omega)}\right]\ , \\
B^F_{22}&=\int_{-1}^1\!\! ds \int_{-1}^1 \!\! dt { (\sigma L^2)\over [(s-t)^2
L^2 + i\epsilon]^{2\omega -3}}\left[{1 -2\omega \over 2 (3 -
2\omega)}\right]\ , \\
B^F_{13}&=\int_{-1}^1\!\! ds \int_{-1}^1 \!\! dt { (-\sigma T^2)\over [-(s+t)^2
T^2 +4L^2+ i\epsilon]^{2\omega -3}}\left[{ (s+t)^2 T^2\over 4L^2 - (s+t)^2
T^2 + i \epsilon} - {2\omega - 5\over 2 (3 - 2\omega)} \right]\ ,\\
B^F_{24}&=\int_{-1}^1\!\! ds \int_{-1}^1 \!\! dt { (-\sigma L^2)\over [(s+t)^2
L^2 -4T^2+ i\epsilon]^{2\omega -3}}\left[{ (s+t)^2 L^2\over (s+t)^2
L^2-4T^2 + i \epsilon} + {2\omega - 5\over 2 (3 - 2\omega)} \right]\ ,\\
B^F_{12}&=\int_{-1}^1\!\! ds \int_{-1}^1 \!\! dt { \sigma L^2T^2
(t-1)(s+1)\over [(s+1)^2  L^2 - T^2 (t-1)^2+ i\epsilon]^{2\omega -2}}\ ,
\end{align}

\normalsize

where

\begin{equation}
\sigma={g^4 C_F C_A\over 16 \pi^{2\omega}}{(
3\omega-1)(2-\omega)(3 - 2 \omega)\Gamma (1-\omega)\Gamma^2(\omega) \Gamma
(2\omega - 4)\over \Gamma(2 \omega)\Gamma(4 - \omega)}
\end{equation}

Again, integration can be performed using Mellin-Barnes techniques leading to

\footnotesize

\begin{multline}
{\cal W}^{(2;se)}_F\!=\!{C_FC_A (2L)^{4 \! -\! 4\omega} (LT)^2 \Gamma^2 (\omega)
(3\omega
\! -\!1) \Gamma(1 \! -\!\omega)\Gamma (2\omega \! -\!2)\over 2 \pi^{2\omega}
\Gamma(2\omega)
\Gamma (4\! -\!\omega)} \times \\
\left\{ e^{\! -\!2i\pi\omega}\beta^{4\omega \! -\!6}\left[ {(2\omega\! -\!
1)\over (4\omega \! -\!6)}
{1 \! -\! (4\omega \! -\!7) (1\! -\!\beta^2)^{4\! -\!2\omega} \! +\!
(4\omega \! -\!8)
_2F_1(2\omega\! -\!3,2\omega \! -\!7/2; 2\omega\! -\!5/2;\beta^2)\over
(2\omega \! -\!4)(4\omega
\! -\!7)}\right.\right. \\
\left.\left. -{2\omega\! -\!1\over (4\omega \! -\!6)(2\omega\! -\!4)}
 [_2F_1(2\omega
\! -\!4,\! -\!1/2;1/2;\beta^2)\! -\!1] \! +\!{1\over 2\omega\! -\!3}
[_2F_1(2\omega
\! -\!3,\! -\!1/2;1/2;\beta^2)\! -\!1] \right.\right. \\
\left.\left. +{1\over
(2\omega\! -\!4)(2\omega\! -\!3)} [(1\! -\!\beta^2)^{4\! -\!2\omega}
\! -\!1] \right]\right. \\
\left. +e^{\! -\!2i\pi\omega} \beta^{4\omega \! -\!4} {2(2\omega
\! -\!3) _2F_1 (2\omega\! -\!2,
2\omega \! -\!5/2; 2\omega \! -\!3/2;\beta^2) \! -\! (4\omega\! -
\!5) (1\! -\!\beta^2)^{3\! -\!2\omega}\over
(2\omega \! -\!3)(4\omega \! -\!5)}\right. \\
\left.+ i\beta{(\omega \! -\!3)\Gamma(1/2)\Gamma (2\omega \!
-\!7/2)\over \Gamma(2\omega
\! -\!2)} \! +\! \beta^2 {3\! -\!\omega\over(\omega\! -\!2)
(4\omega \! -\!7)}\right\} \ .
\label{bolleapp}
\end{multline}

\normalsize

\bigskip
By performing the large $T$ limit in eq. (\ref{bolleapp}) one arrives at eq.
(\ref{bubble}).

\chapter{Spider Diagrams in Feynman Gauge}
\label{appspider}
\markboth{Appendix \ref{appspider}}{Spider Diagrams in Feynman Gauge}

The spider diagrams are by far the most complicated to evaluate.
The sum of  the 6 inequivalent  spider diagrams, with the appropriate weights,
is given by

\footnotesize

\begin{multline}
\label{spider1app}
{\cal
W}^{(2;3g)}_F=2(S_{124}+S_{123}+S_{122}+S_{144}+S_{112}+S_{114})\equiv\\
{C_FC_A\Gamma(2\omega-2)L^2T^{6-4\omega}\over 32\pi^{2\omega}} \int_0^1
d\rho_1 \int_0^1d\rho_2\int_0^1d\rho_3\delta (1-\rho_1-\rho_2-\rho_3)
(\rho_1\rho_2\rho_3)^{\omega-2}\times\\
\left\{ \int_{-1}^1\!\!ds_1\int_{-1}^1\!\!ds_2\int_{-1}^1\!\!ds_3
(\rho_1 (\rho_1-1) + \rho_2(\rho_2-1) - 2\rho_1\rho_2 + (\rho_2-\rho_1)s_3
\rho_3)\times \right.\\
\left. [(\rho_1 \!-\!\rho_2 \!- \! s_3\rho_3)^2\!-\!\beta^2 (s_1
\rho_1 \!-\!s_2\rho_2 \!+\! \rho_3)^2\!-\!\rho_1(1\!-\!\beta^2 s_1^2)\!
-\!\rho_2 (1\!-\!\beta^2 s_2^2)\!
-\!\rho_3(s_3^2\!-\!\beta^2)\!+\!i\epsilon]^{2-2\omega}\right.\\
\left.+\int_{-1}^1\!\!ds_1\int_{-1}^1\!\!ds_2\int_{-1}^1\!\!ds_3 (\rho_1
(\rho_1-1) + \rho_3(\rho_3-1) - 2\rho_1\rho_3 + (\rho_1-\rho_3)s_2
\rho_2)\times\right.\\
\left. [(s_1\rho_1 \!-\!\rho_2 \!-\!
s_3\rho_3)^2\!-\!\beta^2 ( \rho_1 \!+\!s_2\rho_2 \!-\!
\rho_3)^2\!-\!\rho_1(s_1^2\!-\!\beta^2 ) \!-\!\rho_2 (1\!-\!\beta^2 s_2^2)
\!-\!\rho_3(s_3^2\!-\!\beta^2)\!+\!i\epsilon]^{2- 2\omega}\right.\\
\left.+\int_{-1}^1 \!\! ds_1 \int_{-1}^{s_1}\!\! ds_2 \int_{-1}^{1} \!\!ds_3
(\rho_1 (\rho_1-1) - \rho_2(\rho_2-1) + (\rho_1-\rho_2)s_3
\rho_3)\times\right.\\
\left. [(\rho_1 \!+\!\rho_2 \!+\! s_3\rho_3)^2\!-\!\beta^2
(s_1 \rho_1 \!+\!s_2\rho_2 \!-\! \rho_3)^2\!-\!\rho_1(1\!-\!s_1^2\beta^2 )
\!-\!\rho_2 (1\!-\!\beta^2 s_2^2)
\!-\!\rho_3(s_3^2\!-\!\beta^2)\!+\!i\epsilon]^{2-2\omega}\right.\\
+\left.\int_{-1}^1\!\!ds_1\int_{-1}^{s_1}\!\!ds_2\int_{-1}^1\!\!ds_3 (\rho_1
(1-\rho_1) + \rho_2(\rho_2-1)  + (\rho_1-\rho_2)s_3 \rho_3)\right.
\times\\
\left. [(\rho_1 \!+\!\rho_2 \!-\! s_3\rho_3)^2\!-\!\beta^2 (
s_1\rho_1 \!+\!s_2\rho_2 \!+\! \rho_3)^2\!-\!\rho_1(1\!-\!s_1^2\beta^2 )
\!-\!\rho_2 (1\!-\!\beta^2 s_2^2)
\!-\!\rho_3(s_3^2\!-\!\beta^2)\!+\!i\epsilon]^{2-2\omega}\right.\\
+\left.\int_{-1}^1\!\!ds_1\int_{-1}^1\!\!ds_2\int_{-1}^{s_2}\!\!ds_3 (\rho_2
(1-\rho_2) + \rho_3(\rho_3-1)  + (\rho_2-\rho_3)s_1
\rho_1)\times\right.\\
\left. [(\rho_1 \!+\! s_2\rho_2 \!+\!
s_3\rho_3)^2\!-\!\beta^2 ( s_1\rho_1 \!-\!\rho_2 \!-\!
\rho_3)^2\!-\!\rho_1(1\!-\! s_1^2\beta^2 ) \!-\!\rho_2 (s_2^2\!-\!\beta^2 )
\!-\!\rho_3(s_3^2\!-\!\beta^2)\!+\!i\epsilon]^{2- 2\omega}\right.\\
+\left.\int_{-1}^1\!\!ds_1\int_{-1}^1\!\!ds_2\int_{-1}^{s_2}\!\!ds_3 (\rho_2
(\rho_2-1) + \rho_3(1-\rho_3) + (\rho_2-\rho_3)s_1
\rho_1)\right.\times\\
\left. [(\rho_1 \!-\!s_2\rho_2 \!-\!
s_3\rho_3)^2\!-\!\beta^2 ( s_1\rho_1 \!+\!\rho_2 \!+\!
\rho_3)^2\!-\!\rho_1(1\!-\!s_1^2\beta^2 ) \!-\!\rho_2 (s_2^2\!-\!\beta^2)
\!-\!\rho_3(s_3^2\!-\!\beta^2)\!+\!i\epsilon]^{2-2\omega}\right\}.\\
\end{multline}

\normalsize

The above integrals can be more conveniently grouped as
\begin{equation}
{\cal W}^{(2;3g)}_F={C_FC_A(LT)^2(2 T)^{4-4\omega}\over \pi^{2\omega}}
e^{-2i\pi\omega} [I_1(\beta^2)+
I_2(\beta^2)+I_3(\beta^2)+I_4(\beta^2)]
\end{equation}
where
\begin{eqnarray}
I_1(\beta^2)& =& \Gamma(2\omega-3)
2^{4\omega-10}\int_0^1\! [d\rho]\int_{-1}^1\! [ds]
(\rho_1\rho_2\rho_3)^{\omega-2}
{\rho_1-\rho_2\over\rho_1+\rho_2}{\partial\over\partial s_3}\nonumber\\
&&[P_1^{3-2\omega} +\theta(s_1-s_2) P_2^{3-2\omega} + \theta(s_1-s_2)
P_3^{3-2\omega}]\nonumber\\
I_2(\beta^2)& =& \Gamma(2\omega-3)
2^{4\omega-10}\int_0^1\! [d\rho]\int_{-1}^1\! [ds]
(\rho_1\rho_2\rho_3)^{\omega-2} {\rho_2-\rho_1\over
\beta^2(\rho_1+\rho_2)}{\partial\over\partial s_3}\nonumber\\
 &&[P_4^{3-2\omega}
+\theta(s_1-s_2) P_5^{3-2\omega} + \theta(s_1-s_2) P_6^{3-2\omega}]\nonumber\\
I_3(\beta^2)& =& -\Gamma(2\omega-2) 2^{4\omega-7}\int_0^1\!
[d\rho]\int_{-1}^1\! [ds] (\rho_1\rho_2\rho_3)^{\omega-2}
{\rho_1\rho_2\over\rho_1+\rho_2}
P_1^{2-2\omega}\nonumber\\
I_4(\beta^2)& =& -\Gamma(2\omega-2) 2^{4\omega-7}\int_0^1\!
[d\rho]\int_{-1}^1\! [ds] (\rho_1\rho_2\rho_3)^{\omega-2}
{\rho_1\rho_2\over\rho_1+\rho_2}
P_4^{2-2\omega}
\end{eqnarray}

where $[ds]=ds_1ds_2ds_3$,
$[d\rho]=d\rho_1d\rho_2d\rho_3\delta(1-\rho_1-\rho_2-\rho_3)$ and

\footnotesize

\begin{align}
\begin{split}
P_1\! &=\!(\rho_1 \! -\!\rho_2 \! +\! s_3\rho_3)^2\! -\!\beta^2 (s_1
\rho_1 \! -\!s_2\rho_2 \! -\! \rho_3)^2\! -\!\rho_1(1\! -\!s_1^2\beta^2 )
\! -\!\rho_2 (1\! -\!\beta^2 s_2^2)
\! -\!\rho_3(s_3^2\! -\!\beta^2)\! +\!i\epsilon\\
P_2\! &=\!(\rho_1 \! +\!\rho_2 \! +\! s_3\rho_3)^2\! -\!\beta^2 (s_1
\rho_1 \! +\!s_2\rho_2 \! -\! \rho_3)^2\! -\!\rho_1(1\! -\!s_1^2\beta^2 )
\! -\!\rho_2 (1\! -\!\beta^2 s_2^2)
\! -\!\rho_3(s_3^2\! -\!\beta^2)\! +\!i\epsilon\\
P_3\! &=\!(\rho_1 \! +\!\rho_2 \! -\! s_3\rho_3)^2\! -\!\beta^2 (s_1
\rho_1 \! +\!s_2\rho_2 \! +\! \rho_3)^2\! -\!\rho_1(1\! -\!s_1^2\beta^2 )
\! -\!\rho_2 (1\! -\!\beta^2 s_2^2)
\! -\!\rho_3(s_3^2\! -\!\beta^2)\! +\!i\epsilon\\
P_4\! &=\!(s_1\rho_1 \! -\!s_2\rho_2 \! -\!\rho_3)^2\! -\!\beta^2 (
\rho_1 \! -\!\rho_2 \! +\!s_3 \rho_3)^2\! -\!\rho_1( s_1^2\! -\!\beta^2 )
\! -\!\rho_2 (s_2^2\! -\!\beta^2 )
\! -\!\rho_3(1\! -\! s_3^2\beta^2)\! +\!i\epsilon\\
P_5\! &=\!(s_1\rho_1 \! +\!s_2\rho_2 \! +\!\rho_3)^2\! -\!\beta^2 (
\rho_1 \! +\!\rho_2 \! -\!s_3 \rho_3)^2\! -\!\rho_1( s_1^2\! -\!\beta^2 )
 \! -\!\rho_2 (s_2^2\! -\!\beta^2 )
\! -\!\rho_3(1\! -\! s_3^2\beta^2)\! +\!i\epsilon\\
P_6\! &=\!(s_1\rho_1 \! +\!s_2\rho_2 \! -\!\rho_3)^2\! -\!\beta^2 (
\rho_1 \! +\!\rho_2 \! +\!s_3 \rho_3)^2\! -\!\rho_1( s_1^2\! -\!\beta^2 )
\! -\!\rho_2 (s_2^2\! -\!\beta^2 )
\! -\!\rho_3(1\! -\! s_3^2\beta^2)\! +\!i\epsilon
\end{split}
\end{align}

\normalsize

As already anticipated in the main text, the above integrals have not been
evaluated exactly. However, the leading power of $T$ is just the factor
$T^{6-4\omega}$ contained in the overall constant, as can be easily realized by
the fact that the integrals $I_1 ,\cdots ,I_4$ are finite when evaluated for
$\beta=0$. In turn, only $I_1 (\beta^2=0)$ and $I_3 (\beta^2=0) $ can be
evaluated analytically, whereas for $I_2(\beta^2=0)$ and $I_4(\beta^2=0)$
we have only an expansion around $\omega=1$ that, however, is just what
we need. The results are

\footnotesize

\begin{equation}
\begin{split}
I_1(\beta^2=0)&={\Gamma(2\omega-2)\over
3-2\omega}\left[{\Gamma(\omega)\Gamma(\omega+1)\over \omega
(\omega-2)^2\Gamma(2\omega-1)}-{\pi\over
(2\omega-4)\sin(\pi\omega)}\right]\\
I_2(\beta^2=0)&=-{1\over4(\omega-1)^2}+ {\gamma-1/2\over
2(\omega-1)}\\
&+{17\over 4}+{1\over2}\gamma(1-\gamma) + {7\over 24}\pi^2 - \pi^2 \log2 -
{3\over 2}\zeta(3) +{\cal O}(\omega -1)\\
I_3(\beta^2=0)&=
{2\over\Gamma(5-2\omega)}\left[\Gamma(2\omega-2)\Gamma(1-\omega)
\Gamma(3-\omega)+{\pi\over\sin(\pi\omega)}\times\right.\\
&\left.\sum_{n=1}^\infty {1\over n!}\left({\Gamma
(2\omega-2+n)\Gamma(3+n-\omega) \over
(2n+1)\Gamma(n+\omega)}-{\Gamma(n+\omega-1)\Gamma(4-2\omega +n)
\over(2n+3-2\omega)\Gamma(2-\omega+n)}\right)\right]\\
I_4(\beta^2=0)&=-{1\over 2(\omega-1)^2} +{\gamma -4\over (\omega
-1)}\\
&+8\gamma -\gamma^2 - 22 +{\pi^2\over 12} +\pi^2\log2 +{3\over 2}\zeta(3)
+{\cal O} (\omega -1) \ .\\
\end{split} 
\end{equation}

\normalsize

\bigskip
By expanding $I_1$ and $I_3$ one can easily get eq. (\ref{ragnilaurent}).

\chapter{Crossed Diagrams in CPV Formulation}
\label{appcrossedcpv}
\markboth{Appendix \ref{appcrossedcpv}}{Crossed Diagrams in CPV Formulation}

\noindent
In order to understand why crossed diagrams cannot contribute in the
CPV case, we first exhibit the quantities $E^{CPV}_{ij}(s,t)$.
Only two of them are independent, thanks to eq. (\ref{syml}), and
different from zero: 
\begin{equation}
E^{CPV}_{12}(s,t)={{iL^2}\over 2}(1+t)\delta(1-s-\beta (1+t)),
\end{equation}
and 
\begin{equation}
E^{CPV}_{13}(s,t)={{iL^2}\over {\beta}}\delta(s+t+2\beta),
\end{equation}
(we are considering the case $\beta<1$).

Let us look at the first diagram in Fig.\protect\ref{fig4}; its contribution would be
$C^{CPV}_{(13)(13)}$, according to the notation developed in eq. 
(\ref{doppiml}). In this case the integration domain would be
constrained by the product $\delta(s+t+2\beta) \ \delta(u+v+2\beta)$,
with the conditions $t>u$ and $s>v$, to produce the crossing.
These conditions clearly cannot be fulfilled.

Another independent possibility would be $C^{CPV}_{(12)(13)}$.
The constraint now would be given by 
$\delta(1-s-\beta (1+t)) \ \delta(u+v+2\beta)$
with the crossing condition $u>s$, which is clearly impossible.

Finally $C^{CPV}_{(12)(12)}$ would be affected by the constraint
$\delta(1-s-\beta (1+t)) \ \delta(1-u-\beta(1+v))$ with the
conditions $s>u$ and $t>v$, which again are clearly impossible.

In higher orders the argument can be repeated considering the
propagators pairwise. On the other hand 
the conclusion on the vanishing of crossed diagrams 
would become immediately apparent in a graphical picture.

Therefore only planar diagrams survive, both in the Abelian and in the 
non-Abelian case. But, for planar diagrams, the only difference between the
two cases
is the appearance in the latter of the Casimir constant $C_F$. Hence
the Abelian exponentiation theorem continues to hold, leading to 
eq. (\ref{wlcpv}) (see eqs. (\ref{w2}), (\ref{abexp}) and (\ref{g2cpv})).

\chapter{Crossed Diagrams in ML Formulation ($D=2$)}
\label{appml}
\markboth{Appendix \ref{appml}}{Crossed Diagrams in ML Formulation ($D=2$)}

\noindent
In this appendix we shall give the main sketch for the computation
of the independent diagrams $C^{ML}_{(ij)(kl)}$ needed to 
derive the ${\cal O}(g^4)$ term in the perturbative expansion of the Wilson
loop ${\cal W}_\gamma^{ML}(L,T)$ in the causal formulation of the light-cone
gauge. As already explained in the main text, we can restrict ourselves to
the maximally non-Abelian diagrams, namely those providing a $C_FC_A$ factor.
Such diagrams are those in which the position of the
propagators is crossed, and there are 35 topologically inequivalent diagrams of
this type. However, thanks to the symmetry relations (\ref{syml}), the number of
independent diagrams to be evaluated is 19  (see eq. (\ref{relationsml})).

We first need the $E^{ML}_{ij}(t,s)$ functions defined in eq. (\ref{e}) 
that are appropriate
to the present case: substituting the parameterization of the path (\ref{path})
and the propagator in the causal formulation (\ref{propml}) in eq. (\ref{e}),
we can derive all the functions  $E^{ML}_{ij}(t,s)$. They are given by
\begin{eqnarray}
 E^{ML}_{11}(t,s)\;=\;E^{ML}_{33}(t,s)&=&-\frac{L^2}{4\pi\b^2}\nonumber\\
  E^{ML}_{22}(t,s)\;=\;E^{ML}_{44}(t,s)&=&\frac{L^2}{4\pi}\nonumber\\
  E^{ML}_{12}(t,s)\;=\;E^{ML}_{34}(t,s)&=&\frac{L^2}{4\pi\b}\;
                \frac{1-t+\b(1+s)}{1-t-\b(1+s)-i\e}\nonumber\\
  E^{ML}_{23}(t,s)\;=\;E^{ML}_{41}(t,s)&=&\frac{L^2}{4\pi\b}\;
                \frac{\b(1-t)-(1+s)}{\b(1-t)+1+s}\nonumber\\
  E^{ML}_{13}(t,s)&=&\frac{L^2}{4\pi\b^2}\;
                \frac{t+s-2\b}{t+s+2\b+i\e}\nonumber\\
  E^{ML}_{24}(t,s)&=&-\frac{L^2}{4\pi}\;
                \frac{\b t+\b s+2}{\b t+\b s-2}
\label{eml}
\end{eqnarray}
where,  $\beta=L/T$ and the symmetry relations  (\ref{syml}) have been
taken into account. The position  
(and the appearance) of poles in the above functions
 clearly
depends on the magnitude of $\beta$. Being interested in the large--$T$ limit,
we shall always consider the domain $\beta <1$. Consequently, the 
functions $ E^{ML}_{23}(t,s)$, $E^{ML}_{41}(t,s)$ and $ E^{ML}_{24}(t,s)$ 
do not present poles: this is the reason why in
eq. (\ref{eml}) we omitted the  prescription for those functions as
irrelevant (to this purpose, remember that $s,t\in[-1,1]$, 
see eq. (\ref{path})).

The diagrams $C^{ML}_{(ij)(kl)}$ are then defined in eq. (\ref{doppiml}) as multiple
integrals of functions $E^{ML}_{ij}(s,t)$. The notation is such that $C^{ML}_{(ij)(kl)}$
denotes the diagram with two {\it crossed} propagators, the first joining the
segments $(\gamma_i , \gamma_j)$ and the second joining the segments
$(\gamma_k , \gamma_l)$. 
 Once one diagram $C^{ML}_{(ij)(kl)}$ is evaluated, its value
has to be multiplied by a factor 8, which is the number of permutations 
of the indices $(ij)(kl)$ that maintains the position of the two
propagators crossed: 
this is a consequence of the first equation in (\ref{syml}). More
explicitly, this means 
\begin{equation}
C^{ML}_{( i j )( k l )}\! =\!
C^{ML}_{( ji )( kl )} \! = \!
C^{ML}_{( ij )( lk )}\! = \!
C^{ML}_{( ji )( lk )}\! = \!
C^{ML}_{( kl )( ij )}\! = \!
C^{ML}_{( lk )( ij )}\! = \!
C^{ML}_{( kl )( ji )}\! = \!
C^{ML}_{( lk )( ji )}\,.
\label{c}
\end{equation}
 To preserve
crossing,   the integration extrema have to be carefully chosen, and the
integration variables   $t,s,u,v$ have to be suitably nested.
Just as an example,  in the  diagram $C_{(11)(11)}$ the integration variables 
have to be such that $1>v>s>u>t>-1$ (see Fig. \protect\ref{fig3}
at page~\pageref{fig3}). Consequently, once
$t\in [-1,1]$, all the other integration extrema are fixed by the nesting, i.e.
$u\in [t,1]$, 
 $s\in [u,1]$,  $v\in [s,1]$. 

In the following calculation, we shall omit, for brevity, the factor $L^2/4\pi$,
which is common to all the propagators (\ref{eml}), defining $${\cal
E}^{ML}_{(ij)(kl)}(t,s)=(4\pi/L^2)E^{ML}_{(ij)(kl)}(t,s).$$ The corresponding diagrams
will obviously rescale by a factor $(L^2/4\pi)^2$, and will be denoted by
${\cal C}^{ML}_{(ij)(kl)}$, namely 
${\cal C}^{ML}_{(ij)(kl)}=(4\pi/L^2)^2C^{ML}_{(ij)(kl)}$.

Although in principle the evaluation of the 19 independent (rescaled) diagrams
is now clear, the practical calculation is rather cumbersome. We shall list here
the final results.   

\begin{equation}
{\cal C}^{ML}_{(\! 1 \! 1 \!)(\! 1 \! 1 \!)}=\!\!
\int_{-1}^1\dif t \int_t^1\dif u \int_u^1\dif s
                  \int_s^1\dif v\;{\cal E}^{ML}_{11}(t,s)\;{\cal E}^{ML}_{11}(u,v)
              =\frac{2}{3\b^4}\;\;\label{first}
\end{equation}

\begin{equation}
{\cal C}^{ML}_{(\! 2 \! 2 \!)(\! 2 \! 2 \!)}=\!\!
\int_{-1}^1\dif t \int_t^1\dif u \int_u^1\dif s
                  \int_s^1\dif v\;{\cal E}^{ML}_{22}(t,s)\;{\cal E}^{ML}_{22}(u,v)
              =\frac{2}{3}\;\;
\end{equation}

\begin{multline}
{\cal C}^{ML}_{(\! 1 \! 1 \!)(\! 1 \! 3 \!)}=\!\!
\int_{-1}^1\dif u \int_{-1}^1\dif v\;{\cal E}^{ML}_{13}(u,v)
                  \int_{-1}^u\dif t \int_u^1\dif s\;{\cal E}^{ML}_{11}(t,s)\\
 =-\frac{8}{3\b^4}+\frac{64}{3\b^2}+\left(-\frac{16}{3\b^3}
   +\frac{16}{\b}-\frac{32}{3}\right)i\pi+\frac{64}{3}\ln(\b)+\\
            +\left(\frac{16}{3\b^3}
   -\frac{16}{\b}-\frac{32}{3}\right)\ln(1+\b)+\left(-\frac{16}{3\b^3}
   +\frac{16}{\b}-\frac{32}{3}\right)\ln(1-\b)
\end{multline}

\begin{multline}
{\cal C}^{ML}_{(\! 2 \! 2 \!)(\! 2 \! 4 \!)}=\!\!
\int_{-1}^1\dif u \int_{-1}^1\dif v\;{\cal E}^{ML}_{24}(u,v)
                  \int_{-1}^u\dif t \int_u^1\dif s\;{\cal E}^{ML}_{22}(t,s)\\
             =\frac{64}{3\b^2}-\frac{8}{3}+\\
+\left(-\frac{32}{3\b^4}
   -\frac{16}{\b^3}+\frac{16}{3\b}\right)\ln(1+\b)
             +\left(-\frac{32}{3\b^4}
   +\frac{16}{\b^3}-\frac{16}{3\b}\right)\ln(1-\b)\;\;
\end{multline}

\begin{multline}
{\cal C}^{ML}_{(\! 1 \! 1 \!)(\! 1 \! 2 \!)}=\!\!
\int_{-1}^1\dif u \int_{-1}^1\dif v\;{\cal E}^{ML}_{12}(u,v)
                  \int_{-1}^u\dif t \int_u^1\dif s\;{\cal E}^{ML}_{11}(t,s)\\
           =-\frac{20}{3\b^2}+\frac{8}{\b}
               +\left(-\frac{32}{3\b}+8\right)i\pi
               +\left(\frac{32}{3\b}-8\right)\ln(\b)+\\
            +\left(\frac{8}{3\b^4}
               -\frac{32}{3\b}+8\right)\ln(1-\b)\;\;
\end{multline}

\begin{multline}
{\cal C}^{ML}_{(\! 2 \! 2 \!)(\! 2 \! 3 \!)}=\!\!
\int_{-1}^1\dif u \int_{-1}^1\dif v\;{\cal E}^{ML}_{23}(u,v)
                  \int_{-1}^u\dif t \int_u^1\dif s\;{\cal E}^{ML}_{22}(t,s)\\
              =-\frac{8}{\b^3}-\frac{20}{3\b^2}-\frac{8}{3}\ln(\b)+
                 \left(\frac{8}{\b^4}+\frac{32}{3\b^3}+\frac{8}{3}
                  \right)\ln(1+\b)\;\;
\end{multline}

\begin{multline}
{\cal C}^{ML}_{(\! 1 \! 1 \!)(\! 1 \! 4 \!)}=\!\!
\int_{-1}^1\dif u \int_{-1}^1\dif v\;{\cal E}^{ML}_{14}(u,v)
                      \int_{-1}^u\dif t \int_u^1\dif s\;{\cal E}^{ML}_{11}(t,s)\\
               =-\frac{20}{3\b^2}-\frac{8}{\b}-\left(\frac{32}{3\b}+8
                      \right)\ln(\b)+\left(\frac{8}{3\b^4}+\frac{32}{3\b}
                      +8\right)\ln(1+\b)\;\;
\end{multline}

\begin{multline}
{\cal C}^{ML}_{(\! 2 \! 2 \!)(\! 1 \! 2 \!)}=\!\!
\int_{-1}^1\dif u \int_{-1}^1\dif v\;{\cal E}^{ML}_{21}(u,v)
                  \int_{-1}^u\dif t \int_u^1\dif s\;{\cal E}^{ML}_{22}(t,s)\\
              =\frac{8}{\b^3}-\frac{20}{3\b^2}
                  +\frac{8}{3}\,i\pi-\frac{8}{3}\ln(\b)
                  +\left(\frac{8}{\b^4}-\frac{32}{3\b^3}
                  +\frac{8}{3}\right)\ln(1-\b)\;\;
\end{multline}

\footnotesize

\begin{multline}
{\cal C}^{ML}_{(\! 1 \! 2 \!)(\! 1 \! 3 \!)}=\!\!
\int_{-1}^1\dif u \int_{-1}^1\dif v\;{\cal E}^{ML}_{12}(u,v)
                  \int_u^1\dif t \int_{-1}^1\dif s\;{\cal E}^{ML}_{13}(t,s)\\
  =\left(\frac{16\pi^2}{9}-\frac{8}{3}\right)\frac{1}{\b^3}
   +\left(\frac{16\pi^2}{3} -16\pi i\right)\frac{1}{\b^2}+\\
      +\left(-\frac{16\pi^2}{3} +\frac{16\pi}{3}\,i
   \right)\frac{1}{\b}-16\pi i
       +\left(\frac{32}{3\b^2}+\frac{16}{3\b}
   +32+32\pi i\right)\ln(\b)+\\
      +\left[\frac{16}{3\b^2}-\frac{32}{3\b} -16
   +\bigg(\frac{32\pi}{3\b^3}
   +\frac{32\pi}{\b^2}-\frac{64\pi}{3}\bigg)i\right]\ln(1+\b)
-32\ln^2(\b)+\\
      +\left[-\frac{32}{3\b^4}+\frac{16}{\b^3}
   +\frac{16}{3\b^2}+\frac{16}{3\b}-16+\!\bigg(\!\!-\frac{64\pi}{3\b^3}
   +\frac{32\pi}{\b^2}-\frac{32\pi}{3}\bigg)i\right]\ln(1-\b)+\\
      +\left[\frac{16}{3\b^3}+\frac{16}{\b^2}
   +\frac{32}{\b}+\frac{32}{3}\right]\ln^2(1+\b)
       +\left[-\frac{64}{3\b^3}+\frac{32}{\b^2}-\frac{32}{3}\right]
   \ln^2(1-\b)+\\
      +\left[-\frac{32}{3\b^3}-\frac{32}{\b^2}-\frac{32}{\b}+
   \frac{32}{3}\right]\ln(\b)\ln(1+\b)+\\
      +\left[-\frac{32}{3\b^3}-\frac{64}{\b^2}+\frac{32}{\b}
   +\frac{160}{3}\right]\ln(\b)\ln(1-\b)+\\
      +\left[\frac{32}{\b^3}
   +\frac{32}{\b^2}-\frac{32}{\b}-32\right]\ln(1+\b)\ln(1-\b)+\\
      +\left[-\frac{32}{3\b^3}-\frac{32}{\b^2}\right]\dil{\b}
       +\left[\frac{32}{3\b^3}+\frac{32}{\b^2}
   +\frac{32}{\b}+\frac{32}{3}\right]\dil{\frac{\b}{1+\b}}-\\
      -\,\frac{32}{3}\;\dil{1-\frac{1}{\b}}
       +\left[-\frac{32}{3\b^3}-\frac{32}{\b^2}\right]
                                          \dil{\frac{1}{1+\b}}\;\;
\end{multline}

\begin{multline}
{\cal C}^{ML}_{(\! 2 \! 3 \!)(\! 2 \! 4 \!)}=\!\!
\int_{-1}^1\dif u \int_{-1}^1\dif v\;{\cal E}^{ML}_{23}(u,v)
  \int_u^1\dif t \int_{-1}^1\dif s\;{\cal E}^{ML}_{24}(t,s)\\
     =-\,\frac{16\pi^2}{3\b^3}+\frac{16\pi^2}{3\b^2}+
  \left(\frac{8}{3}-\frac{16\pi^2}{9}\right)\frac{1}{\b}
     +\left[-\frac{64}{3\b^2}+\frac{16}{\b}+\frac{32}{3}\right]
  \ln(\b)+\\
    +\left[-\frac{16}{\b^4}-\frac{16}{3\b^3}+\frac{16}{3\b^2}
  -\frac{16}{\b}-\frac{32}{3}\right]      \ln(1+\b)
+\left[-\frac{32}{\b^2}-\frac{64}{3\b}\right]    
  \ln(\b)\ln(1+\b)+\\
 +\left[-\frac{16}{\b^4}+\frac{32}{3\b^3}+\frac{16}{3\b^2}
  \right]      \ln(1-\b)
     +\left[-\frac{16}{3\b^4}+\frac{32}{\b^2}+\frac{64}{3\b}
  \right]      \ln^2(1+\b)+\\
    +\left[\frac{16}{3\b^4}-\frac{16}{\b^3}+\frac{16}{\b^2}
  -\frac{16}{3\b}\right]     \ln^2(1-\b)  
     +\left[-\frac{64}{\b^2}+\frac{128}{3\b}\right]
  \ln(\b)\ln(1-\b)+\\
    +\left[-\frac{32}{\b^4}+\frac{32}{\b^3}+\frac{32}{\b^2}
  -\frac{32}{\b}\right]    \ln(1+\b)\ln(1-\b)+\\
    +\left[-\frac{32}{3\b^4}+\frac{32}{\b^3}-\frac{32}{\b^2}
  +\frac{32}{3\b}\right]   \dil{1-\b}
     +\left[\frac{32}{\b^2}-\frac{32}{3\b}\right] 
  \dil{1-\frac{1}{\b}}+\\
    +\,\frac{32}{3\b^4}\; \dil{\frac{1}{1+\b}}
     +\left[-\frac{32}{\b^2}+\frac{32}{3\b}\right]
  \dil{-\frac{1}{\b}}\;\;
\end{multline}

\begin{multline}
{\cal C}^{ML}_{(\! 1 \! 3 \!)(\! 1 \! 4 \!)}=\!\!
\int_{-1}^1\dif u \int_{-1}^1\dif v\;{\cal E}^{ML}_{14}(u,v)
  \int_{-1}^u\dif t \int_{-1}^1\dif s\;{\cal E}^{ML}_{13}(t,s)\\
     =\left(\frac{8}{3}+\frac{16\pi^2}{9}\right)\frac{1}{\b^3}
  +\left(-\frac{16\pi^2}{3}+\frac{16\pi}{3}\,i\right)\frac{1}{\b^2}
     +\left(\frac{16\pi^2}{3}+\frac{32\pi}{3}\,i\right)\frac{1}{\b}
  -\frac{32\pi^2}{9}+\\
    +\left[\frac{32}{3\b^2}-\frac{16}{3\b}+32+\frac{64\pi}{3}\,i
  \right]\ln(\b)-16\pi i  -\frac{80}{3}\ln^2(\b)+\\
    +\left[-\frac{32}{3\b^4}\!-\!\frac{16}{\b^3}\!+\!\frac{16}{3\b^2}\!
  -\!\frac{16}{3\b}\!-\!16\!+\!\!\bigg(\!\!\!-\!\frac{32\pi}{\b^3}\!+\!\frac{32\pi}{\b^2}\!
  +\!\frac{32\pi}{\b}\!-\!32\pi\bigg)i\right]\!\ln(1+\b)+\\
    +\left[\frac{16}{3\b^2}+\frac{32}{3\b}-16+\bigg(
  -\frac{32\pi}{3\b^3}+\frac{32\pi}{\b^2}-\frac{32\pi}{\b}
  +\frac{32\pi}{3}\bigg)i\right]\ln(1-\b)-\\
     +\left[\frac{64}{3\b^3}+\frac{32}{\b^2}-\frac{16}{3}
  \right]\ln^2(1+\b)
     +\left[-\frac{32}{3\b^3}+\frac{32}{\b^2}-\frac{32}{\b}
  +\frac{32}{3}\right]\ln^2(1-\b)+\\
    +\left[\frac{32}{3\b^3}-\frac{64}{\b^2}-\frac{32}{\b}
  +\frac{128}{3}\right]\ln(\b)\ln(1+\b)+\\
    +\left[\frac{32}{3\b^3}-\frac{32}{\b^2}+\frac{32}{\b}
  +\frac{32}{3}\right]\ln(\b)\ln(1-\b)+\\
    +\left[-\frac{32}{\b^3}+\frac{32}{\b^2}+\frac{32}{\b}
  -32\right]\ln(1+\b)\ln(1-\b)+\\
    +\left[\frac{32}{3\b^3}-\frac{32}{\b^2}\right]\dil{-\b}
     +\,\frac{32}{3}\;\dil{\frac{\b}{1+\b}}+\\
    +\left[-\frac{32}{3\b^3}+\frac{32}{\b^2}-\frac{32}{\b}
  +\frac{32}{3}\right]\dil{-\frac{\b}{1-\b}}
     +\left[-\frac{32}{3\b^3}+\frac{32}{\b^2}\right]\dil{1-\b}\;\;
\end{multline}

\begin{multline}
{\cal C}^{ML}_{(\! 1 \! 2 \!)(\! 2 \! 4 \!)}=\!\!
\int_{-1}^1\dif u \int_{-1}^1\dif v\;{\cal E}^{ML}_{21}(u,v)
  \int_{-1}^u\dif t \int_{-1}^1\dif s\;{\cal E}^{ML}_{24}(t,s)\\
    =-\,\frac{16\pi^2}{3\b^3}+\left(-\frac{16\pi^2}{3}
  +\frac{64\pi}{3}\,i\right)\frac{1}{\b^2}
     +\left(-\frac{8}{3}-\frac{16\pi^2}{9}+16\pi i\right)
  \frac{1}{\b}-\frac{32\pi}{3}\,i+\\
+\left[-\frac{16}{\b^4}-\frac{32}{3\b^3}+\frac{16}{3\b^2}
  +\bigg(-\frac{32\pi}{3\b^4}-\frac{32\pi}{\b^3}+\frac{64\pi}{3\b}
  \bigg)i\right]\ln(1+\b)+\\
    +\left[-\frac{16}{\b^4}+\frac{16}{3\b^3}+\frac{16}{3\b^2}
  +\frac{16}{\b}-\frac{32}{3}\!+\!\bigg(\!\!-\frac{32\pi}{3\b^4}
  +\frac{32\pi}{\b^2}-\frac{64\pi}{3\b}\bigg)i\right]\ln(1-\b)+\\
    +\left[\frac{32}{3\b^4}+\frac{32}{\b^3}+\frac{16}{\b^2}
  +\frac{16}{3\b}\right]\ln^2(1+\b)  +\left[-\frac{64}{3\b^2}
   -\frac{16}{\b}+\frac{32}{3}\right]
  \ln(\b)+\\
    +\left[-\frac{32}{3\b^4}+\frac{32}{\b^2}-\frac{64}{3\b}
  \right]\ln^2(1-\b)
     +\left[-\frac{32}{\b^2}-\frac{32}{\b}\right]\ln(\b)\ln(1+\b)+\\
   +\left[-\frac{32}{\b^2}+\frac{64}{3\b}\right]\ln(\b)\ln(1-\b)
     +\left[-\frac{32}{\b^4}-\frac{32}{\b^3}+\frac{32}{\b^2}
  +\frac{32}{\b}\right]\ln(1+\b)\ln(1-\b)+\\
    +\left[\frac{32}{\b^2}+\frac{32}{3\b}\right]\dil{\b}
     +\left[-\frac{32}{\b^2}-\frac{32}{3\b}\right]
  \dil{\frac{\b}{1+\b}}-
     \frac{32}{3\b^4}\;\dil{1-\b}+\\
    +\left[\frac{32}{3\b^4}+\frac{32}{\b^3}+\frac{32}{\b^2}
  +\frac{32}{3\b}\right]\dil{\frac{1}{1+\b}}\;\;
\end{multline}

\normalsize

\begin{multline}
{\cal C}^{ML}_{(\! 1 \! 2 \!)(\! 1 \! 4 \!)}=\!\!
\int_{-1}^1\dif u \int_{-1}^1\dif v\;{\cal E}^{ML}_{12}(u,v)
    \int_u^1\dif t \int_{-1}^1\dif s\;{\cal E}^{ML}_{14}(t,s)\\
    =\frac{4\pi^2}{9\b^4}+\left(\frac{8}{3}-\frac{16\pi^2}{3}
   +\frac{8\pi}{3}\,i\right)\frac{1}{\b^2}
     +\frac{16\pi}{\b}\,i+\frac{8\pi^2}{3}+\frac{28\pi}{3}\,i+
     \left[-\frac{16}{3\b^2}-\frac{56}{3}\right]\ln(\b)+\\
    +\left[\frac{8}{3\b^3}-\frac{44}{3\b^2}-\frac{8}{\b}
   +\frac{28}{3}+\bigg(\frac{8\pi}{3\b^4}-\frac{32\pi}{\b^2}
   -\frac{64\pi}{3\b}+8\pi\bigg)i\right]\ln(1+\b)+\\
    +\left[-\frac{8}{3\b^3}-\frac{44}{3\b^2}+\frac{8}{\b}
   +\frac{28}{3}+\bigg(\frac{8\pi}{3\b^4}-\frac{16\pi}{\b^2}
   +\frac{64\pi}{3\b}-8\pi\bigg)i\right]\ln(1-\b)+\\
+\left[-\frac{8}{3\b^4}+\frac{16}{\b^2}+\frac{128}{3\b}\right]
   \ln(\b)\ln(1+\b)+ 
\left[\frac{4}{3\b^4}-\frac{64}{3\b}-8\right]\ln^2(1+\b)+\\
    +\left[-\frac{16}{3\b^4}+\frac{48}{\b^2}-\frac{128}{3\b}
   \right]\ln(\b)\ln(1-\b)+
\left[\frac{8}{3\b^4}-\frac{16}{\b^2}+\frac{64}{3\b}-8
   \right]\ln^2(1-\b)+\\ 
    +\left[\frac{16}{\b^4}-\frac{32}{\b^2}+16\right]
   \ln(1+\b)\ln(1-\b)
     +\left[-\frac{8}{3\b^4}+\frac{32}{\b^2}\right]\dil{\b}+\\
    +\left[\frac{8}{3\b^4}-\frac{16}{\b^2}-\frac{64}{3\b}-8\right]
   \dil{\frac{\b}{1+\b}}+
\left[-\frac{8}{3\b^4}+\frac{32}{\b^2}\right]
   \dil{\frac{1}{1+\b}}+\\
    +\left[\frac{8}{3\b^4}-\frac{16}{\b^2}+\frac{64}{3\b}-8\right]
   \dil{-\frac{\b}{1-\b}} \;\;
\end{multline}

\begin{multline}
{\cal C}^{ML}_{(\! 1 \! 2 \!)(\! 2 \! 3 \!)}=\!\!
\int_{-1}^1\dif u \int_{-1}^1\dif v\;{\cal E}^{ML}_{23}(u,v)
                   \int_u^1\dif t \int_{-1}^1\dif s\;{\cal E}^{ML}_{21}(t,s)\\
    =\left(\frac{64\pi^2}{9}-8\pi i\right)\frac{1}{\b^3}
  +\left(\frac{8}{3}-\frac{52\pi}{3}\,i\right)\frac{1}{\b^2}
    -\frac{8\pi}{3\b}\,i+\frac{4\pi^2}{9}+
 \frac{40}{3}\ln^2(\b)+\\
   +\left[\frac{28}{3\b^4}-\frac{8}{\b^3}-\frac{44}{3\b^2}
  +\frac{8}{3\b}+\bigg(\frac{8\pi}{\b^4}+\frac{64\pi}{3\b^3}
  +\frac{40\pi}{3}\bigg)i\right]\ln(1+\b)+\\
   +\left[\frac{28}{3\b^4}+\frac{8}{\b^3}-\frac{44}{3\b^2}
  -\frac{8}{3\b}\right]\ln(1-\b)+
\left[\frac{104}{3\b^2}-\frac{40\pi}{3}\,i\right]\ln(\b)+\\
  +\left[-\frac{8}{\b^4}-\frac{64}{3\b^3}+\frac{4}{3}\right]
  \ln^2(1+\b)
    +\left[-\frac{4}{\b^4}+\frac{32}{3\b^3}-\frac{8}{\b^2}
  +\frac{4}{3}\right]\ln^2(1-\b)+\\
   +\left[\frac{16}{\b^2}-16\right]\ln(\b)\ln(1+\b)
    +\left[\frac{16}{\b^2}-16\right]\ln(\b)\ln(1-\b)+\\
   +\left[\frac{16}{\b^4}-\frac{32}{\b^2}+16\right]
  \ln(1+\b)\ln(1-\b)
    +\left[-\frac{32}{\b^2}+\frac{8}{3}\right]\dil{\b}+\\
  +\left[\frac{32}{\b^2}-\frac{8}{3}\right]\dil{\frac{\b}{1+\b}}
    +\left[\frac{8}{\b^4}-\frac{64}{3\b^3}+\frac{16}{\b^2}
  -\frac{8}{3}\right]\dil{1-\b}+\\
   +\left[-\frac{8}{\b^4}-\frac{64}{3\b^3}-\frac{16}{\b^2}
  +\frac{8}{3}\right]\dil{\frac{1}{1+\b}}\;\;
\end{multline}

\begin{multline}
{\cal C}^{ML}_{(\! 1 \! 2 \!)(\! 1 \! 2 \!)}=\!\!
\int_{-1}^1\dif u \int_{-1}^1\dif v\;{\cal E}^{ML}_{12}(u,v)
                   \int_u^1\dif t \int_v^1\dif s\;{\cal E}^{ML}_{12}(t,s)\\
     =-\frac{4\pi^2}{3\b^4}+\left(-\frac{4}{3}+8\pi i
   \right)\frac{1}{\b^3}+\left(-\frac{4}{3}+4\pi i\right)
   \frac{1}{\b^2}
     -\left(\frac{4}{3}+\frac{8\pi i}{3}\right)\frac{1}{\b}
   -\frac{4\pi^2}{3}-\frac{28\pi}{3}\,i-\\
     -\left[\frac{8}{\b^3}+\frac{4}{\b^2}-\frac{8}{3\b}
   -\frac{28}{3}+8\pi i\right]\ln(\b)+
     8\ln^2(\b)+
     \left[\frac{8}{\b^4}-\frac{16}{\b^2}+8\right]\ln^2(1-\b)+\\
     +\left[-\frac{28}{3\b^4}+\frac{16}{3\b^3}+\frac{8}{\b^2}
   +\frac{16}{3\b}-\frac{28}{3}+\bigg(\frac{8\pi}{\b^4}
   -\frac{16\pi}{\b^2}+8\pi\bigg)i\right]\ln(1-\b)+\\
    +\left[\frac{16}{\b^2}-16\right]\ln(\b)\ln(1-\b)+
     \frac{8}{\b^4}\;\dil{1-\b}+
     8\,\dil{1-\frac{1}{\b}}\;\;
\end{multline}

\begin{multline}
{\cal C}^{ML}_{(\! 2 \! 3 \!)(\! 2 \! 3 \!)}=\!\!
\int_{-1}^1\dif u \int_{-1}^1\dif v\;{\cal E}^{ML}_{23}(u,v)
                   \int_u^1\dif t \int_v^1\dif s\;{\cal E}^{ML}_{23}(t,s)\\
     =\frac{4\pi^2}{3\b^4}+\frac{4}{3\b^3}-\frac{4}{3\b^2}
   +\frac{4}{3\b}+\frac{4\pi^2}{3}
     +\left[\frac{8}{\b^3}-\frac{4}{\b^2}-\frac{8}{3\b}+\frac{28}{3}
   \right]\ln(\b)+\\
    +\left[-\frac{28}{3\b^4}-\frac{16}{3\b^3}+\frac{8}{\b^2}
   -\frac{16}{3\b}-\frac{28}{3}\right]\ln(1+\b)+
     4\ln^2(\b)+\\
   +\left[\frac{4}{\b^4}-\frac{16}{\b^2}+4\right]\ln^2(1+\b)
     +\left[\frac{16}{\b^2}-8\right]\ln(\b)\ln(1+\b)-\\
    -\,8\,\dil{\frac{\b}{1+\b}}
     -\frac{8}{\b^4}\;\dil{\frac{1}{1+\b}}\;\;\label{intermediate}
\end{multline}

\begin{multline}
{\cal C}^{ML}_{(\! 1 \! 3 \!)(\! 1 \! 3 \!)}=\!\!
\int_{-1}^1\dif u \int_{-1}^1\dif v\;{\cal E}^{ML}_{13}(u,v)
                   \int_u^1\dif t \int_v^1\dif s\;{\cal E}^{ML}_{13}(t,s)\\
     =\frac{4}{\b^4}+\frac{32\pi}{3\b^3}\,i
   -\frac{32}{3\b^2}-\frac{64\pi}{\b}\,i+\frac{160\pi}{3}\,i-
     \left[\frac{320}{3}+64\pi i\right]\ln(\b)+\\
    +\left[-\frac{32}{3\b^3}-\frac{64\pi}{\b^2}\,i+\frac{64}{\b}
   +\frac{160}{3}+64\pi i\right]\ln(1+\b)+
 64\ln^2(\b)+\\
    +\left[\frac{32}{3\b^3}-\frac{64}{\b}+\frac{160}{3}
   \right]\ln(1-\b)
    +\left[\frac{64}{\b^2}-64\right]\ln(\b)\ln(1+\b)-\\
    -\,64\ln(\b)\ln(1-\b)+
     \left[-\frac{64}{\b^2}+64\right]\ln(1+\b)\ln(1-\b)+\\
    +\,\frac{64}{\b^2}\;\dil{-\b}-
    \,\frac{64}{\b^2}\;\dil{1-\b}\;\;
\end{multline}

\begin{multline}
{\cal C}^{ML}_{(\! 2 \! 4 \!)(\! 2 \! 4 \!)}=\!\!
\int_{-1}^1\dif u \int_{-1}^1\dif v\;{\cal E}^{ML}_{24}(u,v)
                   \int_u^1\dif t \int_v^1\dif s\;{\cal E}^{ML}_{24}(t,s)\\
    =-\,\frac{32}{3\b^2}+4 +\frac{32}{\b^2}\ln^2(1+\b)
   +\left[\frac{64}{\b^4}-\frac{64}{\b^2}\right]\ln(1+\b)\ln(1-\b)+\\
    +\left[\frac{160}{3\b^4}+\frac{64}{\b^3}-\frac{32}{3\b}\right]
       \ln(1+\b)+
     \left[\frac{160}{3\b^4}-\frac{64}{\b^3}+\frac{32}{3\b}\right]
       \ln(1-\b)-\\
    -\,\frac{64}{\b^2}\;\dil{\b}+
     \frac{64}{\b^2}\;\dil{\frac{\b}{1+\b}}\;\;
\end{multline}

\begin{multline}
{\cal C}^{ML}_{(\! 1 \! 3 \!)(\! 2 \! 4 \!)}=\!\!
\int_{-1}^1\dif t \int_{-1}^1\dif s\;{\cal E}^{ML}_{13}(t,s)
                   \int_{-1}^1\dif u \int_{-1}^1\dif v\;{\cal E}^{ML}_{24}(u,v)\\
     =-\,\frac{16}{\b^2}-\frac{32\pi}{\b}\,i+32\pi i-
      \,64\ln(\b)+\\
     +\left[\frac{32}{\b^4}+\frac{32}{\b^3}+\frac{32}{\b}
   +32+\bigg(\frac{64\pi}{\b^3}-\frac{64\pi}{\b}\bigg)i\right]\ln(1+\b)+\\
     +\left[\frac{32}{\b^4}-\frac{32}{\b^3}-\frac{32}{\b}
   +32+\bigg(\frac{64\pi}{\b^3}-\frac{128\pi}{\b^2}+\frac{64\pi}{\b}\bigg)i
   \right]\ln(1-\b)+\\
     +\left[-\frac{64}{\b^3}-\frac{128}{\b^2}-\frac{64}{\b}\right]
   \ln^2(1+\b)
      +\left[\frac{64}{\b^3}-\frac{128}{\b^2}+\frac{64}{\b}\right]
   \ln^2(1-\b)+\\
     +\left[\frac{128}{\b^2}+\frac{128}{\b}\right]\ln(\b)\ln(1+\b)+
    \left[\frac{128}{\b^2}-\frac{128}{\b}\right]\ln(\b)\ln(1-\b)\;\;\ .
\label{last}
\end{multline}

Some technical details on the dilogarithm function ${\rm
Li}_2(z)$ and on its analytic continuations are in order.
As is well known, ${\rm Li}_2(z)$ can be defined through its integral
representation
\begin{equation}
\dil{z}=\int_z^0\frac{\ln(1-\zeta)}{\zeta}\;\diff\zeta\;\;,
\label{dil}
\end{equation}
where the path joining $z$ and $0$ is arbitrary, provided it does not intersect 
the half-line $]1,+\infty[$~, which is the branch-cut of the integrand function.
On its branch-point the dilogarithm is finite, and takes the value
$\dil{1}=\pi^2/6$. 

If $\beta <1$, the calculation of the above diagrams 
involves dilogarithmic functions, with arguments bounded by the region $-\infty
<{\rm Re}\, z<1$. Eventually, we shall be 
interested in taking the limit $\beta \to
0$ (i.e. large $T$). The arguments of the dilogarithms arising from a first 
integrations of eqs. (\ref{first})--(\ref{last}) 
can tend to $0$, $1$ and $-\infty$
as $\beta\to 0$.
 On the other hand, the simplest expansion of the dilogarithm 
is around the point $z=0$, where a simple series representation holds
\begin{equation}
\dil{z}=\sum_{k=1}^{\infty}\frac{z^k}{k^2} \ ,\qquad \ \ |z|<1\ \ .
\label{dil2}
\end{equation}
Consequently, we need analytic continuation to convert dilogarithms with
arguments tending to $1$ and $-\infty$ into dilogarithms with arguments tending
to $0$, for $\beta\to 0$. These are given by
\begin{eqnarray}
\dil{-z}&=&-\frac{\pi^2}{6}+\frac{1}{2}\ln^2(1+z)-\ln(z)\ln(1+z)
           +\dil{\frac{1}{1+z}} ,\label{ac1}\\
\dil{z}&=&\frac{\pi^2}{6}-\ln(z)\ln(1-z)-\dil{1-z}.
\label{ac2}
\end{eqnarray}

\noindent
To get the final result for the maximally non-Abelian ${\cal O}(g^4)$ terms in
the causal formulation, we have 
\begin{enumerate}
\item to sum all the results (\ref{first})--(\ref{last}); 
the integrals from eqs.
(\ref{first}) to (\ref{intermediate}) have to be multiplied by an extra 
 factor 2 to take into account the 16 relations (\ref{relationsml});
\item to multiply the result by  $L^4/(4\pi)^2$ to take into account the rescaling
from ${\cal C}^{ML}_{(ij)(kl)}$ to ${C}^{ML}_{(ij)(kl)}$;
\item to multiply by a factor 8 to take into account permutations of indices as in
eq. (\ref{c});
\item to multiply by a factor $-C_AC_F/16$ as shown in eq. (\ref{doppiml}).
\end{enumerate}
\noindent 

Following the above points (1) to (4) one arrives at eq. (\ref{wmlg42}):
using eqs.~(\ref{ac1}), (\ref{ac2}), the dependence on the dimensionless ratio $\beta= L/T$ \emph{exactly}
cancels, leading to a pure area behavior for any value of $T$.

\chapter{Bubble Diagrams in ML Formulation}
\label{bubbleml}
\markboth{Appendix \ref{bubbleml}}{Bubble Diagrams in ML Formulation}

Writing explicitly all the possible bubble diagrams $B^{ML}_{ij}$, it is not
difficult to realize that  $B^{ML}_{11}=B^{ML}_{33}$, $B^{ML}_{22}=B^{ML}_{44}$, $B^{ML}_{12}=B^{ML}_{34}$, $B^{ML}_{14}=B^{ML}_{23}$, so that
eq.~(\ref{bub1}) follows.
The six independent bubbles, following eq.~(\ref{bubbola}), are then

\begin{align}
B^{ML}_{11}&=\int_{-1}^1\!\! ds \int_{-1}^1 \!\! dt 
{ (\delta T^4)(s-t)^2 \over [-(s-t)^2
T^2 + i\epsilon]^{2\omega -2}}\ , \\
B^{ML}_{22}&=\int_{-1}^1\!\! ds \int_{-1}^1 \!\! dt { (\delta L^4) (s-t)^2
\over [(s-t)^2
L^2 + i\epsilon]^{2\omega -2}}\ , \\
B^{ML}_{13}&=\int_{-1}^1\!\! ds \int_{-1}^1 \!\! dt 
{ (-\delta T^2) \left[T (s+t) + 2 L \right]^2 \over [-(s+t)^2
T^2 +4L^2+ i\epsilon]^{2\omega -2}},\\
B^{ML}_{24}&=\int_{-1}^1\!\! ds \int_{-1}^1 \!\! dt
{ (-\delta L^2) \left[L (s+t) - 2 T \right]^2 \over [(s+t)^2
L^2 -4T^2+ i\epsilon]^{2\omega -2}} \ ,\\
B^{ML}_{12}&=\int_{-1}^1\!\! ds \int_{-1}^1 \!\! dt { \delta L T
\left[T(1-t)-L(1+s)\right]^2\over [ L^2(s+1)^2  - T^2 (1-t)^2+ i\epsilon]^{2\omega -2}}\ ,\\
B^{ML}_{14}&=\int_{-1}^1\!\! ds \int_{-1}^1 \!\! dt { -\delta L T
\left[T(1-t)+L(1+s)\right]^2 \over [L^2(s+1)^2  - T^2 (1-t)^2+ i\epsilon]^{2\omega -2}}\ ,
\end{align}

where

\begin{equation}
\delta={g^4 C_F C_A\over 64 \pi^{2\omega}}f(\omega) \ ,
\end{equation}

with $f(\omega)$ defined in eq.~(\ref{fd}).

At this stage we could already choose $D=2$, since each bubble gives a finite
contribution, at variance with the Feynman gauge result: in so doing we will
easily recover eq.~(\ref{sestorto}). But we want to study also the behavior
of the bubble contribution for $D>2$. Again, integration can be performed
(using, when necessary, Mellin-Barnes techniques), leading to

\bigskip
\footnotesize

\begin{equation}
B^{ML}_{11}=e^{-2i\pi\omega} T^{8-4\omega}\frac{2^{7-4\omega}}{(7-4\omega)
(2-\omega)} \ ,
\end{equation}

\begin{equation}
B^{ML}_{22}=L^{8-4\omega}\frac{2^{7-4\omega}}{(7-4\omega)
(2-\omega)}\ ,
\end{equation}

\begin{multline}
B^{ML}_{13}=-T^{8-4\omega}2^{9-4\omega}\Bigg\{e^{-2i\pi\omega}\Bigg[
\frac{\beta^2}{5-4\omega}{_2F_1}(2\omega-2,2\omega-5/2;2\omega-3/2;\beta^2)
+\frac{1}{7-4\omega}\\
\times{_2F_1}(2\omega-2,2\omega-7/2;2\omega-5/2;\beta^2)-\frac{1}{2(2\omega-3)}(1-\beta^2)^{3-2\omega}
\frac{\beta^2(5-2\omega)+3-2\omega}{2\omega-4}\Bigg]\\
-\frac{i \beta^{7-4\omega}}{\Gamma(2\omega-2)}\left(\Gamma(3/2)
\Gamma(2\omega-5/2)-\frac{1}{3}\Gamma(5/2)
\Gamma(2\omega-7/2)\right)+\frac{\beta^{8-4\omega}(2\omega-5)}{2(2\omega-3)(2\omega-4)}\Bigg\},
\end{multline}

\begin{equation}
\begin{split}
B^{ML}_{24}&=-e^{-2i\pi\omega} T^{8-4\omega}2^{9-4\omega}\Bigg\{
{_2F_1}(2\omega-2,1/2;3/2;\beta^2)\beta^2 +
{_2F_1}(2\omega-2,3/2;5/2;\beta^2)\frac{\beta^4}{3}\\
&-\frac{1}{2(2\omega-3)}\left[(1-\beta^2)^{3-2\omega}\left(1+\frac{\beta^2(2\omega-3)}{2\omega-4}-\frac{1}{2\omega-4}\right)-1 +\frac{1}{2\omega-4}\right]\Bigg\},\\
\end{split}
\end{equation}

\begin{equation}
B^{ML}_{14}+B^{ML}_{12}= -\frac{2^{8-4\omega}T^{8-4\omega}}{(2\omega-3)
(2\omega-4)}\left[\beta^{8-4\omega}+e^{-2i\pi\omega}
\left(1-(1-\beta^2)^{4-2\omega}\right)\right].
\end{equation}

\normalsize

\bigskip

Summing all contributions according to eq.~(\ref{bub1}), we obtain eq.~(\ref{bub2}).

\chapter{Three Point Green Function in the Limit $D\to 2$}
\label{appgreen}
\markboth{Appendix \ref{appgreen}}{Three Point Green Function in
the Limit $D\to 2$}

In this appendix we show that the three point Green function tends to
zero when $D\to 2$. As explained in the main text, it is sufficient to prove
that the integral in
 eq. (\ref{gamma1}),
with  the constant containing the  simple zero $(\omega -1)$ factorized out, is
convergent when evaluated at $\omega =1$. Such an integral,
 after the change of variables $\alpha = \mu\xi$, $\beta=\eta (1-\mu)$
and after explicit integration over  $d\mu$, reads

\begin{multline}
\label{integral}
I=\int_0^1  d\alpha d\beta \theta (1-\alpha -\beta)\left\{{1-\alpha-\beta\over
1-\alpha}+\beta\log{(1-\beta)(1-\alpha)\over \alpha\beta}\right\}\int_0^\infty
{d\tau\over(1+\tau)}\\
\times{[(x-z)_+ +\beta\tau (x-y)_+]\over [1+\tau (\alpha+\beta)]^3}{[(y-z)_+
+\alpha\tau (y-x)_+]^2\over [-\alpha (x-z)^2 -\beta (y-z)^2 - \alpha\beta \tau
(x-y)^2 + i\epsilon]} \ ,
\end{multline}

\noindent
$\theta$ being the Heavyside function. The most delicate region of this
integral is $\alpha\sim~\beta\sim 0$, so that in order to check
convergence of eq. (\ref{integral}) we can restrict ourselves  to the case when
the curly bracket is replaced by one. After this replacement, we set
$\alpha=\rho\sigma$ and $\beta = \rho (1-\sigma)$.
In the expression obtained after this change of variables, we rescale
$\gamma=\rho\tau$ at fixed $\tau$.
The integral over the $\tau$ variable can be
factorized providing a factor $\log(1 + 1/\gamma)$.
Finally, renaming $\rho=1/\gamma$, eq. (\ref{integral}) with the curly bracket
replaced by one can be equivalently written as

\footnotesize

\begin{equation}
\label{integral2}
{\cal I}=-\int_0^1\!\! d\sigma\int_0^\infty \!\!
{d\rho\over\rho}{\log(1+\rho)\over(1+\rho)^3}{[\rho(x-z) +
(1-\sigma)(x-y)]_+[\rho (y-z) +\sigma (y-x)]_+^2\over [\rho\sigma (x-z)^2 +\rho
(1-\sigma )(y-z)^2 +\sigma(1-\sigma) (x-y)^2-i\epsilon]}\ .
\end{equation}

\normalsize

Dividing the $\rho$ integration domain as $[0,1]\cup [1,\infty)$, we split
${\cal I}$ as ${\cal I}_1 + {\cal I}_2$. In ${\cal I}_1$, $\rho\in[0,1]$ and
therefore we can use the following majorations: $\log(1+\rho)<\rho$ and
$(1+\rho)^{-3}<1$. Thus, integration   in $d\rho$ is straightforward, providing

\footnotesize

\begin{multline}
{\cal I}_1 \sim -\int_0^1 d\sigma {(x-z)_+ (y-z)^2_+\over \sigma (x-z)^2 +
(1-\sigma) (y-z)^2 -i\epsilon} \times \\
\left[ {1\over 3} + {1\over 2} (A-C) +
(B-C)^2 + B + 2AB - AC + (A-C)(B-C)^2 \log\left( {1+C\over C}\right)\right]\ ,
\end{multline}

\normalsize

where $A$, $B$ and $C$ are defined as

\begin{eqnarray}
A &=& (1-\sigma)(x-y)_+/(x-z)_+ \ ,\nonumber\\
B &=& \sigma (x-y)_+/(z-y)_+ \ ,\nonumber\\
C &=& \sigma (1-\sigma)(x-y)^2/[\sigma (x-z)^2 + (1-\sigma)(y-z)^2
-i\epsilon]\ .
\end{eqnarray}

In this form, it is manifest that integration over $\sigma $ is convergent. The
explicit result goes beyond the purpose of the paper, but it can be easily
evaluated providing combinations of rational functions, logarithms and
dilogarithms.

In  ${\cal I}_2$, the $\rho $ integration domain is $[1,\infty)$ and therefore
we can use $(1+\rho)^{-3} < \rho^{-3}$. Thus, the $\rho$ dependent part of the
integrand can be approximated by

\begin{eqnarray}
\label{alg}
&&{(\rho + A)(\rho+B)^2\over (\rho + C)}
{\log(1+\rho)\over \rho^4}=\nonumber\\
&&={\log(1+\rho)\over \rho (\rho+C)}+ {A (\rho +B)^2 + \rho (B^2 + 2 \rho B)\over
(\rho +C)\rho^3} {\log(1+\rho)\over \rho}
\end{eqnarray}

To check convergence, in the second term of the r.h.s. we can replace
$\log(1+\rho)/\rho$ by $1$. Then, integration over $\rho$ becomes
straightforward  and the second term in eq. (\ref{alg}) provides integrals over
$d\sigma$ of the same kind of those in ${\cal I}_1$, where convergence can be
easily checked. The first term in the r.h.s. of eq. (\ref{alg}) is
more delicate. Here  the majoration $\log(1+\rho)<\rho$ is too strong as it
would spoil convergence in  the $\rho$ integration. An explicit integration
over $\rho$ of this term gives

\begin{eqnarray}
\label{iuno}
&&{\cal I}_1^{first}\sim\int_0^1 d\sigma {(x-z)_+(y-z)_+^2\over \sigma (x-z)^2 +
(1-\sigma) (y-z)^2 -i\epsilon}\times\nonumber\\
&&{1\over C}\left[ {\rm Li} \left({ C\over C-1}\right) + {\rm Li} \left(- C
\right) - \log 2 \log \left( {1+C\over 1-C} \right) - {\rm Li}
\left({2C\over C-1}\right)\right]\ , \end{eqnarray}
${\rm Li} (z)$ being the dilogarithm function. Although cumbersome, integration
over $\sigma $ is finite.

\newpage
\markboth{Bibliography}{Bibliography}
\addcontentsline{toc}{chapter}{Bibliography}

\providecommand{\href}[2]{#2}\begingroup\raggedright\endgroup

\end{document}